\documentclass[aps,pre,twocolumn,superscriptaddress]{revtex4-1}

\usepackage{latexsym}
\usepackage{amsmath, amsthm, amssymb}
\usepackage{mathrsfs}
\usepackage{graphicx}
\usepackage{dcolumn}
\usepackage{expl3}
\usepackage{float}
\usepackage{natbib}
\usepackage{hyperref}
\usepackage{placeins}

\allowdisplaybreaks

\begin{document}

\title{Mesoscopic architecture enhances communication across the Macaque connectome revealing structure-function correspondence in the brain}

\author{Anand Pathak}
\affiliation{The Institute of Mathematical Sciences, CIT Campus, 
Taramani, Chennai 600113, India}
\affiliation{Homi Bhabha National Institute, Anushaktinagar, Mumbai 
400 094, India}

\author{Shakti~N.~Menon}
\affiliation{The Institute of Mathematical Sciences, CIT Campus, 
Taramani, Chennai 600113, India}

\author{Sitabhra~Sinha}
\affiliation{The Institute of Mathematical Sciences, CIT Campus, 
Taramani, Chennai 600113, India}
\affiliation{Homi Bhabha National Institute, Anushaktinagar, Mumbai 
400 094, India}

\date{\today}

\begin{abstract}
Analyzing the brain in terms of organizational structures at intermediate scales provides an approach to negotiate the complexity arising from interactions between its large number of components. Focusing on a wiring diagram that spans the cortex, basal ganglia and thalamus of the Macaque brain, we provide a mesoscopic-level description of the topological architecture of one of the most well-studied mammalian connectomes. The robust modules we identify each comprise densely inter-connected cortical and sub-cortical areas that play complementary roles in executing specific cognitive functions. We find that physical proximity between areas is insufficient to explain the modular organization, as similar mesoscopic structures can be obtained even after factoring out the effect of distance constraints on the connectivity. We observe that the distribution profile of brain areas, classified in terms of their intra- and inter-modular connectivity, is conserved across the principal cortical subdivisions, as well as, sub-cortical structures. In particular provincial hubs, which have significantly higher number of connections with members of their module, but relatively less well-connected to other modules, are the only class that exhibits homophily, i.e., a discernible preference to connect to each other. By considering a process of diffusive propagation we demonstrate that this architecture, instead of localizing the activity, facilitates rapid communication across the connectome. By supplementing the topological information about the Macaque connectome with physical locations, volumes and functions of the constituent areas and analyzing this augmented dataset, we reveal a counter-intuitive role played by the modular architecture of the brain in promoting global interaction.
\end{abstract}

\maketitle

\section{Introduction}
Cortical localization, which refers to specific regions of the cerebral cortex being associated with distinct functions
such as vision and language, has long been a dominant paradigm in neuroscience~\cite{Zilles2010}.
As the connectome 
provides the physical substrate for cognition and behavior~\cite{Goulas2019}, it would seem intuitive that such localization 
would be reflected in the structural attributes of the network~\cite{Swanson2010}.
However, brain imaging studies show that a large number of regions become active during any cognitive task, ruling out a simple one-to-one correspondence between a certain set of vertices of the connectome and a particular function~\cite{Fox2012}. This suggests the necessity for a
theoretic framework that investigates the dynamics of the brain in terms of how different areas connect and interact with 
each other~\cite{Friston2002}.
Such an approach should integrate
complementary perspectives that focus on ($a$) dynamics, where {\it distributed} activation of the entire 
network converges to different attractors, and ($b$) computation, in which {\it localized} processing of information occurs in a sequential manner,
allowing us to interpret cognitive processing as dynamical computation~\cite{Deco2008}.

An integrated view of how local and global coordination of activity across the brain can arise may be obtained by adopting
a mesoscopic approach to analyzing the connectome. Such an approach focuses on understanding the interactions within and between communities of densely
inter-connected brain areas ({\it modules}) that have been identified in nervous systems of different organisms~\cite{Scannell1999, Hilgetag2000, Bassett2010, Pan2010, Wang2012, Harriger2012, Shanahan2013, Shih2015, Sporns2016, Chen2020}.
Such structural modularity of the brain is expected from the advantages that such an architecture may confer during evolution 
and development~\cite{Meunier2010,Sporns2016}, such as imparting robustness in the presence of constraints on 
wiring and performance~\cite{Pan2007,Goulas2019}.
Traditionally, modules have been viewed in functional terms, associated with innate, domain-specific mental faculties (such as language) that are 
believed to be relatively independent of each other~\cite{Fodor1983}. 
Examining how such cognitive modules relate to the structural communities
of the connectome addresses the fundamental issue of structure-function correspondence in the brain~\cite{Park2013,Swanson2016}.

In this paper, we focus on the structure-function relation as evident in the 
modular organization of the Macaque connectome, which balances specialized and integrated processing by allowing rapid communication at both 
local and global scales. This is striking in view of the
role that modularity plays in promoting information encapsulation in other network architectures~\cite{Pan2009}.
In performing this analysis, we have added curated spatial and functional information concerning
the brain areas to the existing database of brain connectivity, which can serve as a resource for the community.
The modules revealed by our analysis extends earlier work~\cite{Hilgetag2000,Harriger2012,Chen2020}
by including sub-cortical regions. We show that each module comprises both cortical and sub-cortical components, 
which is intriguing in view of the proposal that the thalamo-cortical loop plays a central role in the computational architecture of the 
neocortex~\cite{Mumford1991}.  
More crucially, we show that the empirically determined pattern of  intra- and inter-modular connectivity facilitates local, as well as global, dissemination,
complementing studies showing that maximizing information flow may cause model networks to evolve towards a modular structure~\cite{Antonopoulos2015,Yamaguti2015}.
Furthermore, while it has been suggested earlier that physical space constraints cannot exclusively account for modules~\cite{Pan2010,Betzel2017,Stiso2018},
our determination of the space-independent modules and their relatively high overlap with the original communities clearly indicate that 
the modularity of the Macaque brain has functional significance, viz., the facilitation of communication across the connectome. 

\section{Materials and Methods}
\subsection{Data}
\noindent
{\em Connectivity.} 
We have used as the basis for reconstructing the Macaque connectome
a directed network of brain regions (cortical and sub-cortical) 
that was compiled 
in Ref.~\cite{Modha2010} using
several hundred tract tracing studies obtained from CoCoMac -
a comprehensive neuroinformatics database~\cite{Stephan2000,Stephan2001,Kotter2004}.
The original network comprised $383$ vertices, representing regions in the
cortex, basal ganglia and thalamus, at different levels of spatial
resolution, and $6602$ directed edges corresponding to tracts, i.e., myelinated
bundles of axons connecting different brain regions, which may span large
distances.
In this hierarchically organized arrangement of subdivisions starting
from the level of the entire brain, the same region may
occur multiple times as a vertex could represent an area that is part
of a larger area corresponding to a different vertex. For example,
the hippocampus is a vertex of the network, as are its subdivisions
CA1, CA3 and Dentate Gyrus. Consequently, there is no unique mapping
between brain regions and vertices of this network. It also leads
to ambiguity in interpreting edges connecting vertices that occur at
any of the levels other than the lowest one in the hierarchy. For instance,
if both vertices A and B link to C, but B is a sub-division of A, it is
unclear if the two edges are distinct. 
These issues make it difficult to interpret any results obtained
by analyzing the original network.

In the connectome we consider here, these issues are avoided by
considering only those nodes that occur at the lowest hierarchical level, i.e.,
corresponding to regions with no further sub-divisions, in the
original network. This yields a network comprising $266$ nodes, representing
brain areas that span a range of spatial scales ranging from the visual
cortex area V1 
(which has a volume of $\sim 2000$ mm$^3$) to the thalamic region PT\#2 
(which has a volume of less than $2$ mm$^3$). The network
that we consider, consequently, consists of the $2602$ directed links 
that occur between these nodes.
Note that this procedure leads to the network having a largest connected
component of $261$ nodes (as the following five regions do not have any 
reported connections to the other areas at the lowest hierarchical 
level: PT\#2, 6b-beta, 4a, 4b and Sub.Th).
Despite the reduction in the size of the network upon removal of the
aforementioned redundancies, the resulting connectome has similar
macroscopic properties as the original network, such as the
exponential nature of the degree distribution (Fig.~S1).\\

\noindent
{\em Spatial Positions.}
As the brain connectome is a spatially embedded network, it is important
to consider geometric information such as physical locations and extent
of the different brain regions, in addition to the connection topology.
As the original network~\cite{Modha2010} did not contain any spatial
information, we have compiled a comprehensive database of the positions
of the areas corresponding to each of the nodes, as well as, the volumes
spanned by them.
We have obtained the stereotaxic coordinates of each 
brain region in our connectome from several sources. Information
about 134 of the 266 regions included in the connectome has been obtained from the 
website~\cite{Paxinoswebsite}
associated with the Paxinos Rhesus Monkey Atlas~\cite{Paxinos2000}.
For the remaining regions, we manually curated the requisite data from
the relevant research literature. The position of a region is identified with the approximate location
of its center obtained from the online three-dimensional visualization platform in the website mentioned above.
The volume spanned by a particular region was estimated by approximating the cross-sectional area
occupied by the region in each of the coronal sections of the brain in which it appears and obtaining
the sum of these areas weighted by the thickness of the sections measured along the rostral-caudal axis.  
Data file $connectome\_nodes.xls$ in the SI 
contains the 3-dimensional coordinates of, and the volumes covered by,
each of the brain regions that we obtained through the above analysis. It also lists the
references that were used to obtain the information about each region.

\subsection{Modularity.} A prominent mesoscopic structural property associated
with many networks that occur in nature is modular organization.
Modules (or communities) are subnetworks that are characterized by a
higher density of connections between the constituent nodes compared
to that between nodes belonging to different modules~\cite{Newman2004b}. 
One of the
most well-known approaches for determining the modules of a network is
to maximize a quantitative measure, $Q$, defined for a given
modular partitioning of the network as,
$Q = L^{-1} \Sigma_{i,j}  B_{ij} \delta_{c_i c_j},$
where $B_{ij} = A_{ij}-(k_i^{\rm in} k_j^{\rm out}/L)$ are elements 
of the modularity matrix {$\mathbf B$}~\cite{Newman2004,Newman2006}.
The adjacency matrix {$\mathbf A$} ($A_{ij} = 1$, if a directed link exists 
from $j$ to $i$, and  $0$, otherwise) specifies the connection topology
of the network, while the number of incoming and outgoing
connections of node $i$ are indicated by the in-degree 
$k_i^{\rm in} = \Sigma_j A_{ij}$ 
and out-degree $k_i^{\rm out} = \Sigma_j A_{ji}$, respectively, with $L$
($= \Sigma_j k_j^{\rm in} = \Sigma_j k_j^{\rm out}$)
being the total number of connections in the network.
The Kronecker delta function $\delta_{ij}$ yields $1$ if the communities
$c_i$ and $c_j$ to which nodes $i$ and $j$ belong respectively, are
identical, and is $0$ otherwise.\\

\noindent
{\em Spectral analysis and its refinement.}
In order to achieve an optimal partitioning of the network through the
maximization of $Q$ we have used the spectral method~\cite{Newman2006}.
Here, we first bisect the network by assigning nodes to one of two
communities according to the sign of the elements of the eigenvector
corresponding to the largest positive eigenvalue of the symmetrized
modularity matrix $\mathbf{B} + \mathbf{B}^{\rm T}$.
Subsequently we refine the partition by swapping the nodes between
communities in order to achieve the highest possible value of $Q$.
The above procedure is carried out recursively on each of the communities 
to further subdivide them until $Q$ cannot be increased 
further~\cite{Newman2006}. This approach yields a maximum value
of $Q$ for a partitioning of the network into $5$ modules with
$Q_{\rm spectral} = 0.485$.\\

\noindent
{\em Robustness of the partitioning.}
To ensure that the modular partitions of the network obtained using the
deterministic
spectral technique (described above) are not sensitively dependent on the 
specific method used for maximizing $Q$, we have used the stochastic simulated
annealing approach to obtain an ensemble of $10^3$ optimal partitions. 
The dissimilarity between the
different partitions generated by each realization of the annealing
technique reflects the extent of degeneracy (and hence, ambiguity) 
inherent in the modular 
decomposition of the network.
Following Refs.~\cite{Clauset2010,Clausetsite},
for each realization of the simulated annealing approach we begin with 
an arbitrary partition of the network and iteratively change the modular
composition by implementing one of three types of operations: (i) move
a randomly chosen node to any other module including a newly created one,
(ii) merge two randomly chosen modules and (iii) split a randomly
chosen module into two parts so as
to minimize the number of connections between the two parts.
Any one of the possible operations (across all types) is chosen
at each step with equal probability. The resulting partition associated
with a change $\Delta Q$ in the modularity is accepted with a probability 
$\exp(-|\Delta Q|/T)$ if $\Delta Q<0$ and $p=1$ otherwise. Here,
the parameter $T$, which is analogous to temperature, is decreased
over time according to a specified cooling schedule. The process terminates
when the number of successive failures at altering the 
modules exceeds a threshold value.
While the $Q$ values corresponding to the partitions obtained for
different realization span a wide range, most of them cluster around
that obtained from the spectral method, $Q_{\rm spectral}$. We focus on the
$291$ partitions whose $Q$ value deviates from $Q_{\rm spectral}$ by less than 
$3\%$. As shown in the SI, the modular membership of
$70\%$ of the nodes remain invariant across all of these partitions,
and are in fact identical to that obtained from
the spectral method,
underlining the robustness of the modular decomposition.
We have also used alternative methods of module identification that
do not rely on maximizing $Q$, viz., the Infomap method~\cite{Rosvall2008},
and have obtained qualitatively
similar results.

\subsection{Classification of brain regions according to their role in the mesoscopic structural organization of the
connectome.}
The importance of a given region within the topological organization
of the Macaque brain network is indicated by
its connectivity within its own module (as defined above), as well
as that across the entire brain, which is evident from its connections to
regions belonging to other modules. These can be quantitatively measured
by the metrics (i) the within module degree $z$-score ($z$) and (ii) 
the participation coefficient ($P$), respectively~\cite{Guimera2005, Guimera2007}. 
To identify regions that have significantly more connections within
their own module, we determine a within module degree $z$-score:
\begin{equation}
z_i = \frac{k^i_{c_i}-\langle k^j_{c_i}\rangle_{j\in c_i}}{\sqrt{\langle (k^j_{c_i})^2\rangle_{j\in c_i}-\langle k^j_{c_i}\rangle^2_{j\in c_i}}},
\end{equation}
where $k^i_{c_i}$ is the number of links between region $i$ and other regions
belonging to its module ($c_i$) and the average $\langle \dots\rangle_{j\in c}$ 
is taken over all regions in a module $c$. 
As in Ref.~\cite{Pan2010}, nodes (regions) having $z>0.7$ are identified
as hubs, the remainder being classified as non-hubs.

In order to distinguish between brain regions in terms of their
inter-modular connectivity we calculate the
participation coefficient $P_i$ of region $i$ as:
\begin{equation}
P_i = 1 - \sum_{c=1}^m\left(\frac{k^i_{c}}{k^i}\right)^2,
\end{equation}
where $k^i_{c}$ is number of links that region $i$ has with those regions
belonging to module $c$ and $k^i = \sum_c k^i_c$ is the total
degree of the $i$-th node (region). 
A region whose connections are restricted within its
own module has $P_i = 0$ while one whose links are uniformly distributed 
among the different modules has $P_i$ closer to $1$.
Based upon the value of $P_i$, which provides a measure of how well a 
node (region) bridges different modules, the non-hub regions are
classified as ultra-peripheral (R1, $p\leq0.05$),
peripheral (R2, $0.05<p\leq0.62$), satellite connectors (R3,
$0.62<p\leq0.8$) and kinless nodes (R4, $p>0.8$), while the hubs can be
demarcated into provincial hubs (R5, $p\leq0.3$), 
connector hubs (R6, $0.3<p\leq0.75$) and global
hubs (R7, $p>0.75$). 


\subsection{Degree- and modularity-preserved network randomization.}
We construct an ensemble of $10^3$ networks obtained by randomizing 
the empirical
network preserving the in-degree and out-degree of each node (region)
as well as the modular organization of the network~\cite{Pan2010}. 
Each network is 
obtained by selecting directed connections, e.g., $i \rightarrow p$ 
and $j \rightarrow q$, such that the source nodes $i,j$ belong to
the same module A and target nodes $p,q$ belong to the same module B (which
could be same as A),
and then rewire them so as to have $i \rightarrow q$ and $j \rightarrow p$.
This procedure is repeated for $10^6$ times for each realization of a 
randomized network. 
To randomize the network preserving the degree alone, we follow the 
same procedure as above with the difference that there is no constraint
on the modular membership of the nodes.

\subsection{Diffusive spreading model.}
As the function of the connectome is to facilitate communication
between the different brain regions, we investigate the role of the
empirically observed pattern of intra- and inter-modular connections
on the diffusion of information across the system. For this purpose, 
we consider discrete random walks that, starting
from a given node on the network, proceeds at each time step from one node 
to a randomly chosen node that receives an outwardly directed link from 
the former.
The rate at which spreading occurs in different parts of the system
can be analyzed by obtaining the 
distribution of first passage times (FPTs) for a random
walk to reach a target node starting from a source node.
For this, we have measured the FPTs $\tau$ to all nodes that are visited 
by a walk initiated from a given node of the network. The process
is repeated $10^3$ times starting from each of the $266$ nodes, with a
walk terminating when either every node has been visited at least once
or a node with no outgoing connections is reached.
Separate distributions for intra-modular FPTs ($\tau^{\rm intra}$) and 
inter-modular FPTs ($\tau^{\rm inter}$) can be obtained
by considering the source and target nodes to be in the same module
or in different modules,
respectively.
For comparison, we also compute the distributions of FPTs $\tau_D$ and 
$\tau_{DM}$ for randomized surrogates
in which either the degrees, or both the degrees and modular
memberships, of the nodes are preserved, respectively. 
In each case, the distribution
is averaged over $20$ network realizations.
The deviation of the empirical FPT distribution from those obtained
from the randomized surrogates by averaging over multiple realizations
is quantified in terms of a $z$-score
measure defined as:
\begin{equation}
z = \frac{P_{emp}(\tau)-\langle P_{rand}(\tau)\rangle}{\langle P_{rand}(\tau)^2\rangle-\langle P_{rand}(\tau)\rangle^2},
\end{equation}
where $P_{emp} (\tau)$ and $P_{rand} (\tau)$ are the empirical and 
randomized surrogate FPT distributions, respectively.

\subsection{Role of spatial geometry in the modular organization of the connectome.}
The physical distance $d_{ij}$ between two brain regions $i$ and $j$,
whose centers
are indicated by the vectors {\bf x} and {\bf y}, respectively, has 
been measured in terms of the Euclidean metric $d$({\bf x},{\bf y}) and
scaled by the geometric mean of the radii $r_i, r_j$ of the two regions 
(the radius of each region being estimated from the its volume, see SI).\\

%
\noindent
{\em Space-independent partitioning of the network into communities.} 
For networks whose nodes are embedded in a space associated with a metric,
it can be argued that the network properties, such as modularity, could be
a consequence of the constraints imposed by the underlying geometry.
We therefore need to modify the method for determining the modular structure
of a network outlined above, in order to take into account the role of the 
physical space in which the network is embedded. This is done by re-defining
the modularity matrix {\bf B} in the definition of the quantity $Q$ 
(given above), so that the expectation of a pair of nodes ($i,j$, say) 
being connected by chance in the null model incorporates the physical
distance ($d_{ij}$) between the nodes. 
Thus, following Ref.~\cite{Expert2011}, we
re-define $B_{ij} = A_{ij}-(k_i^{\rm in} k_j^{\rm out} f(d_{ij})/L),$ where
$f(d) = \Sigma_{d_{ij}=d} A_{ij}/ (k^{in}_i k^{out}_j)$ is referred
to as the {\em deterrence function}. This function, which is estimated
from empirical data for the network, contains information about how the
physical distance between a pair of nodes modulates their connection 
probability. Note that if the communities in the network arise
entirely because of spatial dependence, measuring $Q$ taking into account
the physical distance between nodes does not yield any modular structure.
Moreover, comparing the space-independent modular decomposition of 
the network obtained using this technique with the communities
determined using exclusively information about the connection topology 
(as described earlier), we can infer whether the observed modularity
is primarily driven by physical distance constraints (see SI). The similarity
between the communities obtained using the two methods is quantified
using {\em normalized mutual information}.\\
 
\noindent
{\em Normalized mutual information.}
To quantify the similarity between two modular decompositions 
$\{c^A_i\}^{M_A}_{i=1}$ and $\{c^B_j\}^{M_B}_{j=1}$ resulting
from different partitionings $A$ and $B$ of a network (that yield
$M_A$ and $M_B$ modules, respectively) we have used the normalized
mutual information~\cite{Mackay2003} 
\begin{equation}
I_{\rm norm} (A,B) = \frac{2 \sum_{i} \sum_{j} P({c^{A}_i},c^{B}_j) \ln
[P ({c^{A}_i},c^{B}_j)/ P({c^{A}_i}) P(c^{B}_j)]}{-\sum_{i} P({c^{A}_i})
\ln P({c^{A}_i}) - \sum_{j} P(c^{B}_j) \ln P(c^{B}_j)},
\end{equation}
where $P({c^{A}_i})$ is the probability that a randomly chosen node
lies in module $c^A_i$ in partition $A$, $P(c^{B}_j)$ is the probability 
that a randomly chosen node lies in module module $c^B_j$ in
partition $B$, and $P({c^{A}_i},c^{B}_j)$ is the
joint probability that a randomly chosen node belongs to 
module $c^A_i$ in partition $A$, as well as, to module $c^B_j$ in
partition $B$ ($i=1, \ldots, M_A$, and $j=1, \ldots, M_B$).
Each of the probabilities can be estimated from the ratio of the 
community sizes to the size of the entire network.\\
\begin{figure*}[t!]
\includegraphics[height=8.6cm]{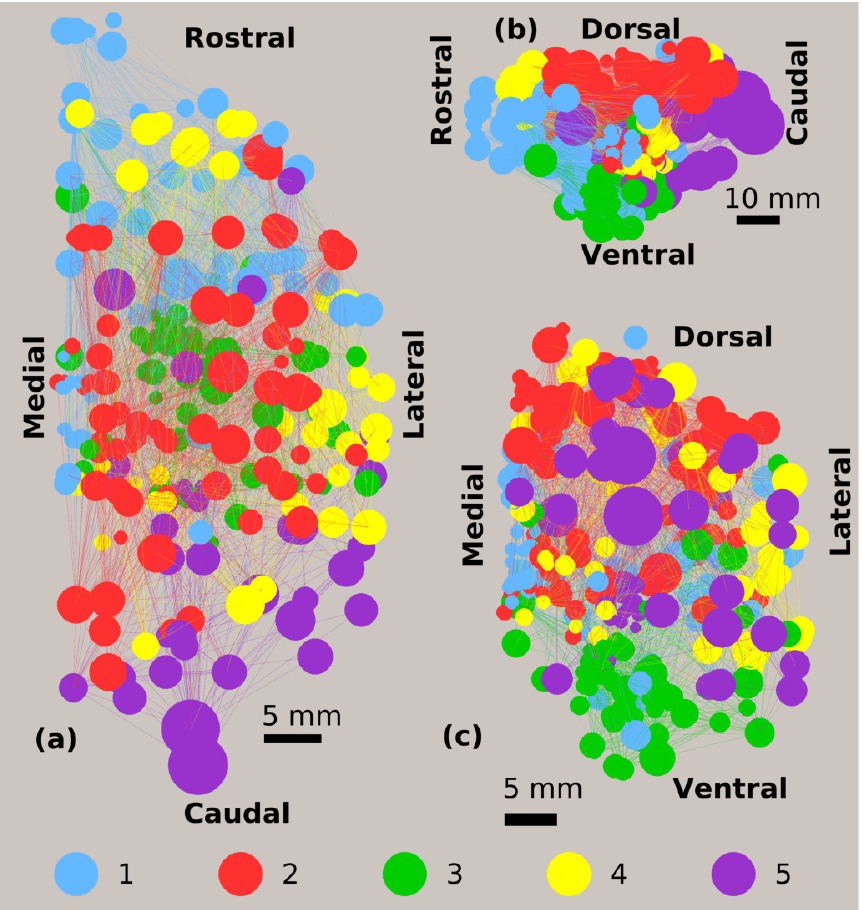}
\includegraphics[height=8.6cm]{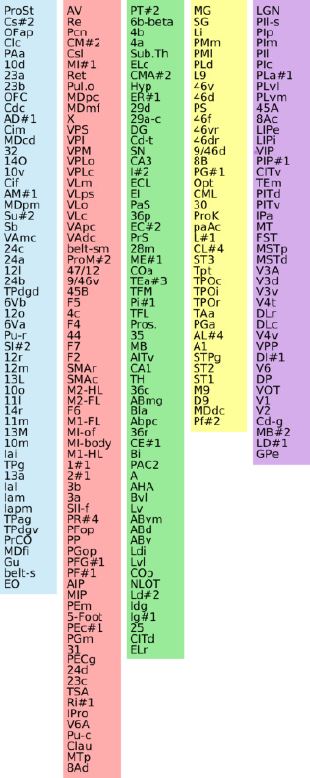}
\includegraphics[height=8.6cm]{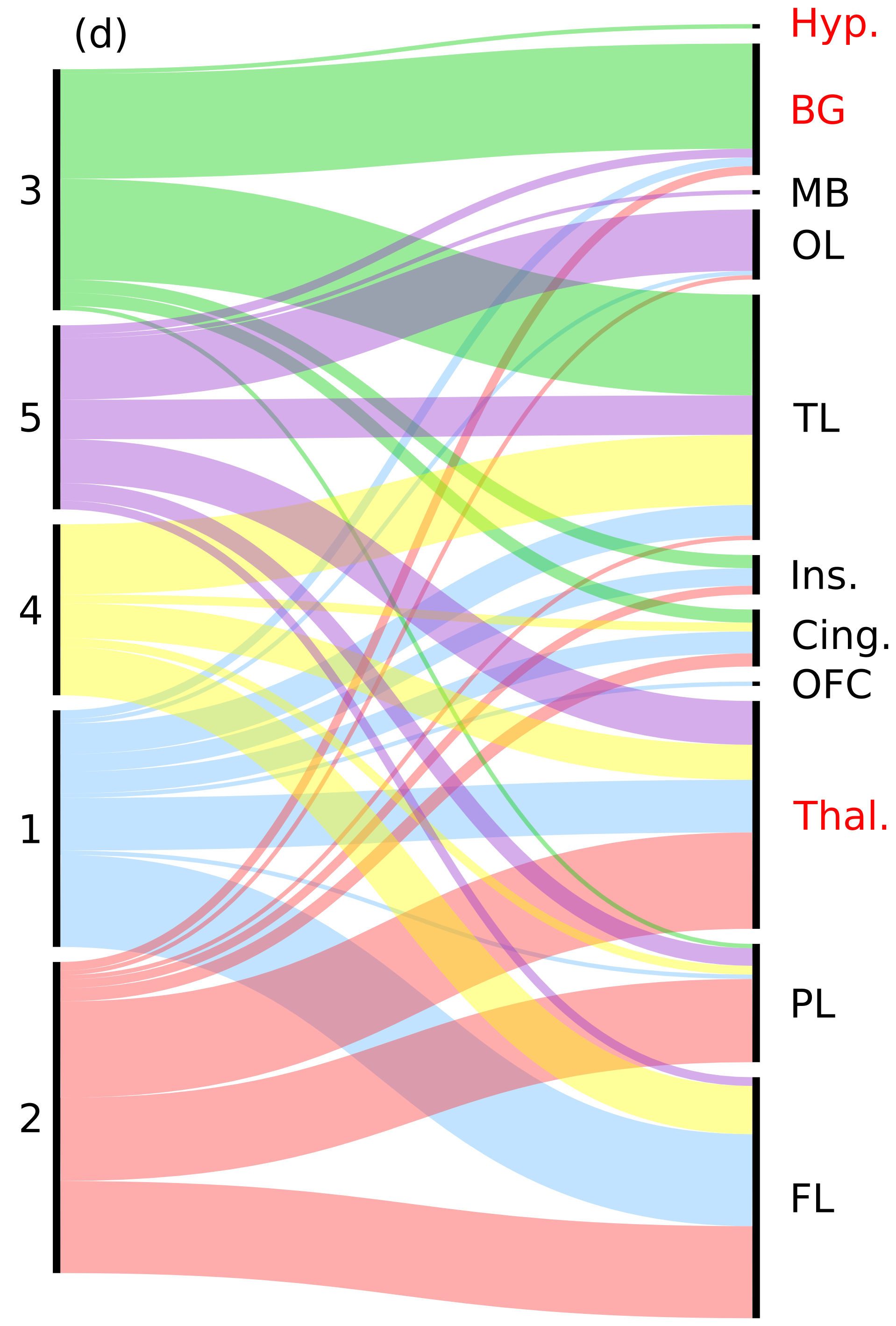}
\centering
\caption{\textbf{Mesoscopic organization of the Macaque brain.} 
The network of brain regions, shown in (a) horizontal, (b) sagittal and 
(c) coronal projections, clearly indicate that the nodes (filled circles)
are organized into five modules, each characterized by dense intra-connectivity.
The modular membership of each node is represented by its color 
(see color key to the right, containing the 
list of brain regions in each module), while node sizes provide
a representation of the relative volumes of the corresponding brain regions
(the spatial scale being indicated by the horizontal bar in each panel).
The spatial positions of the nodes are specified by the three-dimensional
stereotaxic coordinates of the corresponding regions (see Methods).
Links indicate the directed nerve tracts connecting pairs of brain
regions, and are colored in accordance with their source nodes.
For details of each of the brain regions see SI.
(d) Visual representation of the association between the network modules 
and cortical (in black), as well as, sub-cortical (in red) subdivisions of 
the brain, viz., 
\textit{FL}: Frontal Lobe, \textit{PL}: Parietal Lobe, \textit{TL}:
Temporal Lobe, \textit{OL}: Occipital Lobe, \textit{Cing.}: Cingulate,
\textit{Ins.}: Insula, \textit{BG}: Basal Ganglia, \textit{Thal.}:
Thalamus, \textit{Hyp.}: Hypothalamus, \textit{OFC}: Olfactory
complex, and \textit{MB}: Mid-brain.
For a detailed breakdown of the major subdivisions of the brain
in terms of their module membership, see SI.
This alluvial diagram has been created
using the online visualization tool RAW~\cite{Mauri2017}.
}
\label{fig:Fig1}
\end{figure*}

\begin{figure*}[htbp]
\centerline{\includegraphics[width=\linewidth]{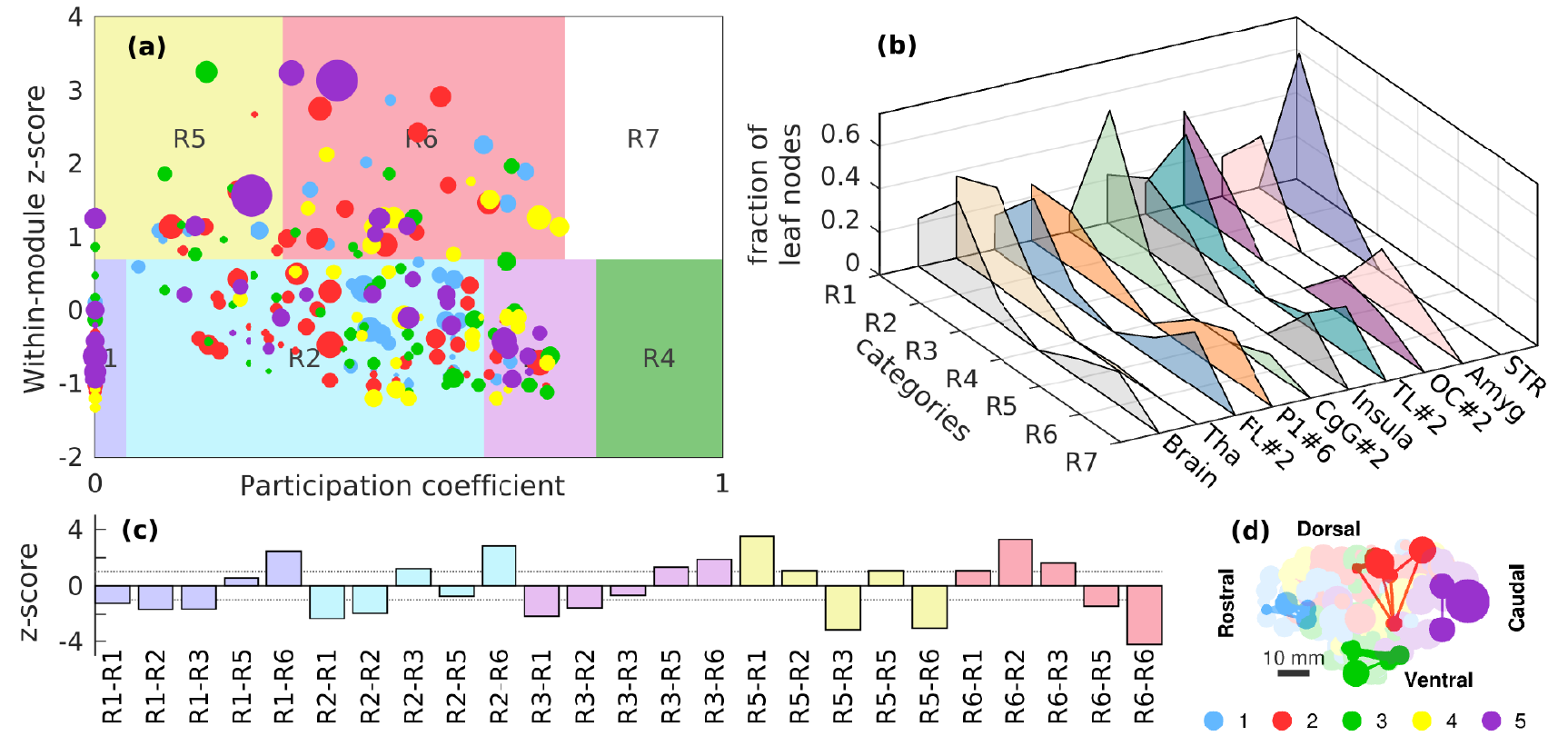}}
\centerline{\includegraphics[width=\linewidth]{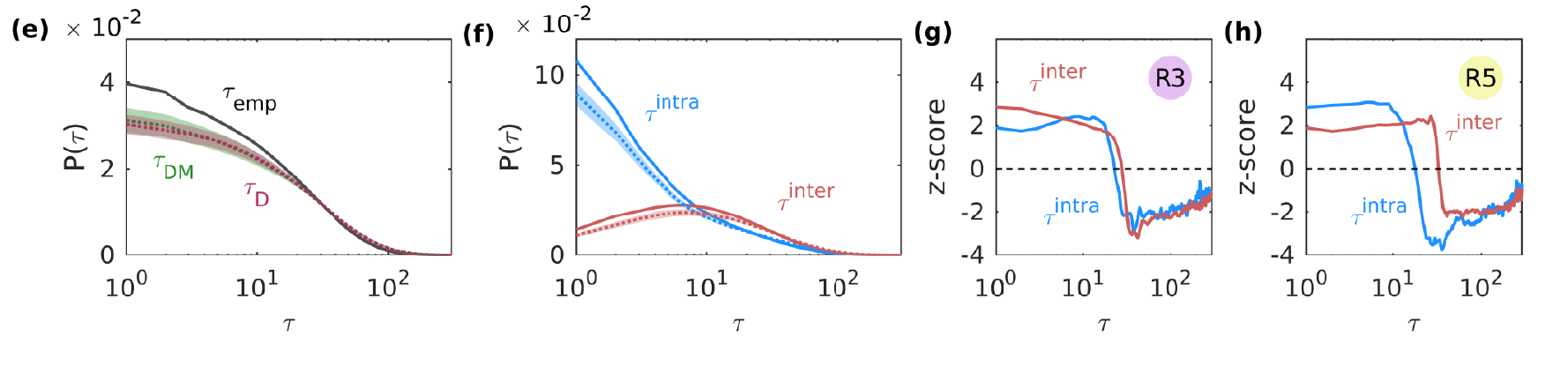}}
\caption{\textbf{Classification of brain regions according
to their intra- and inter-modular connectivity.}
(a) Nodes of the Macaque brain network [colored and scaled as 
per Fig.~\ref{fig:Fig1}~(a-c)] are displayed in accordance with
their within-module degree z-score ($z$)
and participation coefficient ($P$), which provide a measure of
their intra- and inter-modular connectivity respectively. 
This allows the brain regions to be categorized into one of
seven possible categories (see Methods), viz.,
R1: ultra-peripheral,
R2: peripheral, R3: satellite connector,
R4: kinless, R5: provincial hub,
R6: connector hub, and R7: global hub. Note that there are no regions
in the Macaque brain belonging to the categories R4 and R7.
(b) The distribution of the regions of the entire Macaque brain 
across the different categories R1-R7 is similar
to the corresponding distributions observed in several 
anatomical subdivisions, viz.,
\textit{Tha}: Thalamus,
\textit{FL\#2}: Frontal Lobe, \textit{P1\#6}: Parietal Lobe, 
\textit{CgG\#2}: Cingulate Gyrus, Insula, \textit{TL\#2}: Temporal Lobe, 
\textit{OC\#2}: Occipital Lobe, \textit{Amyg}: Amygdala and \textit{STR}:
Striatum.
(c) The connectivity pattern between regions belonging to the different
categories R1-R7 indicated by the $z$-scores for abundance of links 
between each pair of categories (the first symbol in Ri-Rj refers
to the category of the source region and the second to that of the target), 
measured with respect to degree- and 
modularity-preserved randomized ensemble of networks (see Methods).
Large positive (or negative) $z$-scores, i.e., $z>1$ (or $z<-1$), 
indicated by the dotted
lines, imply that the corresponding
connection types occur significantly more (or less) often than
expected from random networks that have
degree sequence and community structure identical to the empirical network. 
(d) Sagittal projection of the network of brain regions [see
Fig.~\ref{fig:Fig1}~(b)] showing that
connections between provincial hubs (highlighted nodes) are localized
within each module.
(e) Temporal evolution of spreading processes, quantified in terms of 
distributions of first passage times ($\tau$) of random walkers 
starting from one node to reach another, contrasted
between the empirical brain network (solid line, $\tau_{emp}$) and randomized
ensembles of networks, generated by preserving either the degrees alone
(red, $\tau_D$), or both the degree and the modular membership of
each node (green, $\tau_{DM}$).
(f) The distribution of $\tau$ differs significantly, depending on whether 
the target and source nodes belong to the same module (blue, 
$\tau^{intra}$) or different modules (red, $\tau^{inter}$). As in (e),
spreading occurs significantly more rapidly in the empirical network
(solid lines) compared to the networks belonging to the randomized ensemble
(obtained by preserving degree and modular membership). 
In both (e) and (f), the dotted lines and the shaded regions around them
represent the mean and standard deviation of $P(\tau)$ calculated over
the randomized ensembles. To see how the different categories R1-R7 of
brain regions allow spreading to occur
faster in the empirical brain network than in equivalent randomized
networks, we focus on the cases where the source nodes are 
either satellite connectors R3 (g) or provincial
hubs R5 (h). 
The z-score indicates that there is a statistically significant shift
in the empirical distribution towards lower values of $\tau$ in both cases.
However, while for R3 the increase in the rate of spreading is similar,
irrespective of whether the target is in the same module or in a different one,
we observe that for R5, there is a relatively larger shift at lower
values for $\tau^{intra}$ as compared to $\tau^{inter}$. This is consistent 
with the connectivity pattern of provincial hubs with the other categories of
nodes [shown in (c)] which particularly favors intra-modular communication.
}
\label{fig:Fig2}
\end{figure*}

\noindent
{\em Surrogate networks.}
In order to explicitly show that the modular organization is not primarily driven by the
constraints imposed by the physical distance $d$ between brain regions, we have demonstrated
how spatial embedding affects the modular decomposition of a network,
using three classes of surrogate random network ensembles (of size $100$ each) 
having different underlying spatial dependences.
The three ensembles, in increasing order of importance of $d$ in governing
the connection probability $P$ between nodes, comprise networks with (a) $P \sim d^0$, (b) $P \sim d^{-1}$,
which is the case in the empirical network, and (c) $P \sim \exp (-d)$, with nodes in each network
occupying the same spatial position as in the empirical network.
Each network (comprising an identical number of nodes and links as in the empirical network) 
was subject to community detection using information about the connection topology alone, as 
well as, space-independent modular decomposition, following the two approaches described above.
The difference between these two sets of partitions provides a measure of the role that spatial
embeddedness of the networks plays in determining the modular nature of their connectivity (see SI).

\section{Results}

\subsection{Mesoscopic organization of brain areas in the Macaque}
Fig.~\ref{fig:Fig1}~(a-c) shows the modular organization of the Macaque
brain network spanning regions from the cortex, basal ganglia and thalamus,
revealed by our analysis (for details see methods). The network is 
seen to comprise $5$ modules, each module $i$ being composed of $m_i$ 
densely inter-connected brain regions
(their numbers ranging between $39$ and $71$, see the color key to the right 
of Fig.~\ref{fig:Fig1}, a-c, containing the list of brain regions in 
each module). The membership of the individual regions in these modules is
seen to be robust (see Methods). Given that the network is embedded in
a specific geometry, namely that of the Macaque brain, it is noteworthy 
that each of the modules are spatially clustered
as is clearly seen from the
projections shown in Fig.~\ref{fig:Fig1}(a-c).
To understand the implications of the spatial location of these modules,
we visually represent the mapping between the modules and the major
anatomical subdivisions of the brain in Fig.~\ref{fig:Fig1}~(d) [see also
SI]. 

We observe that every module comprises sizable number of both 
cortical and sub-cortical regions.
With the exception of \#$3$, the modules have their sub-cortical
components located almost exclusively in the \textit{Thalamus}. 
We note that each of these modules are
associated with different sensory modalities (discussed in detail later),
consistent with one of the primary functions of the Thalamus, namely,
relaying information from the sensory organs to cortical areas for 
further processing. As the Thalamus is also involved in sleep-wake 
regulation coordinated via extensive reciprocal connections with the 
cortex~\cite{Maclean2005, 23Problems, Steriade1993}, it is reasonable to expect that the each of the
network modules will have thalamic components along with cortical ones,
with dense intra-modular connectivity representing thalamo-cortico-thalamic 
circuits~\cite{Guillery1995, Sherman1996}. 
However, none of the sub-cortical components of
module \#$3$ (displayed in green in Fig.~\ref{fig:Fig1}) belong
to the Thalamus and instead constitutes almost the entirety of the 
\textit{Basal Ganglia}.

The locations of the cortical components of the different modules
across the principal lobes of the cortex, viz., \textit{frontal}, 
\textit{temporal}, \textit{parietal} and \textit{occipital}, are
indicated in Fig.~\ref{fig:Fig1}~(d). We observe that there is
no simple correspondence between the modules, which are topological partitions
of the connectome, and the gross anatomical subdivisions of the cortex.
While the regions comprising the frontal and temporal lobes are split 
between several modules, those in the parietal and occipital are dominated 
by single modules (modules \#$2$ and \#$5$, respectively), indicating the
relative homogeneity of the latter lobes in the mesoscopic
organization of the network.
This assumes importance in light of a possible connection between
the modular divisions and functional specialization in the brain - a point
that we discuss below.

As mentioned in the Introduction, the term {\em module} has been 
primarily used in the neuroscience literature to refer to a functionally
integrated set of areas~\cite{Braun2003, Sternberg2011, Gazzaniga2014} that
allows for ``information encapsulation''~\cite{Fodor1983}, whereas we
employ the term in the sense of a specific meso-level structural feature
in the connectome~\cite{Scannell1999, Hilgetag2000, Bassett2010, Pan2010, Wang2012, Harriger2012, Shanahan2013, Shih2015, Sporns2016}.
In analogy with other biological networks where a structure-function
correlation has been established for modules~\cite{Hartwell2000,Guimera2005}, 
we now ask
whether the network modules that appear as separate structural units 
of the brain can be considered as distinct functional units as well.
Using information about the known functions of different cortical 
and sub-cortical areas obtained from decades of experimental
studies,
we have created a mapping between the regions
belonging to each module and the specific functionalities attributed to them
(see SI). A perusal of this information reveals that
the different regions belonging to a module
complement each other in carrying out
various cognitive functions.
For example, several cortical areas in module \#$5$, viz., 45a and 8Ac of the pre-frontal 
cortex, and V1 and V2 of the occipital lobe, are related
through their involvement in vision, even though
they may be part of distinct lobes and have disparate functions
(controlling saccadic eye movements in the case of 45a and 8Ac, and
processing of visual information in the case of V1 and V2).
This suggests a general scheme of organization in which 
the regions associated with each of the principal sensory modalities 
are localized in specific modules, viz., visual in module \#$5$, auditory in
module \#$4$, somatosensory (along with the
principal motor area M1) in module \#$2$ and olfactory (as well as,
gustatory) in module \#$1$.
We show below that
the known behavior of the regions comprising each of the modules 
is consistent with the broad functions attributed to that module.

First, we observe that module \#$5$ (displayed in purple in 
Fig.~\ref{fig:Fig1}) consists of the 
primary visual area in the occipital lobe and association
areas in the parietal (e.g., LIP, VIP etc.) and temporal
lobe (e.g., CIT, PIT, etc.). In addition, its thalamic component 
includes \textit{lateral geniculate nucleus} (LGN), which relays
visual information to the cortex from the retina.
We note that these regions are all involved in various aspects of 
visual cognition, which is consistent with the sensory modality 
associated with this module, viz., vision.
Second, module \#$4$ (displayed in yellow in Fig.~\ref{fig:Fig1}),
consistent with its attributed sensory modality,
is seen to comprise the auditory cortex lying in the 
superior temporal gyrus of the temporal lobe (as well as, the
corresponding association areas), and the 
medial geniculate nucleus in the thalamus, which is the relay for all
auditory information destined for the cortex from the 
brainstem~\cite{Purves2001}.
Third, module \#$2$ (displayed in red in Fig.~\ref{fig:Fig1}), 
contains the primary and secondary somatosensory areas (S1, S2) in the
parietal lobe, while its thalamic component contains all the regions
which together comprise the ventral posterior nucleus that relays
somatosensory information to the cortex. 
Apart from its sensory function, as noted earlier it also consists of primary and
supplementary motor areas which are 
associated with planning, control and execution of voluntary
movements~\cite{Kandel2000}. 
Finally, we note that module \#$1$ (displayed in blue in Fig.~\ref{fig:Fig1}),
has the {\em olfactory complex} and the {\em gustatory cortex},
both located in the frontal lobe, as well as, a few 
other regions (e.g., the olfactory field of the entorhinal cortex, EO, in the
temporal lobe) involved in the sensory processing of smell.
However, the module is dominated by association areas located in the
prefrontal cortex which are involved in high-level
multi-modal sensory integration and decision-making~\cite{Walton2010, Fellows2007, Walton2004, Izquierdo2004, Rolls2000}.


In contrast to the other modules, \#$3$ neither contains motor areas nor
does it include any primary or secondary sensory areas. This is possibly related
to our earlier observation that this module has a distinct structural 
arrangement, in that its sub-cortical
components do not have any contribution from the thalamus, 
but instead comprise regions belonging to the basal ganglia. In particular,
the module contains the entire \textit{amygdala} which is known to regulate
emotional responses and fear-conditioning in mammals~\cite{Weiskrantz1956, Davis1992, Killcross1997, Maren1999}. 
This gains
significance in light of the fact that both the
\textit{Hippocampus} and the \textit{Parahippocampus}, which are
primarily involved in the formation of memory, feature
prominently among this module's cortical components. 
It resonates with the known relation between emotional
state and formation of memories in individuals that have been established
by several studies~\cite{Kensinger2003, Richardson2004, Anderson2006, Tambini2017}.

As the brain is characterized by structures occurring at several scales~\cite{Churchland1988}, 
it is pertinent to ask whether further levels of organization can be 
identified in the connectivity pattern within each of the modules described 
above. Indeed, when we consider module\#$5$, the most robust under 
different realizations of network partitioning (see Methods and SI), and
subject it to further modular decomposition, we observe that it comprises
three communities which we refer to as sub-modules. 
The largest of these contains the visual cortex and almost the entirety
of the sub-cortical components, while the other two (which are
comparable to each other in terms of the number of constituent regions)
are dominated by regions belonging to the superior temporal 
sulcus and the intraparietal sulcus, respectively (see SI). 
Intriguingly, we note that the latter two communities appear to
correspond to regions identified with different visual processing 
pathways, viz., the dorsal  and ventral streams~\cite{Goodale1992,Goodale2018}.

\subsection{Distribution profile of nodes in terms of their
intra- and inter-modular connectivity is conserved across
cortical and sub-cortical subdivisions}
Having described the overall organizational structure of the network at the
mesoscopic level, we now focus on understanding the role played by the
individual brain regions in connecting other regions within their own
module, as well as, across modules. The importance of each region
is quantified in this framework by measuring the within-module degree
$z$-score and the inter-modular 
participation coefficient $P$ (see Methods for details).
As seen in Fig.~\ref{fig:Fig2}~(a),
the $z$-score allows regions to be distinguished between {\em hubs}, i.e.,
those having significantly higher number of connections to other regions
in their module, and {\em non-hubs}, while $P$ further classifies 
the hubs into {\em provincial} (R5), {\em connector} (R6) 
and {\em global} (R7) categories
and the non-hubs into {\em ultra-peripheral} (R1), {\em peripheral} (R2),
{\em satellite connector} (R3) and {\em kinless} (R4) 
classes. We note that regions in
each module have a similar distribution across R1-R3 and R5-R6 (with the
sole exception of module \#$4$ which has no region playing the role of 
a provincial hub, see SI).
Uniformity of this nature can also be
observed in Fig.~\ref{fig:Fig2}~(b) where we compare the 
distributions of constituent regions across the different categories 
for the entire brain with that of the various
subdivisions of the cortex, such as the Frontal (FL\#2), Parietal (P1\#6), 
Temporal (TL\#2) and Occipital (OC\#2) lobes, the Insula and
the Cingulate Gyrus (CgG\#2), as well as, the Amygdala (Amyg) 
which belongs to the basal ganglia. 
However, the Striatum which is also in the basal ganglia, and
the Thalamus (Tha) have the distinctive characteristic of being 
essentially devoid of regions that act as hubs,
indicating a relative lack of heterogeneity in the number of connections
that their constituent regions have with others in their modules.

We have also analyzed the relative frequency with which regions belonging to
the
different categories connect to each other in the Macaque brain, compared to 
the corresponding connectivity pattern observed in surrogate networks 
obtained by degree- and modularity-preserving randomization 
(see Methods)~\cite{Guimera2007}. The profile of connection
preferences between the various categories shown in Fig.~\ref{fig:Fig2}~(c),
with under-representation of connections between R1-R1,
R5-R6 and R6-R6 which has been related to the occurrence of multi-star
structures, resembles other networks involved 
in information propagation~\cite{Guimera2007}. 
As can also be seen from the figure, non-hubs prefer in general to
connect to hubs and vice versa. This is indicative of degree 
disassortativity, i.e., connections between nodes having dissimilar 
characteristics (in this case, the number of connections) are favored.
However, on investigating the connectivity between pairs of these
categories, we notice that source regions belonging to peripheral (R2) 
and provincial hub (R5) categories show a distinct bias in 
their connections in terms of the participation coefficient of the 
target regions. Specifically R2 regions prefer to connect to 
connectors, both hubs (R6) and non-hubs (R3), while avoiding regions
that are localized in their modules (R1, R2 and R5). The trend is reversed
for R5 regions. In particular, they show a slight preference for 
connecting to each other, which is in contrast to the other categories
which exhibit a marked tendency to avoid others of their own kind.

This homophily between provincial hubs could arise from two different patterns
of connectivity between them, viz., one in which connections between
the R5 regions are confined within the same module and another in which
the corresponding regions across different modules are connected.
Fig.~\ref{fig:Fig2}~(d) shows that the empirical evidence supports the former
arrangement where, within each module, provincial hubs connect to each
other preferentially. We note that the three R5 regions indicated in 
module~\#$5$ occur, respectively, in the three different sub-modules that
were identified in the previous subsection.
This intra-modular connectivity within provincial hubs, taken together
with the observation that they preferentially connect to peripheral
regions while avoiding connectors, suggest that they help co-ordinate
activity locally within each module while limiting 
the spread of
information over the network.

\begin{figure*}[htbp]
\centerline{\includegraphics[width=0.5\linewidth]{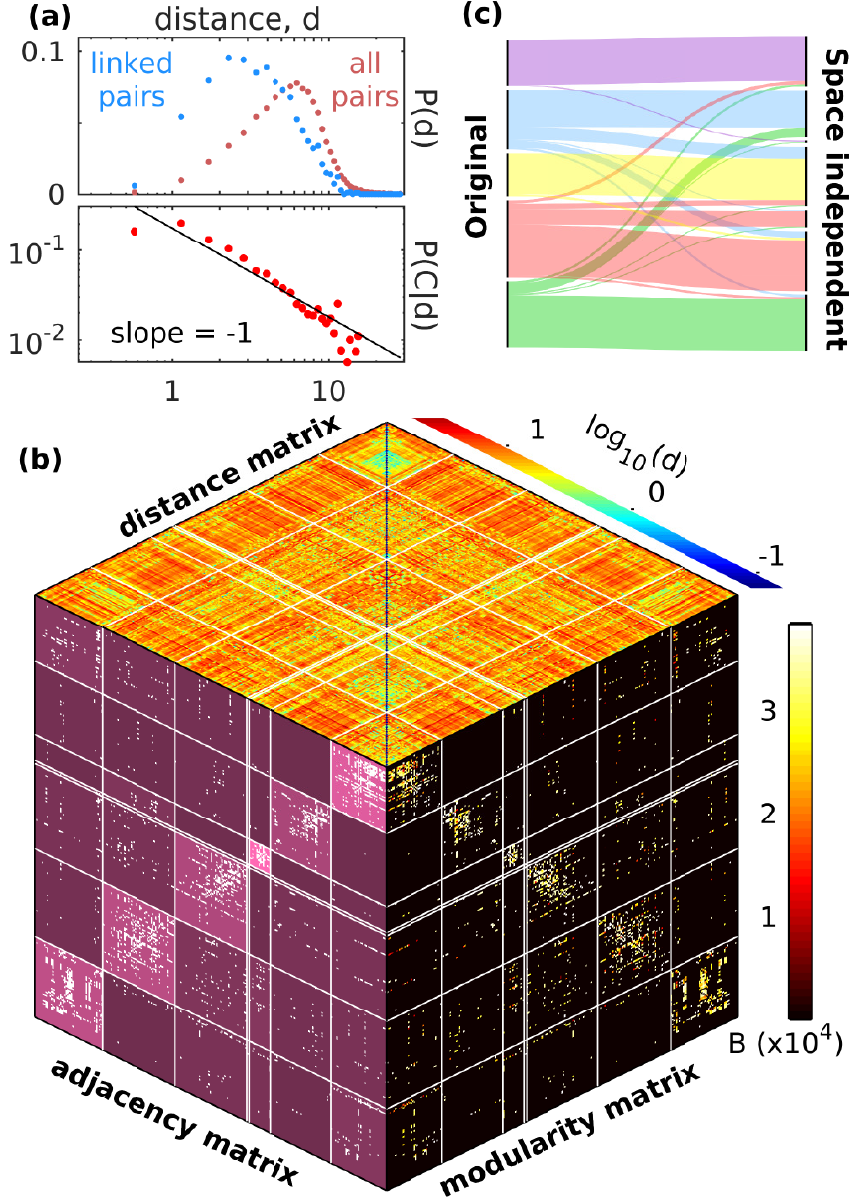}}
\caption{\textbf{Physical distance between brain regions is seen
to constrain their connectivity, but the modular organization of
the network is independent of their three-dimensional spatial arrangement.} 
(a, top) Probability distribution of the physical distances $d$ 
between all pairs of nodes (red) contrasted with that of
connected pairs (blue). 
(a, bottom) The variation with physical distance $d$ of the connection 
probability $P(C|d)$ between a pair of nodes separated by that distance (red).
The empirical data is best fit by the relation $P\sim1/d$ (represented by the
solid line). 
(b) Joint representation of the space-independent modular organization
of the network of brain regions showing the matrices indicating 
adjacency $\{A{ij}\}$ 
(left surface), modularity $\{B_{ij}\}$ (normalized by total number of 
links $L$, right surface) and 
physical distance $\{d_{ij}\}$ 
(top surface) between the different regions.
Note that for matrix $\mathbf A$ the background intensity of each block is 
proportional to the density of connections within that block, and for 
matrix $\mathbf B$ only the values corresponding to linked pairs
of nodes are shown. The nodes are grouped into partitions corresponding
to the space-independent modules of the network with the boundaries indicated
by solid lines.
The relatively large positive values clustered along the diagonal blocks of 
$\mathbf B$ indicate
the occurrence of significantly higher density of connections within 
each module, compared to that expected from the degrees of, and the distance
between, every pair of nodes.
This characteristic signature of modularity is also visible in the
adjacency matrix $\mathbf A$ representing the connection topology, suggesting that
the mesoscopic structure of the brain network is a consequence of factors
beyond the constraints associated with physical distance.
Indeed, this is also true for the spatial clustering of nodes in
each network module seen in Fig.~\ref{fig:Fig1}~(a-c), 
as is apparent from the
distance matrix showing that the modules comprise many nodes
that are spatially proximal even after discounting the effect of distance
in identifying the modules.
(c) Visual representation of the correspondence between the network modules 
determined using exclusively
information about the connection topology (``Original'')
and those obtained from {\em space-independent} partitioning of the network
into communities.
on the right). 
This alluvial diagram has been created
using the online visualization tool RAW~\cite{Mauri2017}. 
}
\label{fig:Fig3}
\end{figure*}
\subsection{Information spreading within the brain is enhanced by the 
specific pattern of intra- and inter-modular connections}
The roles played by regions belonging to different categories in facilitating the transmission of
information within and between modules can be investigated by considering 
a process of diffusive propagation across the network (see Methods).
The distribution of first passage times $\tau$, i.e., the time elapsed
between initiating a random walk from any source node and the earliest arrival
to any given target node, is shown in Fig.~\ref{fig:Fig2}~(e). 
While, in general, presence of modules in networks leads to slower
global diffusion~\cite{Pan2009}, surprisingly we observe that the
distribution for the empirical network is markedly shifted
towards lower values of $\tau$ compared to randomized
networks with an identical degree sequence that may or may not have
modular organization. 
This indicates that, as opposed to information encapsulation, the specific 
pattern of intra- and inter-modular 
connections between brain regions belonging to different categories 
actually promotes faster communication across the network. 
Moreover, as seen from Fig.~\ref{fig:Fig2}~(f), the enhancement of 
the rate of diffusion in the connectome (in comparison to the randomized
surrogates) can be seen both for transmission within a module, as well as, 
between different modules.

We also investigate how nodes having distinct intra- and inter-modular
connectivity roles contribute to enhancing
communication in the network. This is achieved in each case by having
the source node belong to the respective category and comparing the
corresponding distribution of $\tau$ with that obtained from randomized 
surrogates (quantified using $z$-score, see Methods).
Fig.~\ref{fig:Fig2}~(g) shows that starting from a satellite connector R3,
diffusion to other nodes belonging both within its module or to other
modules is significantly faster compared to randomized networks with identical
modular organization and degree distribution. In contrast, as seen from
Fig.~\ref{fig:Fig2}~(h), when starting from a provincial hub R5, the increase
in the rate of diffusion within a module, compared to that
in the surrogate networks, is even higher than the increase in the rate of
diffusion across modules. This resonates with the observation of
homophily between provincial hubs in a module reported earlier (Fig.~\ref{fig:Fig2}~(d)).
When the source node belongs to any of the other 
categories, the difference between the intra- and inter-modular
diffusion time-scales is seen to lie between the range seen for 
these two cases (see SI). This suggests that the
modular character of the mesoscopic organization of the connectome is
further shaped by the distribution of roles played by the different nodes
in allowing information to spread within a module, as well as, across 
different modules.

\subsection{Spatial layout constrains the 
connectivity but does not determine the modular organization of brain regions}
So far we have investigated the modular structure of the network
of brain regions exclusively in terms of the connection topology. However,
the brain is also a physical system that is embedded in three-dimensional
space associated with a distance metric which restricts the
possible connections between its constituent regions. 
Such constraints arise from resource costs related to the spatial volume and
transmission time associated with the connections, and
the rapid energy consumption during synaptic 
transmission~\cite{Bullmore2012, Averbeck2008, Herculano-Houzel2010, Chklovskii2004, Kaiser2004, Buzsaki2004, Niven2008}. 
Thus, given that the
pattern of connections between the regions is a function of
the physical distance between them, 
we can ask
to what extent are the modules a consequence of the brain being a 
spatially embedded network~\cite{Barthelemy2011}.
To investigate the role of spatial constraints on the structure of the 
brain network, we supplement the network topological information with that 
of the physical locations and volumes of each of the regions (shown in 
Fig.~\ref{fig:Fig1}, a-c; for details see Methods).
By comparing the distributions of the physical distances $d$ between all 
possible pairs
of regions (connected or not) with that of only the connected pairs [top
panel of Fig.~\ref{fig:Fig3}~(a)], we can obtain the dependence of the
connection probability between two regions on the distance $d$ between them.
As seen from the bottom panel of Fig.~\ref{fig:Fig3}~(a), this probability
decays linearly with the reciprocal of the distance, i.e., $P(C|d) \sim 1/d$,
explicitly demonstrating the constraint imposed by the spatial layout
of the brain regions on their connectivity.

To see if the restriction on long-range connections implied by the above
constraint is responsible for the mesoscopic organization of the network
we have reported here, we investigate whether the network can be partitioned
into modules even after taking into account the 
distance dependence of the connection probability in the null model (see Methods
for details). Thus, if the modules are exclusively a product
of the distance constraint, the deviation of the empirically obtained
connection probabilities from those of the null model will be minimal,
yielding a single partition comprising the entire network (see SI 
for results on different surrogate networks).
In contrast to the above scenario, we find that applying the method
on the brain network yields an optimal partitioning comprising
seven space-independent modules indicated by the diagonal blocks 
demarcated by white lines in the adjacency matrix shown in
Fig.~\ref{fig:Fig3}~(b) [left surface]. The probability of connections within these modules
deviate strongly from the values expected from the null model as
shown by the modularity matrix [Fig.~\ref{fig:Fig3}~(b), right surface].
The distance matrix [Fig.~\ref{fig:Fig3}~(b), top surface] also appears to
suggest that regions belonging to the same module are, in general, physically
closer to each other than those belonging to different modules.
%
However, this physical proximity cannot provide a causal explanation
for the modular structure
as, even after filtering for spatial effects, the resulting space-independent
modules are substantially similar to those reported in the previous
subsections [see
Fig.~\ref{fig:Fig3}~(c)]. 
The similarity between the results of these two different modular partitionings 
is quantitatively indicated by the corresponding normalized mutual
information $I_{\rm norm}(=0.6$) [see Methods].
Thus, the spatial layout of the brain regions cannot by themselves
explain the mesoscopic organization of the network, and the existence of
the structural modules is a fundamental attribute of the brain.
 
\section{Discussion}
Despite differences in the details of their organization, the
modules that we have identified in the Macaque connectome 
have common structural features. Most notably,
each of them have cortical and thalamic components
with the sole exception of
module \#3, suggesting a distinct functionality of this module.
The sizable thalamic contribution to
modules \#2, \#4 and \#5 can be understood in terms of the roles
that their cortical components play
in processing specific sensory modalities. In particular, the 
information from the corresponding sensory organs arrive 
at the cortical regions belonging to these modules via relay centers 
located in the thalamic component of the respective modules.
This, however, cannot explain the sizable contribution from
thalamic regions to module \#1, as the sensory modalities it is associated
with, namely, olfaction and gustation, do not involve any thalamic relay.
As one of the primary functions of this module is the integration of
information processed in different cortical regions (as mentioned earlier),
it suggests that regions in the thalamic component of this module serve 
as relay centers coordinating inter-cortical communication~\cite{Maclean2005}.

The module with which a particular brain region is associated 
may also alert us to
possible functions of this region that have not yet been identified.
As an example we consider multi-modal
association areas,
which integrate and process inputs from different sensory modalities 
(such as the regions \textit{LIP}, \textit{MIP} and \textit{area 46}). 
Using information about their modular membership, we can identify which
modality or function each of these regions are most strongly associated
with. 
This is illustrated by considering the \textit{LIP},
\textit{VIP}, \textit{AIP} and \textit{MIP} areas of the \textit{Intraparietal
Sulcus}. Although they are all multi-modal association areas, \textit{LIP} and
\textit{VIP} are part of module \#5, whereas areas \textit{AIP} 
and \textit{MIP} are part of module \#2. It is known that \textit{LIP} and
\textit{VIP} are involved in visual attention and saccadic eye
movements~\cite{Gnadt1988,Duhamel1998,Pesaran2002}, which are predominantly 
visual processing tasks
(consistent with the broad function of module \#5). In contrast,
\textit{AIP} and \textit{MIP} coordinate the visual
control of reaching and pointing~\cite{Sakata1995,Eskandar1999,Grefkes2005}, 
which, although guided by visual information, is primarily a motor function 
(consonant with the broad function of module \#2).
Thus, the specific
functionalities of these association areas seem to tie in with
the modules that they belong to. 

We note that the modular nature of the brain has been long recognized, both in terms 
of function and, more recently, in the topological organization of its 
structural connections~\cite{Sporns2016}.
Considerable attention has been focused on the question of structure-function 
convergence in the context of brain modules~\cite{Park2013}. The hypothesis of
``information encapsulation'', whereby it is assumed that the information
processing related to specific functions are relatively unaffected by those
corresponding to other functions, has been suggested as an explanation of
how functional modules can arise from the structural organization of the 
connectome into several communities~\cite{Colombo2013}. Although this may appear intuitive
because spreading processes are generally fast within a module and slow down
during their passage to a different module~\cite{Pan2009}, we find on
the contrary that the specific modular organization of the Macaque connectome
allows signals to spread very fast. In fact, the communication of
information across the empirical network appears to be even faster than that
seen in equivalent networks whose connections are distributed homogeneously.
This is surprising as homogeneous networks tend to exhibit the fastest
speed of propagation globally, which usually tends to reduce once mesoscopic
structural features such as modularity are introduced~\cite{Pan2009}.
We connect this counter-intuitive result to the detailed meso-level
attributes of
the topological organization, specifically the
roles played by different brain regions in terms of their
intra- and inter-modular connections.
By analyzing these connections we
reveal distinctive features of the connectome, namely,
the tendency of provincial hubs within a module to connect to each other,
and the preference shown by connector hubs to link to peripheral nodes 
across different modules.

While the potential of rapid communication between different regions,
made possible by the underlying modular architecture of the network,
suggests a plausible explanation for the evolution of the observed mesoscopic
organization of the macaque brain, it could also plausibly be a consequence
of optimizing for
wiring lengths. However, we have explicitly shown that the constraint
imposed by
the physical distance between the brain regions is insufficient to explain 
the modular partitions observed by us.
Indeed, although the five modules of the connectome that we have identified
comprise brain regions that are, for the most part, spatially proximal,
module \#4 is a prominent exception.
It spans two widely separated locations in the brain, one comprising the
primary and secondary auditory areas which are in the temporal lobe and
the other consisting of association areas located in the prefrontal lobe.
While it is well-established that the temporal lobe regions belonging 
to this module contribute to its associated sensory modality,
viz., auditory processing, it is not entirely clear what role the prefrontal
regions of this module plays in this context. We note, however, that there
are intriguing parallels between these areas and
those occupying corresponding locations in the human brain.
Specifically, the prefrontal and temporal parts of module \#4 that
are known to have a role in social cognition in 
primates~\cite{Sallet2011,Platt2016} correspond to the
Broca's and Wernicke's areas in the human brain, respectively. 
As is well known, the former is responsible for speech production in humans, 
while the latter is critical for language comprehension~\cite{Kandel2000}. 
Although
there is no direct counterpart of language in Macaques, non-human primates are
known to be capable of communicating through signals such as facial expressions
and vocalizations~\cite{Cheney2018}.
This correspondence therefore warrants consideration of
whether some of the areas in
module \#4 of the Macaque brain developed from a common evolutionary
precursor of the apparatus responsible for facilitating language 
in humans. Indeed, this view is supported by recent research~\cite{Wilson2015,Snowdon2017,Rauschecker2018}
that have used language-like behavior
in non-human primates as models for understanding how speech
and language might have evolved in humans~\cite{Kolodny2018}.

\begin{acknowledgments}
We would like to thank Raghavendhra Singh, Sridharan Devarajan and Dipanjan Ray for helpful discussions.
SNM has been supported by the IMSc Complex Systems Project (12th 
Plan), and the Center of Excellence in Complex Systems and Data 
Science, both funded by the Department of Atomic Energy, Government of 
India. The simulations and computations required for this work were 
supported by High Performance Computing facility (Nandadevi and 
Satpura) of The Institute of Mathematical Sciences, which is partially 
funded by DST.
\end{acknowledgments}

\bibliography{references}

\begin{thebibliography}{93}%
\makeatletter
\providecommand \@ifxundefined [1]{%
 \@ifx{#1\undefined}
}%
\providecommand \@ifnum [1]{%
 \ifnum #1\expandafter \@firstoftwo
 \else \expandafter \@secondoftwo
 \fi
}%
\providecommand \@ifx [1]{%
 \ifx #1\expandafter \@firstoftwo
 \else \expandafter \@secondoftwo
 \fi
}%
\providecommand \natexlab [1]{#1}%
\providecommand \enquote  [1]{``#1''}%
\providecommand \bibnamefont  [1]{#1}%
\providecommand \bibfnamefont [1]{#1}%
\providecommand \citenamefont [1]{#1}%
\providecommand \href@noop [0]{\@secondoftwo}%
\providecommand \href [0]{\begingroup \@sanitize@url \@href}%
\providecommand \@href[1]{\@@startlink{#1}\@@href}%
\providecommand \@@href[1]{\endgroup#1\@@endlink}%
\providecommand \@sanitize@url [0]{\catcode `\\12\catcode `\$12\catcode
  `\&12\catcode `\#12\catcode `\^12\catcode `\_12\catcode `\%12\relax}%
\providecommand \@@startlink[1]{}%
\providecommand \@@endlink[0]{}%
\providecommand \url  [0]{\begingroup\@sanitize@url \@url }%
\providecommand \@url [1]{\endgroup\@href {#1}{\urlprefix }}%
\providecommand \urlprefix  [0]{URL }%
\providecommand \Eprint [0]{\href }%
\providecommand \doibase [0]{http://dx.doi.org/}%
\providecommand \selectlanguage [0]{\@gobble}%
\providecommand \bibinfo  [0]{\@secondoftwo}%
\providecommand \bibfield  [0]{\@secondoftwo}%
\providecommand \translation [1]{[#1]}%
\providecommand \BibitemOpen [0]{}%
\providecommand \bibitemStop [0]{}%
\providecommand \bibitemNoStop [0]{.\EOS\space}%
\providecommand \EOS [0]{\spacefactor3000\relax}%
\providecommand \BibitemShut  [1]{\csname bibitem#1\endcsname}%
\let\auto@bib@innerbib\@empty
\bibitem [{\citenamefont {Zilles}\ and\ \citenamefont
  {Amunts}(2010)}]{Zilles2010}%
  \BibitemOpen
  \bibfield  {author} {\bibinfo {author} {\bibfnamefont {K.}~\bibnamefont
  {Zilles}}\ and\ \bibinfo {author} {\bibfnamefont {K.}~\bibnamefont
  {Amunts}},\ }\href {\doibase 10.1038/nrn2776} {\bibfield  {journal} {\bibinfo
   {journal} {Nat Rev Neurosci}\ }\textbf {\bibinfo {volume} {11}},\ \bibinfo
  {pages} {139} (\bibinfo {year} {2010})}\BibitemShut {NoStop}%
\bibitem [{\citenamefont {Goulas}\ \emph {et~al.}(2019)\citenamefont {Goulas},
  \citenamefont {Betzel},\ and\ \citenamefont {Hilgetag}}]{Goulas2019}%
  \BibitemOpen
  \bibfield  {author} {\bibinfo {author} {\bibfnamefont {A.}~\bibnamefont
  {Goulas}}, \bibinfo {author} {\bibfnamefont {R.~F.}\ \bibnamefont {Betzel}},
  \ and\ \bibinfo {author} {\bibfnamefont {C.~C.}\ \bibnamefont {Hilgetag}},\
  }\href {\doibase 10.1126/sciadv.aav9694} {\bibfield  {journal} {\bibinfo
  {journal} {Sci Adv}\ }\textbf {\bibinfo {volume} {5}},\ \bibinfo {pages}
  {eaav9694} (\bibinfo {year} {2019})}\BibitemShut {NoStop}%
\bibitem [{\citenamefont {Swanson}\ and\ \citenamefont
  {Bota}(2010)}]{Swanson2010}%
  \BibitemOpen
  \bibfield  {author} {\bibinfo {author} {\bibfnamefont {L.~W.}\ \bibnamefont
  {Swanson}}\ and\ \bibinfo {author} {\bibfnamefont {M.}~\bibnamefont {Bota}},\
  }\href {\doibase 10.1073/pnas.1015128107} {\bibfield  {journal} {\bibinfo
  {journal} {Proc Natl Acad Sci USA}\ }\textbf {\bibinfo {volume} {107}},\
  \bibinfo {pages} {20610} (\bibinfo {year} {2010})}\BibitemShut {NoStop}%
\bibitem [{\citenamefont {Fox}\ and\ \citenamefont {Friston}(2012)}]{Fox2012}%
  \BibitemOpen
  \bibfield  {author} {\bibinfo {author} {\bibfnamefont {P.~T.}\ \bibnamefont
  {Fox}}\ and\ \bibinfo {author} {\bibfnamefont {K.~J.}\ \bibnamefont
  {Friston}},\ }\href {\doibase 10.1016/j.neuroimage.2011.12.051} {\bibfield
  {journal} {\bibinfo  {journal} {Neuroimage}\ }\textbf {\bibinfo {volume}
  {61}},\ \bibinfo {pages} {407} (\bibinfo {year} {2012})}\BibitemShut
  {NoStop}%
\bibitem [{\citenamefont {Friston}(2002)}]{Friston2002}%
  \BibitemOpen
  \bibfield  {author} {\bibinfo {author} {\bibfnamefont {K.}~\bibnamefont
  {Friston}},\ }\href {\doibase 10.1146/annurev.neuro.25.112701.142846}
  {\bibfield  {journal} {\bibinfo  {journal} {Annu Rev Neurosci}\ }\textbf
  {\bibinfo {volume} {25}},\ \bibinfo {pages} {221} (\bibinfo {year}
  {2002})}\BibitemShut {NoStop}%
\bibitem [{\citenamefont {Deco}\ \emph {et~al.}(2008)\citenamefont {Deco},
  \citenamefont {Jirsa}, \citenamefont {Robinson}, \citenamefont {Breakspear},\
  and\ \citenamefont {Friston}}]{Deco2008}%
  \BibitemOpen
  \bibfield  {author} {\bibinfo {author} {\bibfnamefont {G.}~\bibnamefont
  {Deco}}, \bibinfo {author} {\bibfnamefont {V.~K.}\ \bibnamefont {Jirsa}},
  \bibinfo {author} {\bibfnamefont {P.~A.}\ \bibnamefont {Robinson}}, \bibinfo
  {author} {\bibfnamefont {M.}~\bibnamefont {Breakspear}}, \ and\ \bibinfo
  {author} {\bibfnamefont {K.}~\bibnamefont {Friston}},\ }\href {\doibase
  10.1371/journal.pcbi.1000092} {\bibfield  {journal} {\bibinfo  {journal}
  {PLoS Comput Biol}\ }\textbf {\bibinfo {volume} {4}},\ \bibinfo {pages}
  {e1000092} (\bibinfo {year} {2008})}\BibitemShut {NoStop}%
\bibitem [{\citenamefont {Scannell}\ \emph {et~al.}(1999)\citenamefont
  {Scannell}, \citenamefont {Burns}, \citenamefont {Hilgetag}, \citenamefont
  {O'Neil},\ and\ \citenamefont {Young}}]{Scannell1999}%
  \BibitemOpen
  \bibfield  {author} {\bibinfo {author} {\bibfnamefont {J.~W.}\ \bibnamefont
  {Scannell}}, \bibinfo {author} {\bibfnamefont {G.~A. P.~C.}\ \bibnamefont
  {Burns}}, \bibinfo {author} {\bibfnamefont {C.~C.}\ \bibnamefont {Hilgetag}},
  \bibinfo {author} {\bibfnamefont {M.~A.}\ \bibnamefont {O'Neil}}, \ and\
  \bibinfo {author} {\bibfnamefont {M.~P.}\ \bibnamefont {Young}},\ }\href
  {\doibase 10.1093/cercor/9.3.277} {\bibfield  {journal} {\bibinfo  {journal}
  {Cereb Cortex}\ }\textbf {\bibinfo {volume} {9}},\ \bibinfo {pages} {277}
  (\bibinfo {year} {1999})}\BibitemShut {NoStop}%
\bibitem [{\citenamefont {Hilgetag}\ \emph {et~al.}(2000)\citenamefont
  {Hilgetag}, \citenamefont {Burns}, \citenamefont {O'Neill}, \citenamefont
  {Scannell},\ and\ \citenamefont {Young}}]{Hilgetag2000}%
  \BibitemOpen
  \bibfield  {author} {\bibinfo {author} {\bibfnamefont {C.~C.}\ \bibnamefont
  {Hilgetag}}, \bibinfo {author} {\bibfnamefont {G.~A. P.~C.}\ \bibnamefont
  {Burns}}, \bibinfo {author} {\bibfnamefont {M.~A.}\ \bibnamefont {O'Neill}},
  \bibinfo {author} {\bibfnamefont {J.~W.}\ \bibnamefont {Scannell}}, \ and\
  \bibinfo {author} {\bibfnamefont {M.~P.}\ \bibnamefont {Young}},\ }\href
  {\doibase 10.1098/rstb.2000.0551} {\bibfield  {journal} {\bibinfo  {journal}
  {Philos Trans R Soc London, Ser B}\ }\textbf {\bibinfo {volume} {355}},\
  \bibinfo {pages} {91} (\bibinfo {year} {2000})}\BibitemShut {NoStop}%
\bibitem [{\citenamefont {Bassett}\ \emph {et~al.}(2010)\citenamefont
  {Bassett}, \citenamefont {Greenfield}, \citenamefont {Meyer-Lindenberg},
  \citenamefont {Weinberger}, \citenamefont {Moore},\ and\ \citenamefont
  {Bullmore}}]{Bassett2010}%
  \BibitemOpen
  \bibfield  {author} {\bibinfo {author} {\bibfnamefont {D.~S.}\ \bibnamefont
  {Bassett}}, \bibinfo {author} {\bibfnamefont {D.~L.}\ \bibnamefont
  {Greenfield}}, \bibinfo {author} {\bibfnamefont {A.}~\bibnamefont
  {Meyer-Lindenberg}}, \bibinfo {author} {\bibfnamefont {D.~R.}\ \bibnamefont
  {Weinberger}}, \bibinfo {author} {\bibfnamefont {S.~W.}\ \bibnamefont
  {Moore}}, \ and\ \bibinfo {author} {\bibfnamefont {E.~T.}\ \bibnamefont
  {Bullmore}},\ }\href {\doibase 10.1371/journal.pcbi.1000748} {\bibfield
  {journal} {\bibinfo  {journal} {PLoS Comput Biol}\ }\textbf {\bibinfo
  {volume} {6}},\ \bibinfo {pages} {e1000748} (\bibinfo {year}
  {2010})}\BibitemShut {NoStop}%
\bibitem [{\citenamefont {Pan}\ \emph {et~al.}(2010)\citenamefont {Pan},
  \citenamefont {Chatterjee},\ and\ \citenamefont {Sinha}}]{Pan2010}%
  \BibitemOpen
  \bibfield  {author} {\bibinfo {author} {\bibfnamefont {R.~K.}\ \bibnamefont
  {Pan}}, \bibinfo {author} {\bibfnamefont {N.}~\bibnamefont {Chatterjee}}, \
  and\ \bibinfo {author} {\bibfnamefont {S.}~\bibnamefont {Sinha}},\ }\href
  {\doibase 10.1371/journal.pone.0009240} {\bibfield  {journal} {\bibinfo
  {journal} {PLoS One}\ }\textbf {\bibinfo {volume} {5}},\ \bibinfo {pages}
  {e9240} (\bibinfo {year} {2010})}\BibitemShut {NoStop}%
\bibitem [{\citenamefont {Wang}\ \emph {et~al.}(2012)\citenamefont {Wang},
  \citenamefont {Sporns},\ and\ \citenamefont {Burkhalter}}]{Wang2012}%
  \BibitemOpen
  \bibfield  {author} {\bibinfo {author} {\bibfnamefont {Q.}~\bibnamefont
  {Wang}}, \bibinfo {author} {\bibfnamefont {O.}~\bibnamefont {Sporns}}, \ and\
  \bibinfo {author} {\bibfnamefont {A.}~\bibnamefont {Burkhalter}},\ }\href
  {\doibase 10.1523/JNEUROSCI.6063-11.2012} {\bibfield  {journal} {\bibinfo
  {journal} {J Neurosci}\ }\textbf {\bibinfo {volume} {32}},\ \bibinfo {pages}
  {4386} (\bibinfo {year} {2012})}\BibitemShut {NoStop}%
\bibitem [{\citenamefont {Harriger}\ \emph {et~al.}(2012)\citenamefont
  {Harriger}, \citenamefont {Van Den~Heuvel},\ and\ \citenamefont
  {Sporns}}]{Harriger2012}%
  \BibitemOpen
  \bibfield  {author} {\bibinfo {author} {\bibfnamefont {L.}~\bibnamefont
  {Harriger}}, \bibinfo {author} {\bibfnamefont {M.~P.}\ \bibnamefont {Van
  Den~Heuvel}}, \ and\ \bibinfo {author} {\bibfnamefont {O.}~\bibnamefont
  {Sporns}},\ }\href {\doibase 10.1371/journal.pone.0046497} {\bibfield
  {journal} {\bibinfo  {journal} {PloS One}\ }\textbf {\bibinfo {volume} {7}},\
  \bibinfo {pages} {e46497} (\bibinfo {year} {2012})}\BibitemShut {NoStop}%
\bibitem [{\citenamefont {Shanahan}\ \emph {et~al.}(2013)\citenamefont
  {Shanahan}, \citenamefont {Bingman}, \citenamefont {Shimizu}, \citenamefont
  {Wild},\ and\ \citenamefont {G{\"u}nt{\"u}rk{\"u}n}}]{Shanahan2013}%
  \BibitemOpen
  \bibfield  {author} {\bibinfo {author} {\bibfnamefont {M.}~\bibnamefont
  {Shanahan}}, \bibinfo {author} {\bibfnamefont {V.~P.}\ \bibnamefont
  {Bingman}}, \bibinfo {author} {\bibfnamefont {T.}~\bibnamefont {Shimizu}},
  \bibinfo {author} {\bibfnamefont {M.}~\bibnamefont {Wild}}, \ and\ \bibinfo
  {author} {\bibfnamefont {O.}~\bibnamefont {G{\"u}nt{\"u}rk{\"u}n}},\ }\href
  {\doibase 10.3389/fncom.2013.00089} {\bibfield  {journal} {\bibinfo
  {journal} {Front Comput Neurosci}\ }\textbf {\bibinfo {volume} {7}},\
  \bibinfo {pages} {89} (\bibinfo {year} {2013})}\BibitemShut {NoStop}%
\bibitem [{\citenamefont {Shih}\ \emph {et~al.}(2015)\citenamefont {Shih},
  \citenamefont {Sporns}, \citenamefont {Yuan}, \citenamefont {Su},
  \citenamefont {Lin}, \citenamefont {Chuang}, \citenamefont {Wang},
  \citenamefont {Lo}, \citenamefont {Greenspan},\ and\ \citenamefont
  {Chiang}}]{Shih2015}%
  \BibitemOpen
  \bibfield  {author} {\bibinfo {author} {\bibfnamefont {C.-T.}\ \bibnamefont
  {Shih}}, \bibinfo {author} {\bibfnamefont {O.}~\bibnamefont {Sporns}},
  \bibinfo {author} {\bibfnamefont {S.-L.}\ \bibnamefont {Yuan}}, \bibinfo
  {author} {\bibfnamefont {T.-S.}\ \bibnamefont {Su}}, \bibinfo {author}
  {\bibfnamefont {Y.-J.}\ \bibnamefont {Lin}}, \bibinfo {author} {\bibfnamefont
  {C.-C.}\ \bibnamefont {Chuang}}, \bibinfo {author} {\bibfnamefont {T.-Y.}\
  \bibnamefont {Wang}}, \bibinfo {author} {\bibfnamefont {C.-C.}\ \bibnamefont
  {Lo}}, \bibinfo {author} {\bibfnamefont {R.~J.}\ \bibnamefont {Greenspan}}, \
  and\ \bibinfo {author} {\bibfnamefont {A.-S.}\ \bibnamefont {Chiang}},\
  }\href {\doibase 10.1016/j.cub.2015.03.021} {\bibfield  {journal} {\bibinfo
  {journal} {Curr Biol}\ }\textbf {\bibinfo {volume} {25}},\ \bibinfo {pages}
  {1249} (\bibinfo {year} {2015})}\BibitemShut {NoStop}%
\bibitem [{\citenamefont {Sporns}\ and\ \citenamefont
  {Betzel}(2016)}]{Sporns2016}%
  \BibitemOpen
  \bibfield  {author} {\bibinfo {author} {\bibfnamefont {O.}~\bibnamefont
  {Sporns}}\ and\ \bibinfo {author} {\bibfnamefont {R.~F.}\ \bibnamefont
  {Betzel}},\ }\href {\doibase 10.1146/annurev-psych-122414-033634} {\bibfield
  {journal} {\bibinfo  {journal} {Annu Rev Psychol}\ }\textbf {\bibinfo
  {volume} {67}},\ \bibinfo {pages} {613} (\bibinfo {year} {2016})}\BibitemShut
  {NoStop}%
\bibitem [{\citenamefont {Chen}\ \emph {et~al.}(2020)\citenamefont {Chen},
  \citenamefont {Zhang}, \citenamefont {He},\ and\ \citenamefont
  {Zhou}}]{Chen2020}%
  \BibitemOpen
  \bibfield  {author} {\bibinfo {author} {\bibfnamefont {Y.}~\bibnamefont
  {Chen}}, \bibinfo {author} {\bibfnamefont {Z.-K.}\ \bibnamefont {Zhang}},
  \bibinfo {author} {\bibfnamefont {Y.}~\bibnamefont {He}}, \ and\ \bibinfo
  {author} {\bibfnamefont {C.}~\bibnamefont {Zhou}},\ }\href {\doibase
  10.1093/cercor/bhaa060} {\bibfield  {journal} {\bibinfo  {journal} {Cereb
  Cortex}\ } (\bibinfo {year} {2020}),\ 10.1093/cercor/bhaa060}\BibitemShut
  {NoStop}%
\bibitem [{\citenamefont {Meunier}\ \emph {et~al.}(2010)\citenamefont
  {Meunier}, \citenamefont {Lambiotte},\ and\ \citenamefont
  {Bullmore}}]{Meunier2010}%
  \BibitemOpen
  \bibfield  {author} {\bibinfo {author} {\bibfnamefont {D.}~\bibnamefont
  {Meunier}}, \bibinfo {author} {\bibfnamefont {R.}~\bibnamefont {Lambiotte}},
  \ and\ \bibinfo {author} {\bibfnamefont {E.~T.}\ \bibnamefont {Bullmore}},\
  }\href {\doibase 10.3389/fnins.2010.00200} {\bibfield  {journal} {\bibinfo
  {journal} {Front Neurosci}\ }\textbf {\bibinfo {volume} {4}},\ \bibinfo
  {pages} {200} (\bibinfo {year} {2010})}\BibitemShut {NoStop}%
\bibitem [{\citenamefont {Pan}\ and\ \citenamefont {Sinha}(2007)}]{Pan2007}%
  \BibitemOpen
  \bibfield  {author} {\bibinfo {author} {\bibfnamefont {R.~K.}\ \bibnamefont
  {Pan}}\ and\ \bibinfo {author} {\bibfnamefont {S.}~\bibnamefont {Sinha}},\
  }\href {\doibase 10.1103/PhysRevE.76.045103} {\bibfield  {journal} {\bibinfo
  {journal} {Phys Rev E}\ }\textbf {\bibinfo {volume} {76}},\ \bibinfo {pages}
  {045103} (\bibinfo {year} {2007})}\BibitemShut {NoStop}%
\bibitem [{\citenamefont {Fodor}(1983)}]{Fodor1983}%
  \BibitemOpen
  \bibfield  {author} {\bibinfo {author} {\bibfnamefont {J.~A.}\ \bibnamefont
  {Fodor}},\ }\href@noop {} {\emph {\bibinfo {title} {The Modularity of
  Mind}}}\ (\bibinfo  {publisher} {MIT Press},\ \bibinfo {address} {Cambridge
  MA},\ \bibinfo {year} {1983})\BibitemShut {NoStop}%
\bibitem [{\citenamefont {Park}\ and\ \citenamefont
  {Friston}(2013)}]{Park2013}%
  \BibitemOpen
  \bibfield  {author} {\bibinfo {author} {\bibfnamefont {H.-J.}\ \bibnamefont
  {Park}}\ and\ \bibinfo {author} {\bibfnamefont {K.}~\bibnamefont {Friston}},\
  }\href {\doibase 10.1126/science.1238411} {\bibfield  {journal} {\bibinfo
  {journal} {Science}\ }\textbf {\bibinfo {volume} {342}},\ \bibinfo {pages}
  {1238411} (\bibinfo {year} {2013})}\BibitemShut {NoStop}%
\bibitem [{\citenamefont {Swanson}\ and\ \citenamefont
  {Lichtman}(2016)}]{Swanson2016}%
  \BibitemOpen
  \bibfield  {author} {\bibinfo {author} {\bibfnamefont {L.~W.}\ \bibnamefont
  {Swanson}}\ and\ \bibinfo {author} {\bibfnamefont {J.~W.}\ \bibnamefont
  {Lichtman}},\ }\href {\doibase 10.1146/annurev-neuro-071714-033954}
  {\bibfield  {journal} {\bibinfo  {journal} {Annu Rev Neurosci}\ }\textbf
  {\bibinfo {volume} {39}},\ \bibinfo {pages} {197} (\bibinfo {year}
  {2016})}\BibitemShut {NoStop}%
\bibitem [{\citenamefont {Pan}\ and\ \citenamefont {Sinha}(2009)}]{Pan2009}%
  \BibitemOpen
  \bibfield  {author} {\bibinfo {author} {\bibfnamefont {R.~K.}\ \bibnamefont
  {Pan}}\ and\ \bibinfo {author} {\bibfnamefont {S.}~\bibnamefont {Sinha}},\
  }\href {\doibase 10.1209/0295-5075/85/68006/meta} {\bibfield  {journal}
  {\bibinfo  {journal} {EPL}\ }\textbf {\bibinfo {volume} {85}},\ \bibinfo
  {pages} {68006} (\bibinfo {year} {2009})}\BibitemShut {NoStop}%
\bibitem [{\citenamefont {Mumford}(1991)}]{Mumford1991}%
  \BibitemOpen
  \bibfield  {author} {\bibinfo {author} {\bibfnamefont {D.}~\bibnamefont
  {Mumford}},\ }\href {\doibase 10.1007/BF00202389} {\bibfield  {journal}
  {\bibinfo  {journal} {Biol Cybern}\ }\textbf {\bibinfo {volume} {65}},\
  \bibinfo {pages} {135} (\bibinfo {year} {1991})}\BibitemShut {NoStop}%
\bibitem [{\citenamefont {Antonopoulos}\ \emph {et~al.}(2015)\citenamefont
  {Antonopoulos}, \citenamefont {Srivastava}, \citenamefont {Pinto},\ and\
  \citenamefont {Baptista}}]{Antonopoulos2015}%
  \BibitemOpen
  \bibfield  {author} {\bibinfo {author} {\bibfnamefont {C.~G.}\ \bibnamefont
  {Antonopoulos}}, \bibinfo {author} {\bibfnamefont {S.}~\bibnamefont
  {Srivastava}}, \bibinfo {author} {\bibfnamefont {S.~E. d.~S.}\ \bibnamefont
  {Pinto}}, \ and\ \bibinfo {author} {\bibfnamefont {M.~S.}\ \bibnamefont
  {Baptista}},\ }\href {\doibase 10.1371/journal.pcbi.1004372} {\bibfield
  {journal} {\bibinfo  {journal} {PLoS Comput Biol}\ }\textbf {\bibinfo
  {volume} {11}},\ \bibinfo {pages} {e1004372} (\bibinfo {year}
  {2015})}\BibitemShut {NoStop}%
\bibitem [{\citenamefont {Yamaguti}\ and\ \citenamefont
  {Tsuda}(2015)}]{Yamaguti2015}%
  \BibitemOpen
  \bibfield  {author} {\bibinfo {author} {\bibfnamefont {Y.}~\bibnamefont
  {Yamaguti}}\ and\ \bibinfo {author} {\bibfnamefont {I.}~\bibnamefont
  {Tsuda}},\ }\href {\doibase 10.1016/j.neunet.2014.07.013} {\bibfield
  {journal} {\bibinfo  {journal} {Neural Netw}\ }\textbf {\bibinfo {volume}
  {62}},\ \bibinfo {pages} {3} (\bibinfo {year} {2015})}\BibitemShut {NoStop}%
\bibitem [{\citenamefont {Betzel}\ \emph {et~al.}(2017)\citenamefont {Betzel},
  \citenamefont {Medaglia}, \citenamefont {Papadopoulos}, \citenamefont {Baum},
  \citenamefont {Gur}, \citenamefont {Gur}, \citenamefont {Roalf},
  \citenamefont {Satterthwaite},\ and\ \citenamefont {Bassett}}]{Betzel2017}%
  \BibitemOpen
  \bibfield  {author} {\bibinfo {author} {\bibfnamefont {R.~F.}\ \bibnamefont
  {Betzel}}, \bibinfo {author} {\bibfnamefont {J.~D.}\ \bibnamefont
  {Medaglia}}, \bibinfo {author} {\bibfnamefont {L.}~\bibnamefont
  {Papadopoulos}}, \bibinfo {author} {\bibfnamefont {G.~L.}\ \bibnamefont
  {Baum}}, \bibinfo {author} {\bibfnamefont {R.}~\bibnamefont {Gur}}, \bibinfo
  {author} {\bibfnamefont {R.}~\bibnamefont {Gur}}, \bibinfo {author}
  {\bibfnamefont {D.}~\bibnamefont {Roalf}}, \bibinfo {author} {\bibfnamefont
  {T.~D.}\ \bibnamefont {Satterthwaite}}, \ and\ \bibinfo {author}
  {\bibfnamefont {D.~S.}\ \bibnamefont {Bassett}},\ }\href {\doibase
  10.1162/NETN_a_00002} {\bibfield  {journal} {\bibinfo  {journal} {Network
  Neuroscience}\ }\textbf {\bibinfo {volume} {1}},\ \bibinfo {pages} {42}
  (\bibinfo {year} {2017})}\BibitemShut {NoStop}%
\bibitem [{\citenamefont {Stiso}\ and\ \citenamefont
  {Bassett}(2018)}]{Stiso2018}%
  \BibitemOpen
  \bibfield  {author} {\bibinfo {author} {\bibfnamefont {J.}~\bibnamefont
  {Stiso}}\ and\ \bibinfo {author} {\bibfnamefont {D.~S.}\ \bibnamefont
  {Bassett}},\ }\href {\doibase 10.1016/j.tics.2018.09.007} {\bibfield
  {journal} {\bibinfo  {journal} {Trends Cogn Sci}\ }\textbf {\bibinfo {volume}
  {22}},\ \bibinfo {pages} {1127} (\bibinfo {year} {2018})}\BibitemShut
  {NoStop}%
\bibitem [{\citenamefont {Modha}\ and\ \citenamefont
  {Singh}(2010)}]{Modha2010}%
  \BibitemOpen
  \bibfield  {author} {\bibinfo {author} {\bibfnamefont {D.~S.}\ \bibnamefont
  {Modha}}\ and\ \bibinfo {author} {\bibfnamefont {R.}~\bibnamefont {Singh}},\
  }\href {\doibase 10.1073/pnas.1008054107} {\bibfield  {journal} {\bibinfo
  {journal} {Proc Natl Acad Sci USA}\ }\textbf {\bibinfo {volume} {107}},\
  \bibinfo {pages} {13485} (\bibinfo {year} {2010})}\BibitemShut {NoStop}%
\bibitem [{\citenamefont {Stephan}\ \emph {et~al.}(2000)\citenamefont
  {Stephan}, \citenamefont {Zilles},\ and\ \citenamefont
  {Kötter}}]{Stephan2000}%
  \BibitemOpen
  \bibfield  {author} {\bibinfo {author} {\bibfnamefont {K.~E.}\ \bibnamefont
  {Stephan}}, \bibinfo {author} {\bibfnamefont {K.}~\bibnamefont {Zilles}}, \
  and\ \bibinfo {author} {\bibfnamefont {R.}~\bibnamefont {Kötter}},\ }\href
  {\doibase 10.1098/rstb.2000.0548} {\bibfield  {journal} {\bibinfo  {journal}
  {Philos Trans R Soc London, Ser B}\ }\textbf {\bibinfo {volume} {355}},\
  \bibinfo {pages} {37} (\bibinfo {year} {2000})}\BibitemShut {NoStop}%
\bibitem [{\citenamefont {Stephan}\ \emph {et~al.}(2001)\citenamefont
  {Stephan}, \citenamefont {Kamper}, \citenamefont {Bozkurt}, \citenamefont
  {Burns}, \citenamefont {Young},\ and\ \citenamefont {Kötter}}]{Stephan2001}%
  \BibitemOpen
  \bibfield  {author} {\bibinfo {author} {\bibfnamefont {K.~E.}\ \bibnamefont
  {Stephan}}, \bibinfo {author} {\bibfnamefont {L.}~\bibnamefont {Kamper}},
  \bibinfo {author} {\bibfnamefont {A.}~\bibnamefont {Bozkurt}}, \bibinfo
  {author} {\bibfnamefont {G.~A. P.~C.}\ \bibnamefont {Burns}}, \bibinfo
  {author} {\bibfnamefont {M.~P.}\ \bibnamefont {Young}}, \ and\ \bibinfo
  {author} {\bibfnamefont {R.}~\bibnamefont {Kötter}},\ }\href {\doibase
  10.1098/rstb.2001.0908} {\bibfield  {journal} {\bibinfo  {journal} {Philos
  Trans R Soc London, Ser B}\ }\textbf {\bibinfo {volume} {356}},\ \bibinfo
  {pages} {1159} (\bibinfo {year} {2001})}\BibitemShut {NoStop}%
\bibitem [{\citenamefont {K{\"o}tter}(2004)}]{Kotter2004}%
  \BibitemOpen
  \bibfield  {author} {\bibinfo {author} {\bibfnamefont {R.}~\bibnamefont
  {K{\"o}tter}},\ }\href {\doibase 10.1385/NI:2:2:127} {\bibfield  {journal}
  {\bibinfo  {journal} {Neuroinformatics}\ }\textbf {\bibinfo {volume} {2}},\
  \bibinfo {pages} {127} (\bibinfo {year} {2004})}\BibitemShut {NoStop}%
\bibitem [{Pax()}]{Paxinoswebsite}%
  \BibitemOpen
  \href@noop {} {}\bibinfo {howpublished}
  {\url{https://scalablebrainatlas.incf.org/macaque/PHT00}}\BibitemShut
  {NoStop}%
\bibitem [{\citenamefont {Paxinos}\ \emph {et~al.}(2000)\citenamefont
  {Paxinos}, \citenamefont {Huang},\ and\ \citenamefont {Toga}}]{Paxinos2000}%
  \BibitemOpen
  \bibfield  {author} {\bibinfo {author} {\bibfnamefont {G.}~\bibnamefont
  {Paxinos}}, \bibinfo {author} {\bibfnamefont {X.~F.}\ \bibnamefont {Huang}},
  \ and\ \bibinfo {author} {\bibfnamefont {A.~W.}\ \bibnamefont {Toga}},\
  }\href {https://books.google.co.in/books?id=x-6bQgAACAAJ} {\emph {\bibinfo
  {title} {The Rhesus Monkey Brain in Stereotaxic Coordinates}}}\ (\bibinfo
  {publisher} {Academic Press},\ \bibinfo {address} {San Diego, CA},\ \bibinfo
  {year} {2000})\BibitemShut {NoStop}%
\bibitem [{\citenamefont {Newman}(2004)}]{Newman2004b}%
  \BibitemOpen
  \bibfield  {author} {\bibinfo {author} {\bibfnamefont {M.~E.~J.}\
  \bibnamefont {Newman}},\ }\href {\doibase 10.1140/epjb/e2004-00124-y}
  {\bibfield  {journal} {\bibinfo  {journal} {Eur Phys J B}\ }\textbf {\bibinfo
  {volume} {38}},\ \bibinfo {pages} {321} (\bibinfo {year} {2004})}\BibitemShut
  {NoStop}%
\bibitem [{\citenamefont {Newman}\ and\ \citenamefont
  {Girvan}(2004)}]{Newman2004}%
  \BibitemOpen
  \bibfield  {author} {\bibinfo {author} {\bibfnamefont {M.~E.~J.}\
  \bibnamefont {Newman}}\ and\ \bibinfo {author} {\bibfnamefont
  {M.}~\bibnamefont {Girvan}},\ }\href {\doibase 10.1103/PhysRevE.69.026113}
  {\bibfield  {journal} {\bibinfo  {journal} {Phys Rev E}\ }\textbf {\bibinfo
  {volume} {69}},\ \bibinfo {pages} {026113} (\bibinfo {year}
  {2004})}\BibitemShut {NoStop}%
\bibitem [{\citenamefont {Newman}(2006)}]{Newman2006}%
  \BibitemOpen
  \bibfield  {author} {\bibinfo {author} {\bibfnamefont {M.~E.~J.}\
  \bibnamefont {Newman}},\ }\href {\doibase 10.1073/pnas.0601602103} {\bibfield
   {journal} {\bibinfo  {journal} {Proc Natl Acad Sci USA}\ }\textbf {\bibinfo
  {volume} {103}},\ \bibinfo {pages} {8577} (\bibinfo {year}
  {2006})}\BibitemShut {NoStop}%
\bibitem [{\citenamefont {Good}\ \emph {et~al.}(2010)\citenamefont {Good},
  \citenamefont {De~Montjoye},\ and\ \citenamefont {Clauset}}]{Clauset2010}%
  \BibitemOpen
  \bibfield  {author} {\bibinfo {author} {\bibfnamefont {B.~H.}\ \bibnamefont
  {Good}}, \bibinfo {author} {\bibfnamefont {Y.~A.}\ \bibnamefont
  {De~Montjoye}}, \ and\ \bibinfo {author} {\bibfnamefont {A.}~\bibnamefont
  {Clauset}},\ }\href {\doibase 10.1103/PhysRevE.81.046106} {\bibfield
  {journal} {\bibinfo  {journal} {Phys Rev E}\ }\textbf {\bibinfo {volume}
  {81}},\ \bibinfo {pages} {046106} (\bibinfo {year} {2010})}\BibitemShut
  {NoStop}%
\bibitem [{Cla()}]{Clausetsite}%
  \BibitemOpen
  \href@noop {} {}\bibinfo {howpublished}
  {\url{http://tuvalu.santafe.edu/~aaronc/modularity/}}\BibitemShut {NoStop}%
\bibitem [{\citenamefont {Rosvall}\ and\ \citenamefont
  {Bergstrom}(2008)}]{Rosvall2008}%
  \BibitemOpen
  \bibfield  {author} {\bibinfo {author} {\bibfnamefont {M.}~\bibnamefont
  {Rosvall}}\ and\ \bibinfo {author} {\bibfnamefont {C.~T.}\ \bibnamefont
  {Bergstrom}},\ }\href {\doibase 10.1073/pnas.0706851105} {\bibfield
  {journal} {\bibinfo  {journal} {Proc Natl Acad Sci USA}\ }\textbf {\bibinfo
  {volume} {105}},\ \bibinfo {pages} {1118} (\bibinfo {year}
  {2008})}\BibitemShut {NoStop}%
\bibitem [{\citenamefont {Guimer{\`a}}\ and\ \citenamefont
  {Amaral}(2005)}]{Guimera2005}%
  \BibitemOpen
  \bibfield  {author} {\bibinfo {author} {\bibfnamefont {R.}~\bibnamefont
  {Guimer{\`a}}}\ and\ \bibinfo {author} {\bibfnamefont {L.~A.~N.}\
  \bibnamefont {Amaral}},\ }\href {\doibase 10.1038/nature03288} {\bibfield
  {journal} {\bibinfo  {journal} {Nature (London)}\ }\textbf {\bibinfo {volume}
  {433}},\ \bibinfo {pages} {895} (\bibinfo {year} {2005})}\BibitemShut
  {NoStop}%
\bibitem [{\citenamefont {Guimera}\ \emph {et~al.}(2007)\citenamefont
  {Guimera}, \citenamefont {Sales-Pardo},\ and\ \citenamefont
  {Amaral}}]{Guimera2007}%
  \BibitemOpen
  \bibfield  {author} {\bibinfo {author} {\bibfnamefont {R.}~\bibnamefont
  {Guimera}}, \bibinfo {author} {\bibfnamefont {M.}~\bibnamefont
  {Sales-Pardo}}, \ and\ \bibinfo {author} {\bibfnamefont {L.~A.~N.}\
  \bibnamefont {Amaral}},\ }\href {\doibase 10.1038/nphys489} {\bibfield
  {journal} {\bibinfo  {journal} {Nat Phys}\ }\textbf {\bibinfo {volume} {3}},\
  \bibinfo {pages} {63} (\bibinfo {year} {2007})}\BibitemShut {NoStop}%
\bibitem [{\citenamefont {Expert}\ \emph {et~al.}(2011)\citenamefont {Expert},
  \citenamefont {Evans}, \citenamefont {Blondel},\ and\ \citenamefont
  {Lambiotte}}]{Expert2011}%
  \BibitemOpen
  \bibfield  {author} {\bibinfo {author} {\bibfnamefont {P.}~\bibnamefont
  {Expert}}, \bibinfo {author} {\bibfnamefont {T.~S.}\ \bibnamefont {Evans}},
  \bibinfo {author} {\bibfnamefont {V.~D.}\ \bibnamefont {Blondel}}, \ and\
  \bibinfo {author} {\bibfnamefont {R.}~\bibnamefont {Lambiotte}},\ }\href
  {\doibase 10.1073/pnas.1018962108} {\bibfield  {journal} {\bibinfo  {journal}
  {Proc Natl Acad Sci USA}\ }\textbf {\bibinfo {volume} {108}},\ \bibinfo
  {pages} {7663} (\bibinfo {year} {2011})}\BibitemShut {NoStop}%
\bibitem [{\citenamefont {MacKay}(2003)}]{Mackay2003}%
  \BibitemOpen
  \bibfield  {author} {\bibinfo {author} {\bibfnamefont {D.~J.~C.}\
  \bibnamefont {MacKay}},\ }\href@noop {} {\emph {\bibinfo {title} {Information
  Theory, Inference and Learning Algorithms}}}\ (\bibinfo  {publisher}
  {Cambridge University Press},\ \bibinfo {address} {Cambridge, UK},\ \bibinfo
  {year} {2003})\BibitemShut {NoStop}%
\bibitem [{\citenamefont {Mauri}\ \emph {et~al.}(2017)\citenamefont {Mauri},
  \citenamefont {Elli}, \citenamefont {Caviglia}, \citenamefont {Uboldi},\ and\
  \citenamefont {Azzi}}]{Mauri2017}%
  \BibitemOpen
  \bibfield  {author} {\bibinfo {author} {\bibfnamefont {M.}~\bibnamefont
  {Mauri}}, \bibinfo {author} {\bibfnamefont {T.}~\bibnamefont {Elli}},
  \bibinfo {author} {\bibfnamefont {G.}~\bibnamefont {Caviglia}}, \bibinfo
  {author} {\bibfnamefont {G.}~\bibnamefont {Uboldi}}, \ and\ \bibinfo {author}
  {\bibfnamefont {M.}~\bibnamefont {Azzi}},\ }in\ \href {\doibase
  10.1145/3125571.3125585} {\emph {\bibinfo {booktitle} {Proceedings of the
  12th Biannual Conference on Italian SIGCHI Chapter}}},\ \bibinfo {series and
  number} {CHItaly '17}\ (\bibinfo  {publisher} {ACM},\ \bibinfo {address} {New
  York, NY, USA},\ \bibinfo {year} {2017})\ pp.\ \bibinfo {pages}
  {28:1--28:5}\BibitemShut {NoStop}%
\bibitem [{\citenamefont {MacLean}\ \emph {et~al.}(2005)\citenamefont
  {MacLean}, \citenamefont {Watson}, \citenamefont {Aaron},\ and\ \citenamefont
  {Yuste}}]{Maclean2005}%
  \BibitemOpen
  \bibfield  {author} {\bibinfo {author} {\bibfnamefont {J.~N.}\ \bibnamefont
  {MacLean}}, \bibinfo {author} {\bibfnamefont {B.~O.}\ \bibnamefont {Watson}},
  \bibinfo {author} {\bibfnamefont {G.~B.}\ \bibnamefont {Aaron}}, \ and\
  \bibinfo {author} {\bibfnamefont {R.}~\bibnamefont {Yuste}},\ }\href
  {\doibase 10.1016/j.neuron.2005.09.035} {\bibfield  {journal} {\bibinfo
  {journal} {Neuron}\ }\textbf {\bibinfo {volume} {48}},\ \bibinfo {pages}
  {811} (\bibinfo {year} {2005})}\BibitemShut {NoStop}%
\bibitem [{\citenamefont {Van~Hemmen}\ and\ \citenamefont
  {Sejnowski}(2006)}]{23Problems}%
  \BibitemOpen
  \bibfield  {author} {\bibinfo {author} {\bibfnamefont {L.~J.}\ \bibnamefont
  {Van~Hemmen}}\ and\ \bibinfo {author} {\bibfnamefont {T.~J.}\ \bibnamefont
  {Sejnowski}},\ }\href {https://books.google.co.in/books?id=bk1nDAAAQBAJ}
  {\emph {\bibinfo {title} {23 Problems in Systems Neuroscience}}},\
  Computational Neuroscience Series\ (\bibinfo  {publisher} {Oxford University
  Press, USA},\ \bibinfo {year} {2006})\BibitemShut {NoStop}%
\bibitem [{\citenamefont {Steriade}\ \emph {et~al.}(1993)\citenamefont
  {Steriade}, \citenamefont {McCormick},\ and\ \citenamefont
  {Sejnowski}}]{Steriade1993}%
  \BibitemOpen
  \bibfield  {author} {\bibinfo {author} {\bibfnamefont {M.}~\bibnamefont
  {Steriade}}, \bibinfo {author} {\bibfnamefont {D.~A.}\ \bibnamefont
  {McCormick}}, \ and\ \bibinfo {author} {\bibfnamefont {T.~J.}\ \bibnamefont
  {Sejnowski}},\ }\href {\doibase 10.1126/science.8235588} {\bibfield
  {journal} {\bibinfo  {journal} {Science}\ }\textbf {\bibinfo {volume}
  {262}},\ \bibinfo {pages} {679} (\bibinfo {year} {1993})}\BibitemShut
  {NoStop}%
\bibitem [{\citenamefont {Guillery}(1995)}]{Guillery1995}%
  \BibitemOpen
  \bibfield  {author} {\bibinfo {author} {\bibfnamefont {R.~W.}\ \bibnamefont
  {Guillery}},\ }\href@noop {} {\bibfield  {journal} {\bibinfo  {journal} {J
  Anat}\ }\textbf {\bibinfo {volume} {187}},\ \bibinfo {pages} {583} (\bibinfo
  {year} {1995})}\BibitemShut {NoStop}%
\bibitem [{\citenamefont {Sherman}\ and\ \citenamefont
  {Guillery}(1996)}]{Sherman1996}%
  \BibitemOpen
  \bibfield  {author} {\bibinfo {author} {\bibfnamefont {S.~M.}\ \bibnamefont
  {Sherman}}\ and\ \bibinfo {author} {\bibfnamefont {R.~W.}\ \bibnamefont
  {Guillery}},\ }\href {\doibase 10.1152/jn.1996.76.3.1367} {\bibfield
  {journal} {\bibinfo  {journal} {J Neurophysiol}\ }\textbf {\bibinfo {volume}
  {76}},\ \bibinfo {pages} {1367} (\bibinfo {year} {1996})}\BibitemShut
  {NoStop}%
\bibitem [{\citenamefont {Braun}\ \emph {et~al.}(2003)\citenamefont {Braun},
  \citenamefont {Dumont}, \citenamefont {Duval}, \citenamefont
  {Hamel-H{\'e}bert},\ and\ \citenamefont {Godbout}}]{Braun2003}%
  \BibitemOpen
  \bibfield  {author} {\bibinfo {author} {\bibfnamefont {C.~M.~J.}\
  \bibnamefont {Braun}}, \bibinfo {author} {\bibfnamefont {M.}~\bibnamefont
  {Dumont}}, \bibinfo {author} {\bibfnamefont {J.}~\bibnamefont {Duval}},
  \bibinfo {author} {\bibfnamefont {I.}~\bibnamefont {Hamel-H{\'e}bert}}, \
  and\ \bibinfo {author} {\bibfnamefont {L.}~\bibnamefont {Godbout}},\
  }\href@noop {} {\bibfield  {journal} {\bibinfo  {journal} {J Psychiatry
  Neurosci}\ }\textbf {\bibinfo {volume} {28}},\ \bibinfo {pages} {432}
  (\bibinfo {year} {2003})}\BibitemShut {NoStop}%
\bibitem [{\citenamefont {Sternberg}(2011)}]{Sternberg2011}%
  \BibitemOpen
  \bibfield  {author} {\bibinfo {author} {\bibfnamefont {S.}~\bibnamefont
  {Sternberg}},\ }\href {\doibase 10.1080/02643294.2011.557231} {\bibfield
  {journal} {\bibinfo  {journal} {Cogn Neuropsychol}\ }\textbf {\bibinfo
  {volume} {28}},\ \bibinfo {pages} {156} (\bibinfo {year} {2011})}\BibitemShut
  {NoStop}%
\bibitem [{\citenamefont {Gazzaniga}(2014)}]{Gazzaniga2014}%
  \BibitemOpen
  \bibfield  {author} {\bibinfo {author} {\bibfnamefont {M.~S.}\ \bibnamefont
  {Gazzaniga}},\ }in\ \href@noop {} {\emph {\bibinfo {booktitle} {Self and
  Consciousness}}}\ (\bibinfo  {publisher} {Psychology Press, Abingdon, UK},\
  \bibinfo {year} {2014})\ pp.\ \bibinfo {pages} {96--110}\BibitemShut
  {NoStop}%
\bibitem [{\citenamefont {Hartwell}\ \emph {et~al.}(1999)\citenamefont
  {Hartwell}, \citenamefont {Hopfield}, \citenamefont {Leibler},\ and\
  \citenamefont {Murray}}]{Hartwell2000}%
  \BibitemOpen
  \bibfield  {author} {\bibinfo {author} {\bibfnamefont {L.~H.}\ \bibnamefont
  {Hartwell}}, \bibinfo {author} {\bibfnamefont {J.~J.}\ \bibnamefont
  {Hopfield}}, \bibinfo {author} {\bibfnamefont {S.}~\bibnamefont {Leibler}}, \
  and\ \bibinfo {author} {\bibfnamefont {A.~W.}\ \bibnamefont {Murray}},\
  }\href {\doibase https://doi.org/10.1038/35011540} {\bibfield  {journal}
  {\bibinfo  {journal} {Nature (London)}\ }\textbf {\bibinfo {volume} {402}},\
  \bibinfo {pages} {C47} (\bibinfo {year} {1999})}\BibitemShut {NoStop}%
\bibitem [{\citenamefont {Purves}\ \emph {et~al.}(2001)\citenamefont {Purves},
  \citenamefont {Augustine}, \citenamefont {Fitzpatrick}, \citenamefont {Hall},
  \citenamefont {LaMantia}, \citenamefont {McNamara},\ and\ \citenamefont
  {White}}]{Purves2001}%
  \BibitemOpen
  \bibfield  {author} {\bibinfo {author} {\bibfnamefont {D.~E.}\ \bibnamefont
  {Purves}}, \bibinfo {author} {\bibfnamefont {G.~J.}\ \bibnamefont
  {Augustine}}, \bibinfo {author} {\bibfnamefont {D.~E.}\ \bibnamefont
  {Fitzpatrick}}, \bibinfo {author} {\bibfnamefont {W.~C.}\ \bibnamefont
  {Hall}}, \bibinfo {author} {\bibfnamefont {A.-S.~E.}\ \bibnamefont
  {LaMantia}}, \bibinfo {author} {\bibfnamefont {J.~O.}\ \bibnamefont
  {McNamara}}, \ and\ \bibinfo {author} {\bibfnamefont {L.~E.}\ \bibnamefont
  {White}},\ }\href@noop {} {\emph {\bibinfo {title} {Neuroscience}}}\
  (\bibinfo  {publisher} {Sinauer Associates, Sunderland (MA)},\ \bibinfo
  {address} {Sunderland, MA},\ \bibinfo {year} {2001})\BibitemShut {NoStop}%
\bibitem [{\citenamefont {Kandel}\ \emph {et~al.}(2000)\citenamefont {Kandel},
  \citenamefont {Schwartz},\ and\ \citenamefont {Jessell}}]{Kandel2000}%
  \BibitemOpen
  \bibfield  {author} {\bibinfo {author} {\bibfnamefont {E.}~\bibnamefont
  {Kandel}}, \bibinfo {author} {\bibfnamefont {J.}~\bibnamefont {Schwartz}}, \
  and\ \bibinfo {author} {\bibfnamefont {T.}~\bibnamefont {Jessell}},\
  }\href@noop {} {\emph {\bibinfo {title} {Principles of Neural Science, Fourth
  Edition}}},\ Vol.~\bibinfo {volume} {4}\ (\bibinfo  {publisher} {McGraw-hill,
  New York},\ \bibinfo {year} {2000})\BibitemShut {NoStop}%
\bibitem [{\citenamefont {Walton}\ \emph {et~al.}(2010)\citenamefont {Walton},
  \citenamefont {Behrens}, \citenamefont {Buckley}, \citenamefont {Rudebeck},\
  and\ \citenamefont {Rushworth}}]{Walton2010}%
  \BibitemOpen
  \bibfield  {author} {\bibinfo {author} {\bibfnamefont {M.~E.}\ \bibnamefont
  {Walton}}, \bibinfo {author} {\bibfnamefont {T.~E.~J.}\ \bibnamefont
  {Behrens}}, \bibinfo {author} {\bibfnamefont {M.~J.}\ \bibnamefont
  {Buckley}}, \bibinfo {author} {\bibfnamefont {P.~H.}\ \bibnamefont
  {Rudebeck}}, \ and\ \bibinfo {author} {\bibfnamefont {M.~F.~S.}\ \bibnamefont
  {Rushworth}},\ }\href {\doibase 10.1016/j.neuron.2010.02.027} {\bibfield
  {journal} {\bibinfo  {journal} {Neuron}\ }\textbf {\bibinfo {volume} {65}},\
  \bibinfo {pages} {927} (\bibinfo {year} {2010})}\BibitemShut {NoStop}%
\bibitem [{\citenamefont {Fellows}(2007)}]{Fellows2007}%
  \BibitemOpen
  \bibfield  {author} {\bibinfo {author} {\bibfnamefont {L.~K.}\ \bibnamefont
  {Fellows}},\ }\href {\doibase 10.1196/annals.1401.023} {\bibfield  {journal}
  {\bibinfo  {journal} {Ann N Y Acad Sci}\ }\textbf {\bibinfo {volume}
  {1121}},\ \bibinfo {pages} {421} (\bibinfo {year} {2007})}\BibitemShut
  {NoStop}%
\bibitem [{\citenamefont {Walton}\ \emph {et~al.}(2004)\citenamefont {Walton},
  \citenamefont {Devlin},\ and\ \citenamefont {Rushworth}}]{Walton2004}%
  \BibitemOpen
  \bibfield  {author} {\bibinfo {author} {\bibfnamefont {M.~E.}\ \bibnamefont
  {Walton}}, \bibinfo {author} {\bibfnamefont {J.~T.}\ \bibnamefont {Devlin}},
  \ and\ \bibinfo {author} {\bibfnamefont {M.~F.~S.}\ \bibnamefont
  {Rushworth}},\ }\href {\doibase 10.1038/nn1339} {\bibfield  {journal}
  {\bibinfo  {journal} {Nat Neurosci}\ }\textbf {\bibinfo {volume} {7}},\
  \bibinfo {pages} {1259} (\bibinfo {year} {2004})}\BibitemShut {NoStop}%
\bibitem [{\citenamefont {Izquierdo}\ \emph {et~al.}(2004)\citenamefont
  {Izquierdo}, \citenamefont {Suda},\ and\ \citenamefont
  {Murray}}]{Izquierdo2004}%
  \BibitemOpen
  \bibfield  {author} {\bibinfo {author} {\bibfnamefont {A.}~\bibnamefont
  {Izquierdo}}, \bibinfo {author} {\bibfnamefont {R.~K.}\ \bibnamefont {Suda}},
  \ and\ \bibinfo {author} {\bibfnamefont {E.~A.}\ \bibnamefont {Murray}},\
  }\href {\doibase 10.1523/JNEUROSCI.1921-04.2004} {\bibfield  {journal}
  {\bibinfo  {journal} {J Neurosci}\ }\textbf {\bibinfo {volume} {24}},\
  \bibinfo {pages} {7540} (\bibinfo {year} {2004})}\BibitemShut {NoStop}%
\bibitem [{\citenamefont {Rolls}(2000)}]{Rolls2000}%
  \BibitemOpen
  \bibfield  {author} {\bibinfo {author} {\bibfnamefont {E.~T.}\ \bibnamefont
  {Rolls}},\ }\href {\doibase 10.1093/cercor/10.3.284} {\bibfield  {journal}
  {\bibinfo  {journal} {Cereb Cortex}\ }\textbf {\bibinfo {volume} {10}},\
  \bibinfo {pages} {284} (\bibinfo {year} {2000})}\BibitemShut {NoStop}%
\bibitem [{\citenamefont {Weiskrantz}(1956)}]{Weiskrantz1956}%
  \BibitemOpen
  \bibfield  {author} {\bibinfo {author} {\bibfnamefont {L.}~\bibnamefont
  {Weiskrantz}},\ }\href@noop {} {\bibfield  {journal} {\bibinfo  {journal} {J.
  Comp. Physiol. Psychol.}\ }\textbf {\bibinfo {volume} {49}},\ \bibinfo
  {pages} {381} (\bibinfo {year} {1956})}\BibitemShut {NoStop}%
\bibitem [{\citenamefont {Davis}(1992)}]{Davis1992}%
  \BibitemOpen
  \bibfield  {author} {\bibinfo {author} {\bibfnamefont {M.}~\bibnamefont
  {Davis}},\ }\href {\doibase 10.1146/annurev.ne.15.030192.002033} {\bibfield
  {journal} {\bibinfo  {journal} {Annu Rev Neurosci}\ }\textbf {\bibinfo
  {volume} {15}},\ \bibinfo {pages} {353} (\bibinfo {year} {1992})}\BibitemShut
  {NoStop}%
\bibitem [{\citenamefont {Killcross}\ \emph {et~al.}(1997)\citenamefont
  {Killcross}, \citenamefont {Robbins},\ and\ \citenamefont
  {Everitt}}]{Killcross1997}%
  \BibitemOpen
  \bibfield  {author} {\bibinfo {author} {\bibfnamefont {S.}~\bibnamefont
  {Killcross}}, \bibinfo {author} {\bibfnamefont {T.~W.}\ \bibnamefont
  {Robbins}}, \ and\ \bibinfo {author} {\bibfnamefont {B.~J.}\ \bibnamefont
  {Everitt}},\ }\href {\doibase 10.1038/41097} {\bibfield  {journal} {\bibinfo
  {journal} {Nature (London)}\ }\textbf {\bibinfo {volume} {388}},\ \bibinfo
  {pages} {377} (\bibinfo {year} {1997})}\BibitemShut {NoStop}%
\bibitem [{\citenamefont {Maren}(1999)}]{Maren1999}%
  \BibitemOpen
  \bibfield  {author} {\bibinfo {author} {\bibfnamefont {S.}~\bibnamefont
  {Maren}},\ }\href {\doibase 10.1016/S0166-2236(99)01465-4} {\bibfield
  {journal} {\bibinfo  {journal} {Trends Neurosci}\ }\textbf {\bibinfo {volume}
  {22}},\ \bibinfo {pages} {561} (\bibinfo {year} {1999})}\BibitemShut
  {NoStop}%
\bibitem [{\citenamefont {Kensinger}\ and\ \citenamefont
  {Corkin}(2003)}]{Kensinger2003}%
  \BibitemOpen
  \bibfield  {author} {\bibinfo {author} {\bibfnamefont {E.~A.}\ \bibnamefont
  {Kensinger}}\ and\ \bibinfo {author} {\bibfnamefont {S.}~\bibnamefont
  {Corkin}},\ }\href {\doibase 10.3758/BF03195800} {\bibfield  {journal}
  {\bibinfo  {journal} {Mem Cogn}\ }\textbf {\bibinfo {volume} {31}},\ \bibinfo
  {pages} {1169} (\bibinfo {year} {2003})}\BibitemShut {NoStop}%
\bibitem [{\citenamefont {Richardson}\ \emph {et~al.}(2004)\citenamefont
  {Richardson}, \citenamefont {Strange},\ and\ \citenamefont
  {Dolan}}]{Richardson2004}%
  \BibitemOpen
  \bibfield  {author} {\bibinfo {author} {\bibfnamefont {M.~P.}\ \bibnamefont
  {Richardson}}, \bibinfo {author} {\bibfnamefont {B.~A.}\ \bibnamefont
  {Strange}}, \ and\ \bibinfo {author} {\bibfnamefont {R.~J.}\ \bibnamefont
  {Dolan}},\ }\href {\doibase 10.1038/nn1190} {\bibfield  {journal} {\bibinfo
  {journal} {Nat Neurosci}\ }\textbf {\bibinfo {volume} {7}},\ \bibinfo {pages}
  {278} (\bibinfo {year} {2004})}\BibitemShut {NoStop}%
\bibitem [{\citenamefont {Anderson}\ \emph {et~al.}(2006)\citenamefont
  {Anderson}, \citenamefont {Wais},\ and\ \citenamefont
  {Gabrieli}}]{Anderson2006}%
  \BibitemOpen
  \bibfield  {author} {\bibinfo {author} {\bibfnamefont {A.~K.}\ \bibnamefont
  {Anderson}}, \bibinfo {author} {\bibfnamefont {P.~E.}\ \bibnamefont {Wais}},
  \ and\ \bibinfo {author} {\bibfnamefont {J.~D.~E.}\ \bibnamefont
  {Gabrieli}},\ }\href {\doibase 10.1073/pnas.0506308103} {\bibfield  {journal}
  {\bibinfo  {journal} {Proc Natl Acad Sci USA}\ }\textbf {\bibinfo {volume}
  {103}},\ \bibinfo {pages} {1599} (\bibinfo {year} {2006})}\BibitemShut
  {NoStop}%
\bibitem [{\citenamefont {Tambini}\ \emph {et~al.}(2017)\citenamefont
  {Tambini}, \citenamefont {Rimmele}, \citenamefont {Phelps},\ and\
  \citenamefont {Davachi}}]{Tambini2017}%
  \BibitemOpen
  \bibfield  {author} {\bibinfo {author} {\bibfnamefont {A.}~\bibnamefont
  {Tambini}}, \bibinfo {author} {\bibfnamefont {U.}~\bibnamefont {Rimmele}},
  \bibinfo {author} {\bibfnamefont {E.~A.}\ \bibnamefont {Phelps}}, \ and\
  \bibinfo {author} {\bibfnamefont {L.}~\bibnamefont {Davachi}},\ }\href
  {\doibase 10.1038/nn.4468} {\bibfield  {journal} {\bibinfo  {journal} {Nat
  Neurosci}\ }\textbf {\bibinfo {volume} {20}},\ \bibinfo {pages} {271}
  (\bibinfo {year} {2017})}\BibitemShut {NoStop}%
\bibitem [{\citenamefont {Churchland}\ and\ \citenamefont
  {Sejnowski}(1988)}]{Churchland1988}%
  \BibitemOpen
  \bibfield  {author} {\bibinfo {author} {\bibfnamefont {P.~S.}\ \bibnamefont
  {Churchland}}\ and\ \bibinfo {author} {\bibfnamefont {T.~J.}\ \bibnamefont
  {Sejnowski}},\ }\href {\doibase 10.1126/science.3055294} {\bibfield
  {journal} {\bibinfo  {journal} {Science}\ }\textbf {\bibinfo {volume}
  {242}},\ \bibinfo {pages} {741} (\bibinfo {year} {1988})}\BibitemShut
  {NoStop}%
\bibitem [{\citenamefont {Goodale}\ and\ \citenamefont
  {Milner}(1992)}]{Goodale1992}%
  \BibitemOpen
  \bibfield  {author} {\bibinfo {author} {\bibfnamefont {M.~A.}\ \bibnamefont
  {Goodale}}\ and\ \bibinfo {author} {\bibfnamefont {A.~D.}\ \bibnamefont
  {Milner}},\ }\href {\doibase 10.1016/0166-2236(92)90344-8} {\bibfield
  {journal} {\bibinfo  {journal} {Trends Neurosci}\ }\textbf {\bibinfo {volume}
  {15}},\ \bibinfo {pages} {20} (\bibinfo {year} {1992})}\BibitemShut {NoStop}%
\bibitem [{\citenamefont {Goodale}\ and\ \citenamefont
  {Milner}(2018)}]{Goodale2018}%
  \BibitemOpen
  \bibfield  {author} {\bibinfo {author} {\bibfnamefont {M.~A.}\ \bibnamefont
  {Goodale}}\ and\ \bibinfo {author} {\bibfnamefont {A.~D.}\ \bibnamefont
  {Milner}},\ }\href {\doibase 10.1016/j.cortex.2017.12.002} {\bibfield
  {journal} {\bibinfo  {journal} {Cortex}\ }\textbf {\bibinfo {volume} {98}},\
  \bibinfo {pages} {283} (\bibinfo {year} {2018})}\BibitemShut {NoStop}%
\bibitem [{\citenamefont {Bullmore}\ and\ \citenamefont
  {Sporns}(2012)}]{Bullmore2012}%
  \BibitemOpen
  \bibfield  {author} {\bibinfo {author} {\bibfnamefont {E.}~\bibnamefont
  {Bullmore}}\ and\ \bibinfo {author} {\bibfnamefont {O.}~\bibnamefont
  {Sporns}},\ }\href {\doibase 10.1038/nrn3214} {\bibfield  {journal} {\bibinfo
   {journal} {Nat Rev Neurosci}\ }\textbf {\bibinfo {volume} {13}},\ \bibinfo
  {pages} {336} (\bibinfo {year} {2012})}\BibitemShut {NoStop}%
\bibitem [{\citenamefont {Averbeck}\ and\ \citenamefont
  {Seo}(2008)}]{Averbeck2008}%
  \BibitemOpen
  \bibfield  {author} {\bibinfo {author} {\bibfnamefont {B.~B.}\ \bibnamefont
  {Averbeck}}\ and\ \bibinfo {author} {\bibfnamefont {M.}~\bibnamefont {Seo}},\
  }\href {\doibase 10.1371/journal.pcbi.1000050} {\bibfield  {journal}
  {\bibinfo  {journal} {PLoS Comput Biol}\ }\textbf {\bibinfo {volume} {4}},\
  \bibinfo {pages} {e1000050} (\bibinfo {year} {2008})}\BibitemShut {NoStop}%
\bibitem [{\citenamefont {Herculano-Houzel}\ \emph {et~al.}(2010)\citenamefont
  {Herculano-Houzel}, \citenamefont {Mota}, \citenamefont {Wong},\ and\
  \citenamefont {Kaas}}]{Herculano-Houzel2010}%
  \BibitemOpen
  \bibfield  {author} {\bibinfo {author} {\bibfnamefont {S.}~\bibnamefont
  {Herculano-Houzel}}, \bibinfo {author} {\bibfnamefont {B.}~\bibnamefont
  {Mota}}, \bibinfo {author} {\bibfnamefont {P.}~\bibnamefont {Wong}}, \ and\
  \bibinfo {author} {\bibfnamefont {J.~H.}\ \bibnamefont {Kaas}},\ }\href
  {\doibase 10.1073/pnas.1012590107} {\bibfield  {journal} {\bibinfo  {journal}
  {Proc Natl Acad Sci USA}\ }\textbf {\bibinfo {volume} {107}},\ \bibinfo
  {pages} {19008} (\bibinfo {year} {2010})}\BibitemShut {NoStop}%
\bibitem [{\citenamefont {Chklovskii}(2004)}]{Chklovskii2004}%
  \BibitemOpen
  \bibfield  {author} {\bibinfo {author} {\bibfnamefont {D.~B.}\ \bibnamefont
  {Chklovskii}},\ }\href {\doibase 10.1162/0899766041732422} {\bibfield
  {journal} {\bibinfo  {journal} {Neural Comput}\ }\textbf {\bibinfo {volume}
  {16}},\ \bibinfo {pages} {2067} (\bibinfo {year} {2004})}\BibitemShut
  {NoStop}%
\bibitem [{\citenamefont {Kaiser}\ and\ \citenamefont
  {Hilgetag}(2004)}]{Kaiser2004}%
  \BibitemOpen
  \bibfield  {author} {\bibinfo {author} {\bibfnamefont {M.}~\bibnamefont
  {Kaiser}}\ and\ \bibinfo {author} {\bibfnamefont {C.~C.}\ \bibnamefont
  {Hilgetag}},\ }\href {\doibase 10.1016/j.neucom.2004.01.059} {\bibfield
  {journal} {\bibinfo  {journal} {Neurocomputing}\ }\textbf {\bibinfo {volume}
  {58}},\ \bibinfo {pages} {297} (\bibinfo {year} {2004})}\BibitemShut
  {NoStop}%
\bibitem [{\citenamefont {Buzs{\'a}ki}\ \emph {et~al.}(2004)\citenamefont
  {Buzs{\'a}ki}, \citenamefont {Geisler}, \citenamefont {Henze},\ and\
  \citenamefont {Wang}}]{Buzsaki2004}%
  \BibitemOpen
  \bibfield  {author} {\bibinfo {author} {\bibfnamefont {G.}~\bibnamefont
  {Buzs{\'a}ki}}, \bibinfo {author} {\bibfnamefont {C.}~\bibnamefont
  {Geisler}}, \bibinfo {author} {\bibfnamefont {D.~A.}\ \bibnamefont {Henze}},
  \ and\ \bibinfo {author} {\bibfnamefont {X.-J.}\ \bibnamefont {Wang}},\
  }\href {\doibase 10.1016/j.tins.2004.02.007} {\bibfield  {journal} {\bibinfo
  {journal} {Trends Neurosci}\ }\textbf {\bibinfo {volume} {27}},\ \bibinfo
  {pages} {186} (\bibinfo {year} {2004})}\BibitemShut {NoStop}%
\bibitem [{\citenamefont {Niven}\ and\ \citenamefont
  {Laughlin}(2008)}]{Niven2008}%
  \BibitemOpen
  \bibfield  {author} {\bibinfo {author} {\bibfnamefont {J.~E.}\ \bibnamefont
  {Niven}}\ and\ \bibinfo {author} {\bibfnamefont {S.~B.}\ \bibnamefont
  {Laughlin}},\ }\href {\doibase 10.1242/jeb.017574} {\bibfield  {journal}
  {\bibinfo  {journal} {J Exp Biol}\ }\textbf {\bibinfo {volume} {211}},\
  \bibinfo {pages} {1792} (\bibinfo {year} {2008})}\BibitemShut {NoStop}%
\bibitem [{\citenamefont {Barthélemy}(2011)}]{Barthelemy2011}%
  \BibitemOpen
  \bibfield  {author} {\bibinfo {author} {\bibfnamefont {M.}~\bibnamefont
  {Barthélemy}},\ }\href {\doibase 10.1016/j.physrep.2010.11.002} {\bibfield
  {journal} {\bibinfo  {journal} {Phys Rep}\ }\textbf {\bibinfo {volume}
  {499}},\ \bibinfo {pages} {1} (\bibinfo {year} {2011})}\BibitemShut {NoStop}%
\bibitem [{\citenamefont {Gnadt}\ and\ \citenamefont
  {Andersen}(1988)}]{Gnadt1988}%
  \BibitemOpen
  \bibfield  {author} {\bibinfo {author} {\bibfnamefont {J.~W.}\ \bibnamefont
  {Gnadt}}\ and\ \bibinfo {author} {\bibfnamefont {R.~A.}\ \bibnamefont
  {Andersen}},\ }\href {\doibase 10.1007/BF00271862} {\bibfield  {journal}
  {\bibinfo  {journal} {Exp Brain Res}\ }\textbf {\bibinfo {volume} {70}},\
  \bibinfo {pages} {216} (\bibinfo {year} {1988})}\BibitemShut {NoStop}%
\bibitem [{\citenamefont {Duhamel}\ \emph {et~al.}(1998)\citenamefont
  {Duhamel}, \citenamefont {Colby},\ and\ \citenamefont
  {Goldberg}}]{Duhamel1998}%
  \BibitemOpen
  \bibfield  {author} {\bibinfo {author} {\bibfnamefont {J.-R.}\ \bibnamefont
  {Duhamel}}, \bibinfo {author} {\bibfnamefont {C.~L.}\ \bibnamefont {Colby}},
  \ and\ \bibinfo {author} {\bibfnamefont {M.~E.}\ \bibnamefont {Goldberg}},\
  }\href {\doibase 10.1152/jn.1998.79.1.126} {\bibfield  {journal} {\bibinfo
  {journal} {J Neurophysiol}\ }\textbf {\bibinfo {volume} {79}},\ \bibinfo
  {pages} {126} (\bibinfo {year} {1998})}\BibitemShut {NoStop}%
\bibitem [{\citenamefont {Pesaran}\ \emph {et~al.}(2002)\citenamefont
  {Pesaran}, \citenamefont {Pezaris}, \citenamefont {Sahani}, \citenamefont
  {Mitra},\ and\ \citenamefont {Andersen}}]{Pesaran2002}%
  \BibitemOpen
  \bibfield  {author} {\bibinfo {author} {\bibfnamefont {B.}~\bibnamefont
  {Pesaran}}, \bibinfo {author} {\bibfnamefont {J.~S.}\ \bibnamefont
  {Pezaris}}, \bibinfo {author} {\bibfnamefont {M.}~\bibnamefont {Sahani}},
  \bibinfo {author} {\bibfnamefont {P.~P.}\ \bibnamefont {Mitra}}, \ and\
  \bibinfo {author} {\bibfnamefont {R.~A.}\ \bibnamefont {Andersen}},\ }\href
  {\doibase 10.1038/nn890} {\bibfield  {journal} {\bibinfo  {journal} {Nat
  Neurosci}\ }\textbf {\bibinfo {volume} {5}},\ \bibinfo {pages} {805}
  (\bibinfo {year} {2002})}\BibitemShut {NoStop}%
\bibitem [{\citenamefont {Sakata}\ \emph {et~al.}(1995)\citenamefont {Sakata},
  \citenamefont {Taira}, \citenamefont {Murata},\ and\ \citenamefont
  {Mine}}]{Sakata1995}%
  \BibitemOpen
  \bibfield  {author} {\bibinfo {author} {\bibfnamefont {H.}~\bibnamefont
  {Sakata}}, \bibinfo {author} {\bibfnamefont {M.}~\bibnamefont {Taira}},
  \bibinfo {author} {\bibfnamefont {A.}~\bibnamefont {Murata}}, \ and\ \bibinfo
  {author} {\bibfnamefont {S.}~\bibnamefont {Mine}},\ }\href {\doibase
  10.1093/cercor/5.5.429} {\bibfield  {journal} {\bibinfo  {journal} {Cereb
  Cortex}\ }\textbf {\bibinfo {volume} {5}},\ \bibinfo {pages} {429} (\bibinfo
  {year} {1995})}\BibitemShut {NoStop}%
\bibitem [{\citenamefont {Eskandar}\ and\ \citenamefont
  {Assad}(1999)}]{Eskandar1999}%
  \BibitemOpen
  \bibfield  {author} {\bibinfo {author} {\bibfnamefont {E.~N.}\ \bibnamefont
  {Eskandar}}\ and\ \bibinfo {author} {\bibfnamefont {J.~A.}\ \bibnamefont
  {Assad}},\ }\href {\doibase 10.1038/4594} {\bibfield  {journal} {\bibinfo
  {journal} {Nat Neurosci}\ }\textbf {\bibinfo {volume} {2}},\ \bibinfo {pages}
  {88} (\bibinfo {year} {1999})}\BibitemShut {NoStop}%
\bibitem [{\citenamefont {Grefkes}\ and\ \citenamefont
  {Fink}(2005)}]{Grefkes2005}%
  \BibitemOpen
  \bibfield  {author} {\bibinfo {author} {\bibfnamefont {C.}~\bibnamefont
  {Grefkes}}\ and\ \bibinfo {author} {\bibfnamefont {G.~R.}\ \bibnamefont
  {Fink}},\ }\href {\doibase 10.1111/j.1469-7580.2005.00426.x} {\bibfield
  {journal} {\bibinfo  {journal} {J Anat}\ }\textbf {\bibinfo {volume} {207}},\
  \bibinfo {pages} {3} (\bibinfo {year} {2005})}\BibitemShut {NoStop}%
\bibitem [{\citenamefont {Colombo}(2013)}]{Colombo2013}%
  \BibitemOpen
  \bibfield  {author} {\bibinfo {author} {\bibfnamefont {M.}~\bibnamefont
  {Colombo}},\ }\href {\doibase DOI: 10.1086/670331} {\bibfield  {journal}
  {\bibinfo  {journal} {Philos Sci}\ }\textbf {\bibinfo {volume} {80}},\
  \bibinfo {pages} {356} (\bibinfo {year} {2013})}\BibitemShut {NoStop}%
\bibitem [{\citenamefont {Sallet}\ \emph {et~al.}(2011)\citenamefont {Sallet},
  \citenamefont {Mars}, \citenamefont {Noonan}, \citenamefont {Andersson},
  \citenamefont {O'Reilly}, \citenamefont {Jbabdi}, \citenamefont {Croxson},
  \citenamefont {Jenkinson}, \citenamefont {Miller},\ and\ \citenamefont
  {Rushworth}}]{Sallet2011}%
  \BibitemOpen
  \bibfield  {author} {\bibinfo {author} {\bibfnamefont {J.}~\bibnamefont
  {Sallet}}, \bibinfo {author} {\bibfnamefont {R.~B.}\ \bibnamefont {Mars}},
  \bibinfo {author} {\bibfnamefont {M.~P.}\ \bibnamefont {Noonan}}, \bibinfo
  {author} {\bibfnamefont {J.~L.}\ \bibnamefont {Andersson}}, \bibinfo {author}
  {\bibfnamefont {J.~X.}\ \bibnamefont {O'Reilly}}, \bibinfo {author}
  {\bibfnamefont {S.}~\bibnamefont {Jbabdi}}, \bibinfo {author} {\bibfnamefont
  {P.~L.}\ \bibnamefont {Croxson}}, \bibinfo {author} {\bibfnamefont
  {M.}~\bibnamefont {Jenkinson}}, \bibinfo {author} {\bibfnamefont {K.~L.}\
  \bibnamefont {Miller}}, \ and\ \bibinfo {author} {\bibfnamefont {M.~F.~S.}\
  \bibnamefont {Rushworth}},\ }\href {\doibase 10.1126/science.1210027}
  {\bibfield  {journal} {\bibinfo  {journal} {Science}\ }\textbf {\bibinfo
  {volume} {334}},\ \bibinfo {pages} {697} (\bibinfo {year}
  {2011})}\BibitemShut {NoStop}%
\bibitem [{\citenamefont {Platt}\ \emph {et~al.}(2016)\citenamefont {Platt},
  \citenamefont {Seyfarth},\ and\ \citenamefont {Cheney}}]{Platt2016}%
  \BibitemOpen
  \bibfield  {author} {\bibinfo {author} {\bibfnamefont {M.~L.}\ \bibnamefont
  {Platt}}, \bibinfo {author} {\bibfnamefont {R.~M.}\ \bibnamefont {Seyfarth}},
  \ and\ \bibinfo {author} {\bibfnamefont {D.~L.}\ \bibnamefont {Cheney}},\
  }\href {\doibase 10.1098/rstb.2015.0096} {\bibfield  {journal} {\bibinfo
  {journal} {Phil Trans R Soc B}\ }\textbf {\bibinfo {volume} {371}},\ \bibinfo
  {pages} {20150096} (\bibinfo {year} {2016})}\BibitemShut {NoStop}%
\bibitem [{\citenamefont {Cheney}\ and\ \citenamefont
  {Seyfarth}(2018)}]{Cheney2018}%
  \BibitemOpen
  \bibfield  {author} {\bibinfo {author} {\bibfnamefont {D.~L.}\ \bibnamefont
  {Cheney}}\ and\ \bibinfo {author} {\bibfnamefont {R.~M.}\ \bibnamefont
  {Seyfarth}},\ }\href {\doibase 10.1073/pnas.1717572115} {\bibfield  {journal}
  {\bibinfo  {journal} {Proc Natl Acad Sci USA}\ }\textbf {\bibinfo {volume}
  {115}},\ \bibinfo {pages} {1974} (\bibinfo {year} {2018})}\BibitemShut
  {NoStop}%
\bibitem [{\citenamefont {Wilson}\ \emph {et~al.}(2015)\citenamefont {Wilson},
  \citenamefont {Kikuchi}, \citenamefont {Sun}, \citenamefont {Hunter},
  \citenamefont {Dick}, \citenamefont {Smith}, \citenamefont {Thiele},
  \citenamefont {Griffiths}, \citenamefont {Marslen-Wilson},\ and\
  \citenamefont {Petkov}}]{Wilson2015}%
  \BibitemOpen
  \bibfield  {author} {\bibinfo {author} {\bibfnamefont {B.}~\bibnamefont
  {Wilson}}, \bibinfo {author} {\bibfnamefont {Y.}~\bibnamefont {Kikuchi}},
  \bibinfo {author} {\bibfnamefont {L.}~\bibnamefont {Sun}}, \bibinfo {author}
  {\bibfnamefont {D.}~\bibnamefont {Hunter}}, \bibinfo {author} {\bibfnamefont
  {F.}~\bibnamefont {Dick}}, \bibinfo {author} {\bibfnamefont {K.}~\bibnamefont
  {Smith}}, \bibinfo {author} {\bibfnamefont {A.}~\bibnamefont {Thiele}},
  \bibinfo {author} {\bibfnamefont {T.~D.}\ \bibnamefont {Griffiths}}, \bibinfo
  {author} {\bibfnamefont {W.~D.}\ \bibnamefont {Marslen-Wilson}}, \ and\
  \bibinfo {author} {\bibfnamefont {C.~I.}\ \bibnamefont {Petkov}},\ }\href
  {\doibase 10.1038/ncomms9901} {\bibfield  {journal} {\bibinfo  {journal} {Nat
  Commun}\ }\textbf {\bibinfo {volume} {6}},\ \bibinfo {pages} {8901} (\bibinfo
  {year} {2015})}\BibitemShut {NoStop}%
\bibitem [{\citenamefont {Snowdon}(2017)}]{Snowdon2017}%
  \BibitemOpen
  \bibfield  {author} {\bibinfo {author} {\bibfnamefont {C.~T.}\ \bibnamefont
  {Snowdon}},\ }\href {\doibase 10.1126/science.aam7443} {\bibfield  {journal}
  {\bibinfo  {journal} {Science}\ }\textbf {\bibinfo {volume} {355}},\ \bibinfo
  {pages} {1120} (\bibinfo {year} {2017})}\BibitemShut {NoStop}%
\bibitem [{\citenamefont {Rauschecker}(2018)}]{Rauschecker2018}%
  \BibitemOpen
  \bibfield  {author} {\bibinfo {author} {\bibfnamefont {J.~P.}\ \bibnamefont
  {Rauschecker}},\ }\href {\doibase 10.1016/j.cobeha.2018.06.003} {\bibfield
  {journal} {\bibinfo  {journal} {Curr Opin Behav Sci}\ }\textbf {\bibinfo
  {volume} {21}},\ \bibinfo {pages} {195} (\bibinfo {year} {2018})}\BibitemShut
  {NoStop}%
\bibitem [{\citenamefont {Kolodny}\ and\ \citenamefont
  {Edelman}(2018)}]{Kolodny2018}%
  \BibitemOpen
  \bibfield  {author} {\bibinfo {author} {\bibfnamefont {O.}~\bibnamefont
  {Kolodny}}\ and\ \bibinfo {author} {\bibfnamefont {S.}~\bibnamefont
  {Edelman}},\ }\href {\doibase 10.1098/rstb.2017.0052} {\bibfield  {journal}
  {\bibinfo  {journal} {Phil Trans R Soc B}\ }\textbf {\bibinfo {volume}
  {373}},\ \bibinfo {pages} {20170052} (\bibinfo {year} {2018})}\BibitemShut
  {NoStop}%
\end{thebibliography}%


\begin{thebibliography}{10}

\bibitem{Modha2010}
Modha DS, Singh R (2010) Network architecture of the long-distance pathways in
  the macaque brain.
\newblock {\em Proc Natl Acad Sci USA} 107(30):13485--13490.

\bibitem{Rosvall2008}
Rosvall M, Bergstrom CT (2008) Maps of random walks on complex networks reveal
  community structure.
\newblock {\em Proc Natl Acad Sci USA} 105(4):1118--1123.

\bibitem{Newman2006}
Newman MEJ (2006) Modularity and community structure in networks.
\newblock {\em Proc Natl Acad Sci USA} 103(23):8577--8582.

\bibitem{Mauri2017}
Mauri M, Elli T, Caviglia G, Uboldi G, Azzi M (2017) Rawgraphs: A visualisation
  platform to create open outputs in {\em Proceedings of the 12th Biannual
  Conference on Italian SIGCHI Chapter}, CHItaly '17.
\newblock (ACM, New York, NY, USA), pp. 28:1--28:5.

\bibitem{Clauset2010}
Good BH, De~Montjoye YA, Clauset A (2010) Performance of modularity
  maximization in practical contexts.
\newblock {\em Phys Rev E} 81(4):046106.

\bibitem{Verleysen2007}
Lee JA, Verleysen M (2007) {\em Nonlinear dimensionality reduction}.
\newblock (Springer Science \& Business Media).

\bibitem{Walton2010}
Walton ME, Behrens TEJ, Buckley MJ, Rudebeck PH, Rushworth MFS (2010) Separable
  learning systems in the macaque brain and the role of orbitofrontal cortex in
  contingent learning.
\newblock {\em Neuron} 65(6):927--939.

\bibitem{Fellows2007}
Fellows LK (2007) The role of orbitofrontal cortex in decision making: a
  component process account.
\newblock {\em Ann N Y Acad Sci} 1121(1):421--430.

\bibitem{Walton2004}
Walton ME, Devlin JT, Rushworth MFS (2004) Interactions between decision making
  and performance monitoring within prefrontal cortex.
\newblock {\em Nat Neurosci} 7(11):1259.

\bibitem{Izquierdo2004}
Izquierdo A, Suda RK, Murray EA (2004) Bilateral orbital prefrontal cortex
  lesions in rhesus monkeys disrupt choices guided by both reward value and
  reward contingency.
\newblock {\em J Neurosci} 24(34):7540--7548.

\bibitem{Rolls2000}
Rolls ET (2000) The orbitofrontal cortex and reward.
\newblock {\em Cereb Cortex} 10(3):284--294.

\bibitem{Graziano2007}
Graziano MS, Aflalo TN (2007) Mapping behavioral repertoire onto the cortex.
\newblock {\em Neuron} 56(2):239--251.

\bibitem{Gentilucci1988}
Gentilucci M, et~al. (1988) Functional organization of inferior area 6 in the
  macaque monkey.
\newblock {\em Exp Brain Res} 71(3):475--490.

\bibitem{Carmichael1994}
Carmichael ST, Clugnet MC, Price JL (1994) Central olfactory connections in the
  macaque monkey.
\newblock {\em J Comp Neurol} 346(3):403--434.

\bibitem{Apicella1991}
Apicella P, Ljungberg T, Scarnati E, Schultz W (1991) Responses to reward in
  monkey dorsal and ventral striatum.
\newblock {\em Exp Brain Res} 85(3):491--500.

\bibitem{Baleydier1980}
Baleydier C, Mauguiere F (1980) The duality of the cingulate gyrus in monkey.
  neuroanatomical study and functional hypothesis.
\newblock {\em Brain} 103(3):525--554.

\bibitem{Evrard2012}
Evrard HC, Forro T, Logothetis NK (2012) Von economo neurons in the anterior
  insula of the macaque monkey.
\newblock {\em Neuron} 74(3):482--489.

\bibitem{Schall2004}
Schall JD (2004) On the role of frontal eye field in guiding attention and
  saccades.
\newblock {\em Vis Res} 44(12):1453--1467.

\bibitem{Kandel2000}
Kandel E, Schwartz J, Jessell T (2000) {\em Principles of Neural Science,
  Fourth Edition}.
\newblock (McGraw-hill, New York) Vol.{}~4.

\bibitem{Manzoni1986}
Manzoni T, Conti F, Fabri M (1986) Callosal projections from area sii to si in
  monkeys: anatomical organization and comparison with association projections.
\newblock {\em J Comp Neurol} 252(2):245--263.

\bibitem{Andersen1990}
Andersen RA, Asanuma C, Essick G, Siegel R (1990) Corticocortical connections
  of anatomically and physiologically defined subdivisions within the inferior
  parietal lobule.
\newblock {\em Journal of Comparative Neurology} 296(1):65--113.

\bibitem{Sakata1995}
Sakata H, Taira M, Murata A, Mine S (1995) Neural mechanisms of visual guidance
  of hand action in the parietal cortex of the monkey.
\newblock {\em Cereb Cortex} 5(5):429--438.

\bibitem{Eskandar1999}
Eskandar EN, Assad JA (1999) Dissociation of visual, motor and predictive
  signals in parietal cortex during visual guidance.
\newblock {\em Nat Neurosci} 2(1):88--93.

\bibitem{Caminiti1996}
Caminiti R, Ferraina S, Johnson PB (1996) The sources of visual information to
  the primate frontal lobe: a novel role for the superior parietal lobule.
\newblock {\em Cereb Cortex} 6(3):319--328.

\bibitem{Murray2007}
Murray EA, Bussey TJ, Saksida LM (2007) Visual perception and memory: a new
  view of medial temporal lobe function in primates and rodents.
\newblock {\em Annu Rev Neurosci} 30:99--122.

\bibitem{Malkova2003}
Malkova L, Mishkin M (2003) One-trial memory for object-place associations
  after separate lesions of hippocampus and posterior parahippocampal region in
  the monkey.
\newblock {\em J Neurosci} 23(5):1956--1965.

\bibitem{Matsumura1999}
Matsumura N, et~al. (1999) Spatial-and task-dependent neuronal responses during
  real and virtual translocation in the monkey hippocampal formation.
\newblock {\em J Neurosci} 19(6):2381--2393.

\bibitem{Courellis2019}
Courellis HS, et~al. (2019) Spatial encoding in primate hippocampus during free
  navigation.
\newblock {\em PLoS Biol} 17(12).

\bibitem{Jutras2010}
Jutras MJ, Buffalo EA (2010) Recognition memory signals in the macaque
  hippocampus.
\newblock {\em Proc Natl Acad Sci USA} 107(1):401--406.

\bibitem{Weiskrantz1956}
Weiskrantz L (1956) Behavioral changes associated with ablation of the
  amygdaloid complex in monkeys.
\newblock {\em J. Comp. Physiol. Psychol.} 49(4):381.

\bibitem{Davis1992}
Davis M (1992) The role of the amygdala in fear and anxiety.
\newblock {\em Annu Rev Neurosci} 15(1):353--375.

\bibitem{Petrides1991}
Petrides M (1991) Monitoring of selections of visual stimuli and the primate
  frontal cortex.
\newblock {\em Proc R Soc Lond, Ser B} 246(1317):293--298.

\bibitem{Petrides1995}
Petrides M (1995) Impairments on nonspatial self-ordered and externally ordered
  working memory tasks after lesions of the mid-dorsal part of the lateral
  frontal cortex in the monkey.
\newblock {\em J Neurosci} 15(1):359--375.

\bibitem{Morel1993}
Morel A, Garraghty P, Kaas J (1993) Tonotopic organization, architectonic
  fields, and connections of auditory cortex in macaque monkeys.
\newblock {\em J Comp Neurol} 335(3):437--459.

\bibitem{Romanski2009}
Romanski LM, Averbeck BB (2009) The primate cortical auditory system and neural
  representation of conspecific vocalizations.
\newblock {\em Annu Rev Neurosci} 32:315--346.

\bibitem{Barraclough2005}
Barraclough NE, Xiao D, Baker CI, Oram MW, Perrett DI (2005) Integration of
  visual and auditory information by superior temporal sulcus neurons
  responsive to the sight of actions.
\newblock {\em J Cogn Neurosci} 17(3):377--391.

\bibitem{Goodale1992}
Goodale MA, Milner AD (1992) Separate visual pathways for perception and
  action.
\newblock {\em Trends Neurosci} 15(1):20--25.

\bibitem{Goodale2018}
Goodale MA, Milner AD (2018) Two visual pathways—where have they taken us and
  where will they lead in future?
\newblock {\em Cortex} 98:283--292.

\bibitem{Duhamel1998}
Duhamel JR, Colby CL, Goldberg ME (1998) Ventral intraparietal area of the
  macaque: congruent visual and somatic response properties.
\newblock {\em J Neurophysiol} 79(1):126--136.

\bibitem{Pesaran2002}
Pesaran B, Pezaris JS, Sahani M, Mitra PP, Andersen RA (2002) Temporal
  structure in neuronal activity during working memory in macaque parietal
  cortex.
\newblock {\em Nat Neurosci} 5(8):805--811.

\bibitem{Smelser2001}
Smelser NJ, Baltes PB (2001) {\em International encyclopedia of the social \&
  behavioral sciences}.
\newblock (Elsevier Amsterdam) Vol.{}~11.

\end{thebibliography}

\end{document}


\renewcommand\thefigure{S\arabic{figure}}
\renewcommand\thetable{S\arabic{table}}

\begin{center}
{\Large SUPPLEMENTARY INFORMATION FOR}

\vspace{0.3cm}
{\Large
Mesoscopic architecture enhances communication across the Macaque connectome revealing structure-function correspondence in the brain}

\vspace{0.4cm}
Anand Pathak$^{1,2}$, Shakti N. Menon$^{1}$ and Sitabhra Sinha$^{1,2}$

\vspace{0.2cm}
\small{$^1$ The Institute of Mathematical Sciences, CIT Campus, Taramani, Chennai 600113, India \\
$^2$ Homi Bhabha National Institute, Anushaktinagar, Mumbai 400 094, India}
\end{center}


\section{The Structure of the Macaque Connectome}

The data describing the Macaque connectome that we have used in our analysis are included in two files that are part of 
the Supplementary Information. 
The file \verb|connectome_nodes.xls| is a spreadsheet comprising $11$ columns which
contain information about each of the $266$ brain regions which correspond to the nodes of the network. The first $7$ columns 
provide the identity of the region in terms of the serial number by which they are identified in our study, the abbreviation, the name, their position 
in the Macaque brain described in terms of the 3-dimensional coordinates as per the Paxinos atlas and the volume that they occupy. The
eighth and ninth columns contain information arising from our analysis, viz., the module they belong to and their role in the mesoscopic organization
(described in the main text),  respectively. The last two columns contain, respectively, the references and the web resource 
from which we have gleaned information about their 
positions and volumes.

The file \verb|connectome_links.dat| is an adjacency list containing information about the directed connections between the different nodes (brain regions)
via anatomical tracts, with the first column
indicating the source node and the second column the target node (both nodes being represented by their serial number as given in the 
file \verb|connectome_nodes.xls|).

\begin{figure}[h!]
\begin{tabular}{cc}
\includegraphics[width=0.48\linewidth]{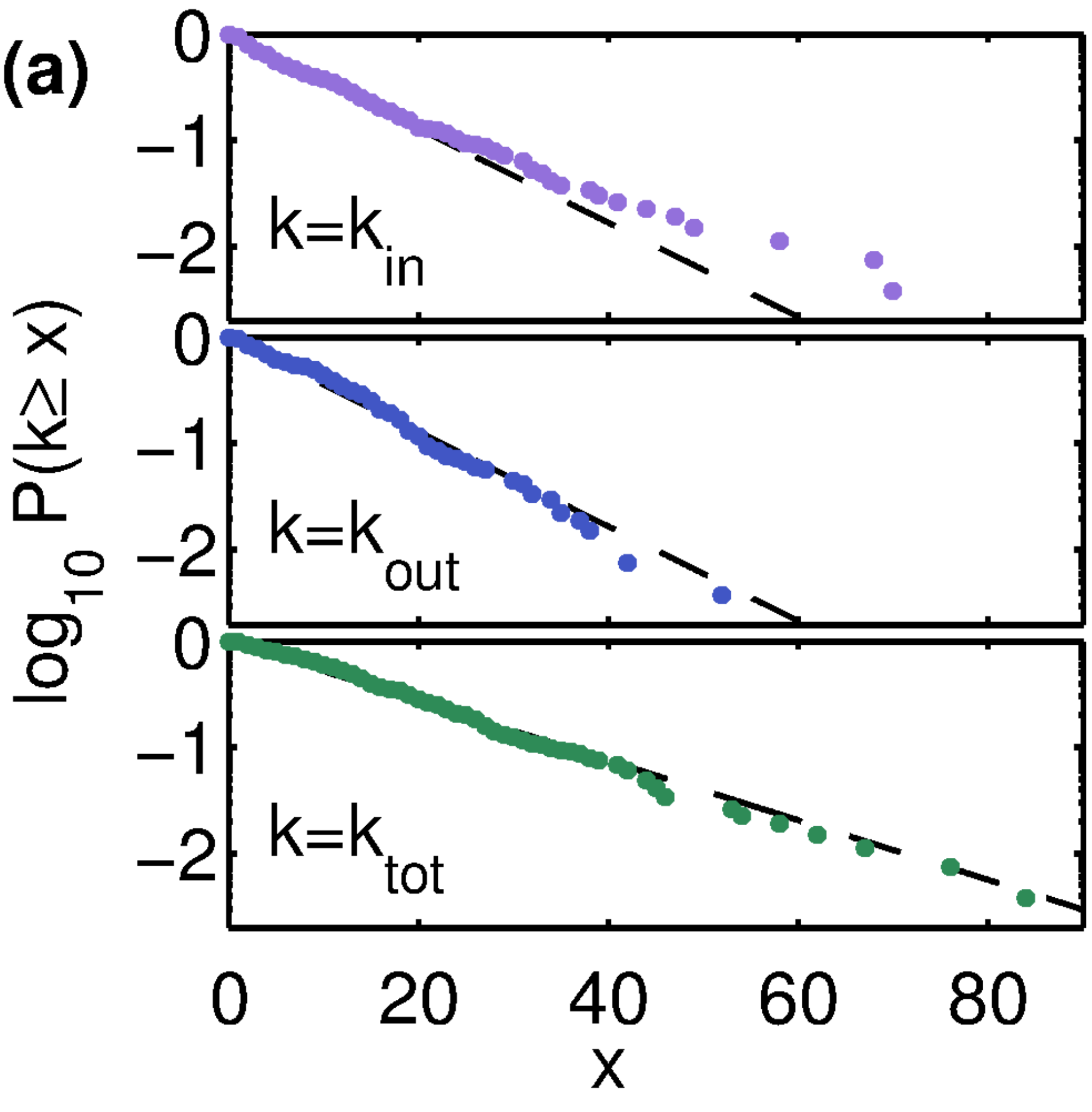}&
\includegraphics[width=0.48\linewidth]{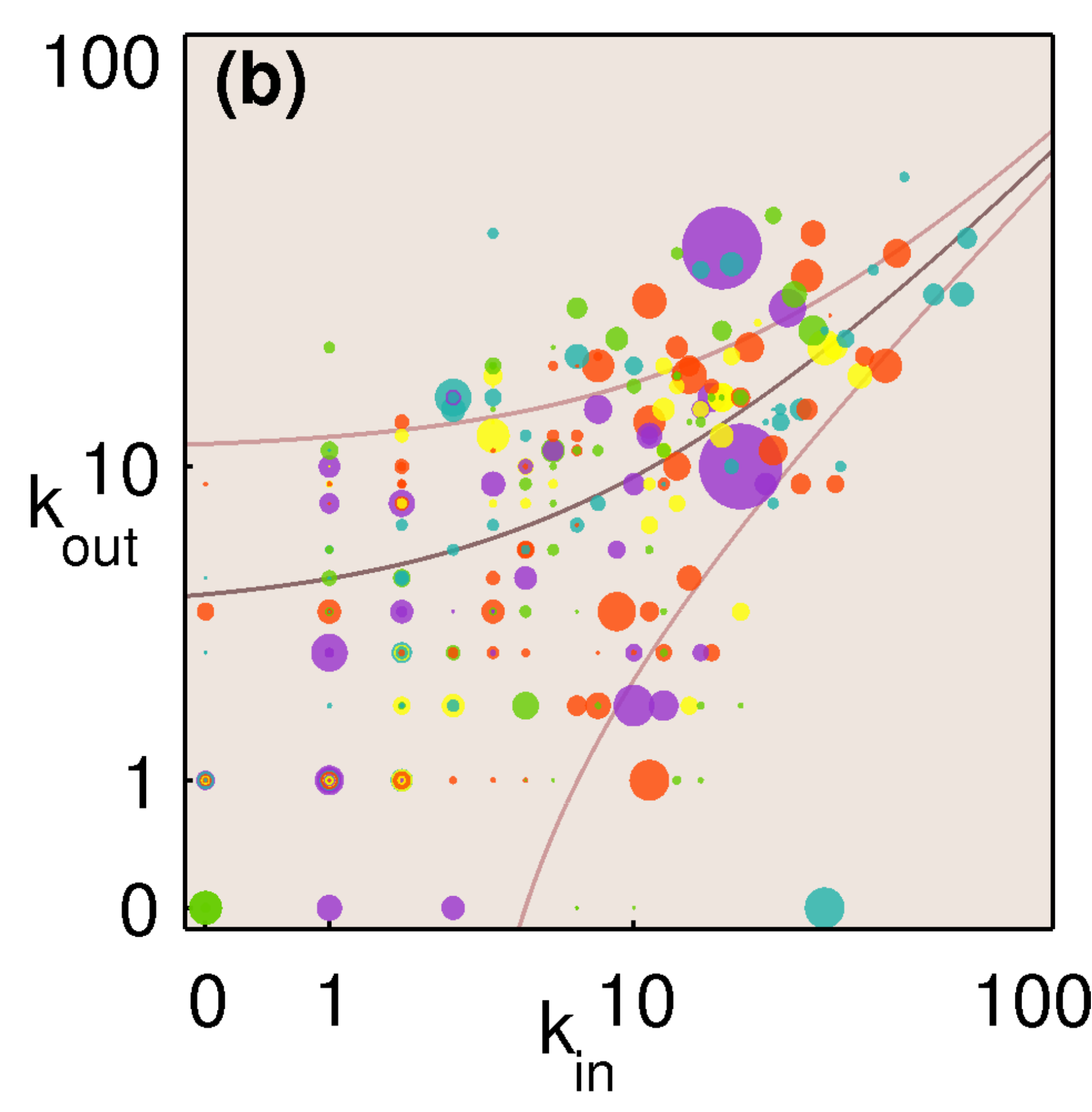}\\
\end{tabular}
\caption{
(a) The distributions of the (top) in-degree, (center) out-degree and (bottom) total degree of the Macaque connectome,
indicating the best-fit exponential distribution (broken line) in each case. (b) Scatter plot indicating the correlation between in-degree and out-degree of the
different nodes in the connectome.
The relatively darker central curve represents the best-fit linear relation between $k_{in}$ and $k_{out}$ (the linear correlation coefficient is $r = 0.62$, 
with a $p$-value of $0$) corresponding to a slope of $0.49$. The two lighter curves on either side indicate the root mean square deviation of
the empirical data from the best-fit linear relation.
The color and sizes of the nodes are same as in Fig.~1 of the main text.
}
\label{fig:FigS1}
\end{figure}

Fig.~\ref{fig:FigS1} shows some of the macroscopic properties of the network, viz., the cumulative distributions for the number of in-coming, out-going
and total connections, and the correlation between the number of in-coming and out-going connections for each node. 
Fig.~\ref{fig:FigS1}~(a) shows that the total degree distribution of the nodes [shown in the bottom panel] follows an exponential distribution, in agreement
with the observed properties of the network investigated by Modha and Singh~\cite{Modha2010}. This suggests that the network
that we have worked with, which has been processed extensively from the original network of $383$ nodes (which contained
many redundancies, as explained in the main text), shares the same macroscopic features as the original network. 
The top and middle panels show the in-degree and out-degree distributions. While both of these appear to also follow an exponential
form, the former shows a deviation in the tail indicating that there exist regions that have more in-coming connections than is expected
given the form of the distribution. 
In particular, the four nodes having the highest in-degrees [top panel] that show the largest deviation from the best-fit exponential distribution 
are all located in the pre-frontal cortex and are also seen to belong to the same module, viz.,$\#1$.
This is in accordance with the known cognitive function of 
prefrontal cortex regions which is high-level multi-modal sensory integration.

In order to see whether regions which attract many in-coming connections also tend to have many out-going
connections, we have looked at the correlation between in- and out-degrees in terms of a scatter plot [Fig.~\ref{fig:FigS1}~(b)]. Here the nodes
are colored according to the module to which they belong, while the relative volumes are represented by the size of the corresponding
markers. We note that while most of the nodes are fit well by a linear relation between in-degree and out-degree, there does
appear to be several nodes which have a disproportionately higher number of out-going connections than is expected from their
in-degree, given the linear relation between the two. 
Specifically, there are 38 nodes
whose out-degree deviate significantly from the value that is expected from the best-fit linear relation with their in-degree, i.e., they are
larger than the upper bound given by the root mean square deviation (the upper lighter curve, see figure caption for details). 
Although the membership of these outliers span across all modules and functional categories (in terms of the role they play in the mesoscopic organization
of the connectome), we note that $32\%$ of all connector hubs (R6) and $26\%$ of all satellite connectors (R3) 
belong to these outliers. As nodes belonging to both of these categories are characterized by having their
connections are distributed over several modules, it suggests a possible functional importance of the outlier nodes in 
coordinating information processing in the Macaque brain.

\FloatBarrier

\newpage
\FloatBarrier
\section{Modular Organization of the Connectome}
\subsection{Establishing the robustness of the modular decomposition}
\begin{figure}[htbp!]
\centerline{\includegraphics[width=1.0\linewidth]{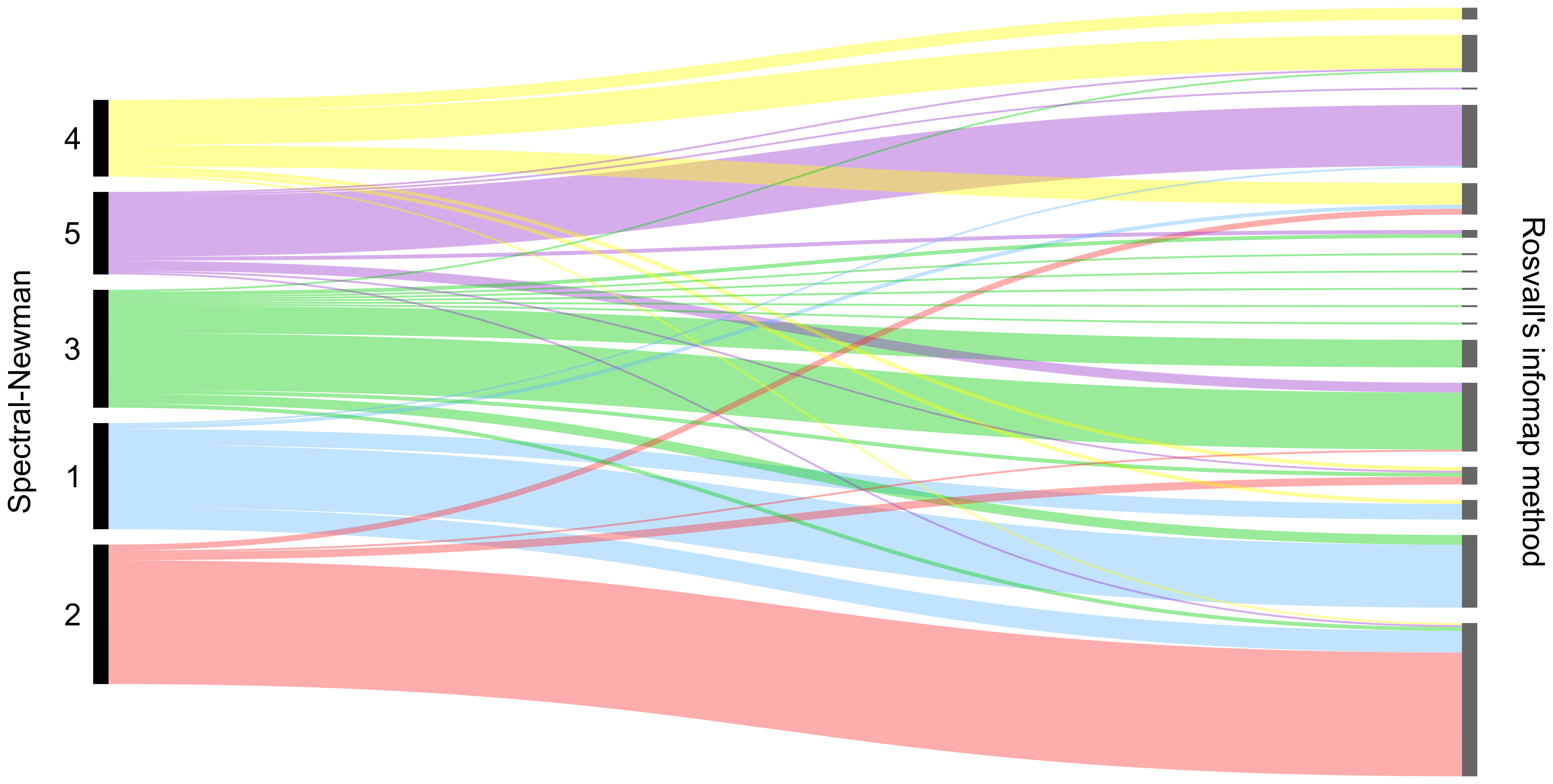}}
\caption{Visual representation of the comparison between the modular decomposition of the Macaque connectome obtained using spectral 
partitioning [left] with that obtained using the Infomap method [right]. 
The modules are represented as vertical bars, connected by bands which are colored according to the module obtained using the spectral method
from which they originate [using the same color scheme as in Fig.~1~(d) of the main text]. This alluvial diagram has been created using the
online visualization tool RAW~\cite{Mauri2017}. 
}
\label{fig:FigS2}
\end{figure}

As described in the Methods section of the main text, we have ensured that the partitioning of the connectome 
is not sensitively dependent on the specific method used for the decomposition. Fig.~\ref{fig:FigS2} shows that the communities 
obtained using the Infomap method~\cite{Rosvall2008}, which is based upon optimally compressing information about dynamic processes on the network, 
have a high degree of overlap with those obtained using a spectral method~\cite{Newman2006} that maximizes the 
modularity $Q$ (for details, see Methods in main text). While the Infomap method generates a larger number of modules (specifically, $17$), not only are many of these extremely small (in some cases comprising only a single node), but several of them
are in fact further subdivisions of the relatively fewer modules (specifically, $5$) obtained using the spectral method. 
The relatively high degree of correspondence between the partitions generated by using techniques that employ completely different
principles suggests that the modular decomposition reported here is an intrinsic property of the network, and is not strongly
affected by the partitioning method used.
\FloatBarrier

\begin{figure}[h!]
\includegraphics[width=0.9\linewidth]{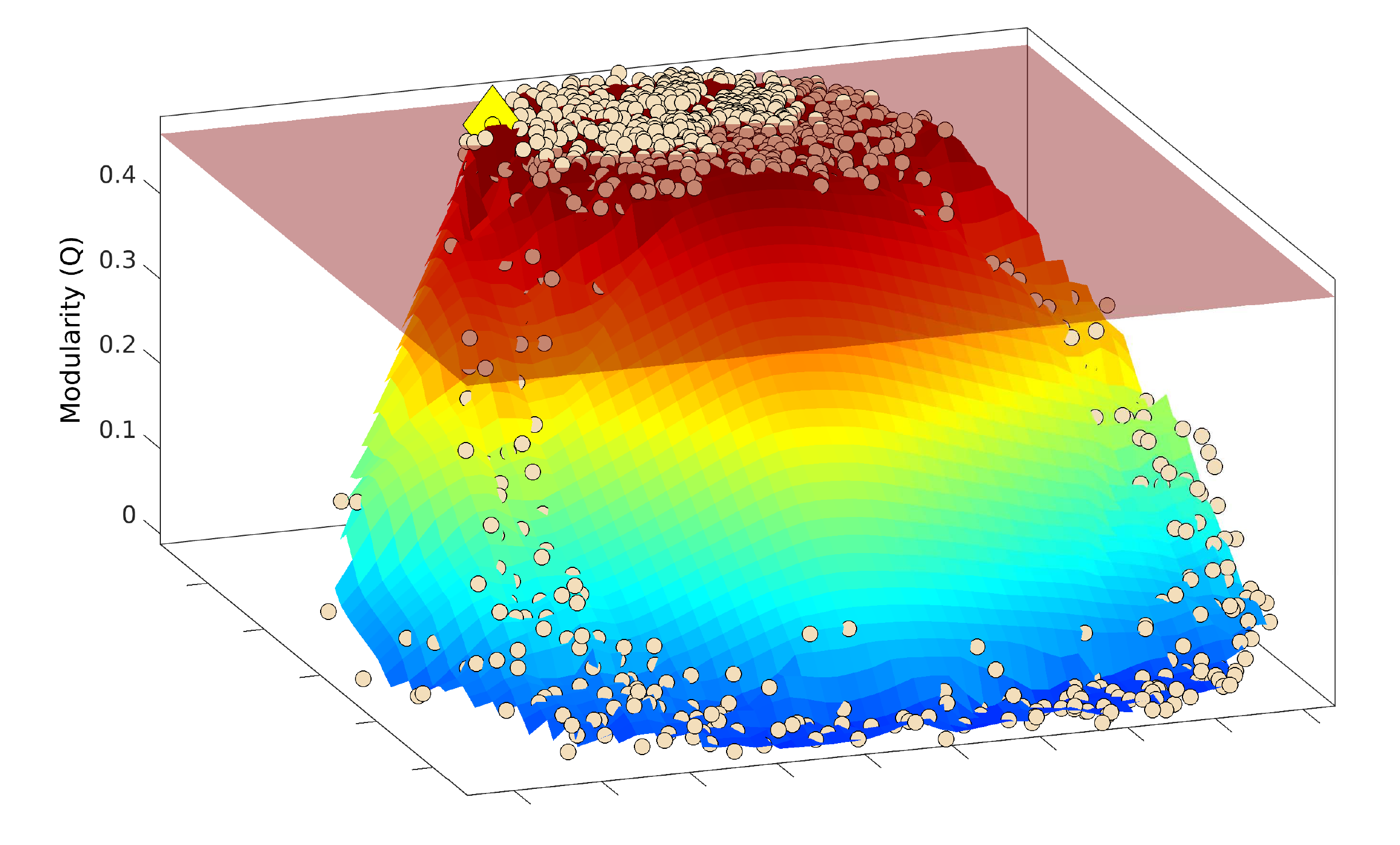}
\caption{Modularity of the Macaque connectome, shown as a function reconstructed from $10^3$ partitionings (circles) obtained through a simulated
annealing method for determining communities~\cite{Clauset2010}. The axes on the horizontal plane orthogonal to the vertical
axis that corresponds to modularity $Q$ represent embedding dimensions that are themselves complex functions of the partition space, such that
the scale of these axes are irrelevant. The distance between the partitionings (whose positions on the horizontal plane are obtained by CCA)
are indicative of the degree of dissimilarity between the corresponding modular partitions of the network.
The partition obtained by the deterministic spectral method yielding a $Q$-value of $Q_{spectral} = 0.485$ (diamond), and which 
has been used for our analysis, is seen to occur in the high-modularity 
plateau comprising a large number of similar partitions, all having a high value of $Q$. 
The $291$ partitionings that occur at the top of this surface, whose $Q$ values differ by less than $3 \%$ from $Q_{spectral} = 0.485$
(specifically, the circles lying above the translucent plane corresponding to $Q = 0.47$ shown in the figure), 
have been used to determine the robustness of the modular identities of the different nodes in the connectome,  as shown in Fig.~\ref{fig:FigS4} (left).
}
\label{fig:FigS3}
\end{figure}

To verify that the method used for maximizing $Q$ does not alter our results significantly, we have performed $10^3$ realizations of
a stochastic simulated annealing algorithm for detecting communities~\cite{Clauset2010}. As mentioned in the Methods (see main text), by comparing
between these large number of optimal partitionings of the network, we can determine the extent to which the modular groupings among
the different nodes is robust. Fig.~\ref{fig:FigS3} shows the Modularity $Q$ values corresponding to these realizations, using a
representation such that similar partitionings (corresponding to the circles) occur close to each other in the two-dimensional plane
orthogonal to the axis representing $Q$. The two-dimensional coordinates of each circle in this plane is obtained by 
Curvilinear Component Analysis (CCA, see Ref.~\cite{Verleysen2007}) as described in Ref.~\cite{Clauset2010}.

\begin{figure}[tbp]
\begin{tabular}{cc}
\includegraphics[width=0.48\linewidth]{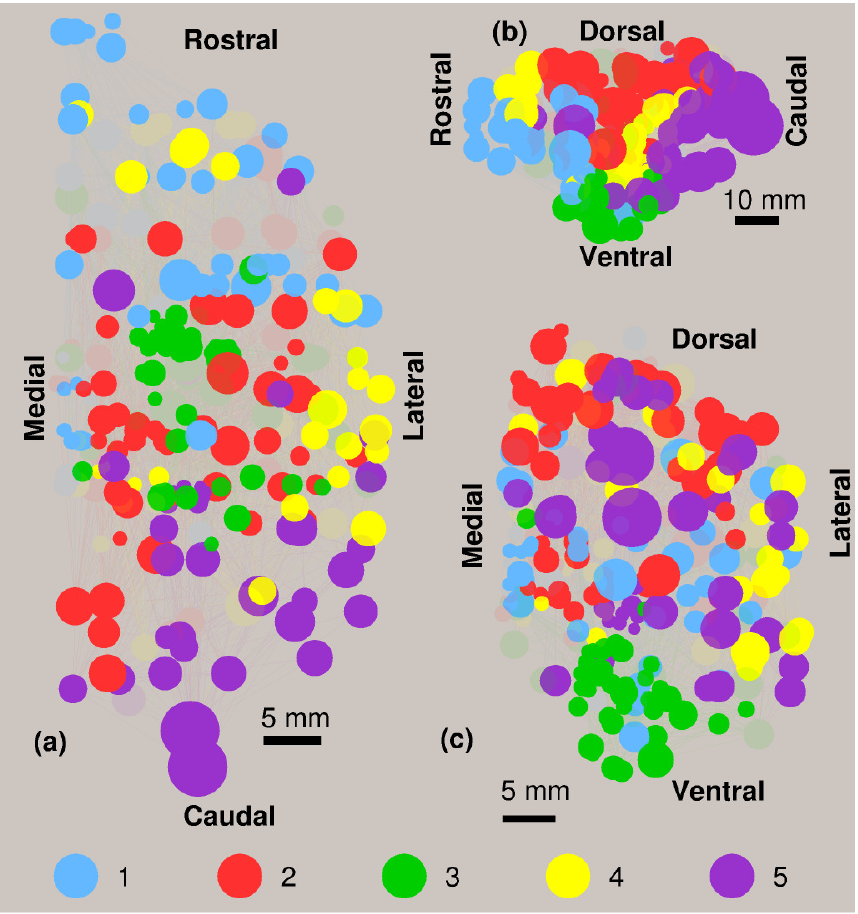}&
\includegraphics[width=0.48\linewidth]{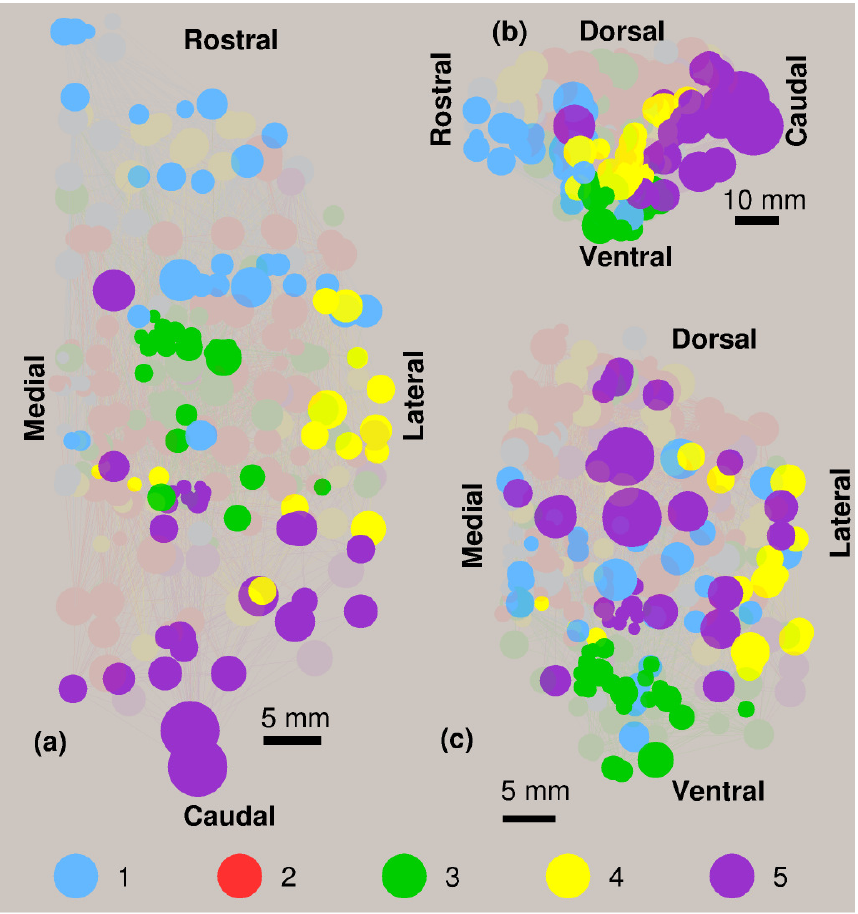}
\end{tabular}
\caption{
The network of brain regions shown in (a) horizontal, (b) sagittal and (c) coronal projections, indicating the regions (highlighted) whose
modular memberships are invariant across the partitionings obtained by the spectral method (used in our analysis) as well as those obtained 
by simulated annealing, whose $Q$ differs by less than (left) $3 \%$ and (right) $7 \%$ from $Q_{spectral} = 0.485$.
As in Fig.~1~(a-c) in the main text, the modular membership of each node is represented by its color (see color key at the bottom of
each panel), the spatial positions of the nodes are specified by the three-dimensional stereotaxic coordinates of the corresponding regions,
and node sizes provide a representation of the relative volumes of the corresponding brain regions (the spatial scale
being indicated by the horizontal bar shown next to each projection).
(Left) Within the $291$ partitionings that have $Q > 0.47$, around $70 \%$ of the $266$ brain regions have the same modular membership as
that seen in the spectral modular decomposition, underlining the robustness of their modular identities.
(Right) For the $625$ partitionings that have $Q > 0.45$, we see a much higher degree of variation in the modular identities of the regions
across the partitionings as a result of including those with much lower values of $Q$.
Specifically, modules $\#1$, $\#3$ and $\#5$ are seen to have several nodes that are robust (i.e., consistently belong to the corresponding module) across
the partitionings, while, for module $\#4$, only some of the nodes belonging to the temporal lobe have this property. 
The nodes belonging to module $\#2$, on the other hand, change from one partitioning to another. 
}
\label{fig:FigS4}
\end{figure}

As can be seen from Fig.~\ref{fig:FigS3}, there are a large number of partitionings having high values of $Q$ that occur close to each
other in the plateau and where the partition obtained from the spectral method (diamond, having a $Q$-value of $Q_{spectral} = 0.485$) 
that has been for our analysis is also seen.
This suggests that the modular decomposition of the nodes in these high $Q$ partitionings are similar to that determined by the spectral
method. Fig.~\ref{fig:FigS4} shows the brain regions whose modular identity is invariant across all the partitionings whose $Q$ differs
by less than $3 \%$  (i.e., $Q> 0.47$, left panel)  and $7 \%$ (i.e., $Q>0.45$, right panel) from $Q_{spectral}$. 
The conserved modular memberships of a large fraction ($\sim 70\%$) of the brain regions across all the different 
partitionings possessing high modularity (highlighted nodes in Fig.~\ref{fig:FigS4}~[left]; see Table ~\ref{tab:tab1} for their 
identities) emphasizes that the modular
mesoscopic organization we have described here does not depend sensitively on the method used to partition the network,
underlining that it is an intrinsic property of the Macaque connectome. 
\FloatBarrier

\FloatBarrier

\textcolor{white}{}
\vspace{-2.5cm}
\begin{longtable}{V{2}C{2.9cm}V{2}C{1.3cm}V{2}C{1.0cm}V{2}C{5.0cm}V{2}}
\caption{Brain regions highlighted in Fig.~\ref{fig:FigS4}~(left) whose modular membership is conserved across all the $291$ distinct partitionings
with $Q > 0.47$, arranged according to the modules, and subsequently into the largest anatomical subdivision (viz., lobe / nuclei), to which they belong.
For each region, the corresponding within-module degree $z$-score and the participation coefficient are displayed in the last two columns
(see Methods for details).} 
\label{tab:tab1}
\endfirsthead
\endhead
\hlineB{2}
\multicolumn{4}{V{2}cV{2}}{\MIa{Module \#1}}\\
\hlineB{2}
\Hd{Lobe/nuclei}&	 \Hd{region}&	 \Hd{zscore}&	 \Hd{pcoeff}\\
\hlineB{2}
\multirow{18}{*}{FL    }
&\MIa{14r}    	 & \MIa{1.154}&		\MIa{0.344}\\
&\MIb{13L}    	 & \MIb{0.481}&		\MIb{0.148}\\
&\MIa{13M}    	 & \MIa{0.443}&		\MIa{0.102}\\
&\MIb{13a}    	 & \MIb{2.837}&		\MIb{0.471}\\
&\MIa{32}    	 & \MIa{3.061}&		\MIa{0.688}\\
&\MIb{10m}    	 & \MIb{-0.230}&	\MIb{0.000}\\
&\MIa{10v}    	 & \MIa{-0.791}&	\MIa{0.000}\\
&\MIb{10d}    	 & \MIb{-0.866}&	\MIb{0.000}\\
&\MIa{10o}    	 & \MIa{0.443}&		\MIa{0.000}\\
&\MIb{12o}    	 & \MIb{2.613}&		\MIb{0.619}\\
&\MIa{12m}    	 & \MIa{0.593}&		\MIa{0.263}\\
&\MIb{12r}    	 & \MIb{0.107}&		\MIb{0.069}\\
&\MIa{12l}    	 & \MIa{2.239}&		\MIa{0.659}\\
&\MIb{11l}    	 & \MIb{0.368}&		\MIb{0.108}\\
&\MIa{11m}    	 & \MIa{1.042}&		\MIa{0.447}\\
&\MIb{PrCO}   	 & \MIb{0.032}&		\MIb{0.559}\\
&\MIa{6Vb}    	 & \MIa{-0.305}&	\MIa{0.457}\\
&\MIb{6Va}    	 & \MIb{-0.267}&	\MIb{0.438}\\
\hlineB{2}\hlineB{2}\multirow{6}{*}{TL    }
&\MIa{TPag}   	 & \MIa{-0.529}&	\MIa{0.000}\\
&\MIb{TPg}    	 & \MIb{-0.305}&	\MIb{0.000}\\
&\MIa{TPdgv}  	 & \MIa{-0.604}&	\MIa{0.000}\\
&\MIb{TPdgd}  	 & \MIb{-0.604}&	\MIb{0.000}\\
&\MIa{Su\#2}   	 & \MIa{-0.754}&	\MIa{0.000}\\
&\MIb{Sb}     	 & \MIb{-0.754}&	\MIb{0.000}\\
\hlineB{2}\hlineB{2}\multirow{1}{*}{cing  }
&\MIa{24b}    	 & \MIa{0.780}&		\MIa{0.549}\\
\hlineB{2}\multirow{4}{*}{Insula}
&\MIb{Ial}    	 & \MIb{-0.267}&	\MIb{0.000}\\
&\MIa{Iam}    	 & \MIa{0.630}&		\MIa{0.373}\\
&\MIb{Iapm}   	 & \MIb{0.705}&		\MIb{0.515}\\
&\MIa{Iai}    	 & \MIa{1.752}&		\MIa{0.432}\\
\hlineB{2}\multirow{7}{*}{Thal  }
&\MIb{AM\#1}   	 & \MIb{-0.492}&	\MIb{0.569}\\
&\MIa{Cim}    	 & \MIa{-0.828}&	\MIa{0.000}\\
&\MIb{Cif}    	 & \MIb{-0.716}&	\MIb{0.444}\\ 	
&\MIa{Cdc}    	 & \MIa{-0.754}&	\MIa{0.560}\\
&\MIb{MDcd}   	 & \MIb{-0.828}&	\MIb{0.000}\\
&\MIa{MDpm}   	 & \MIa{-0.754}&	\MIa{0.000}\\
&\MIb{MDfi}   	 & \MIb{-0.679}&	\MIb{0.000}\\
\hlineB{2}\hlineB{2}\multirow{2}{*}{BG    }
&\MIa{SI\#2}   	 & \MIa{0.219}&	 \MIa{0.410}\\
&\MIb{Pu\_r}   	 & \MIb{0.219}&	 \MIb{0.429}\\
\hlineB{2}\hlineB{2}\multirow{1}{*}{OFC   }
&\MIa{OFC}    	 & \MIa{-0.679}&	 \MIa{0.408}\\
\hlineB{2}\multicolumn{4}{V{2}cV{2}}{\MIIa{Module \#2}}\\
\hlineB{2}
\Hd{Lobe/nuclei}&	 \Hd{region}&	 \Hd{zscore}&	 \Hd{pcoeff}\\
\hlineB{2}
\multirow{9}{*}{FL    }
&\MIIa{F5}     	 & \MIIa{2.706}&	 \MIIa{0.515}\\
&\MIIb{4c}     	 & \MIIb{-0.575}&	 \MIIb{0.180}\\
&\MIIa{F4}     	 & \MIIa{0.807}&	 \MIIa{0.355}\\
&\MIIb{F7}     	 & \MIIb{3.512}&	 \MIIb{0.551}\\
&\MIIa{F2}     	 & \MIIa{2.418}&	 \MIIa{0.360}\\
&\MIIb{M2-FL}  	 & \MIIb{-0.862}&	 \MIIb{0.480}\\
&\MIIa{F6}     	 & \MIIa{1.267}&	 \MIIa{0.512}\\
&\MIIb{M1-FL}  	 & \MIIb{2.073}&	 \MIIb{0.256}\\
&\MIIa{MI-of}  	 & \MIIa{-0.402}&	 \MIIa{0.272}\\
\hlineB{2}\multirow{17}{*}{PL    }
&\MIIb{1\#1}  	 & \MIIb{0.691}&	 \MIIb{0.174}\\
&\MIIa{2\#1}   	 & \MIIa{0.634}&	 \MIIa{0.121}\\
&\MIIb{3b}     	 & \MIIb{0.346}&	 \MIIb{0.322}\\
&\MIIa{3a}     	 & \MIIa{0.173}&	 \MIIa{0.227}\\
&\MIIb{SII-f}  	 & \MIIb{-0.920}&	 \MIIb{0.000}\\
&\MIIa{PR\#4}  	 & \MIIa{-0.690}&	 \MIIa{0.000}\\
&\MIIb{PFop}   	 & \MIIb{-0.805}&	 \MIIb{0.000}\\
&\MIIa{PGop}   	 & \MIIa{-0.690}&	 \MIIa{0.000}\\
&\MIIb{PFG\#1} 	 & \MIIb{0.404}&	 \MIIb{0.140}\\
&\MIIa{PF\#1}  	 & \MIIa{0.807}&	 \MIIa{0.353}\\
&\MIIb{AIP}    	 & \MIIb{-0.517}&	 \MIIb{0.165}\\
&\MIIa{MIP}    	 & \MIIa{0.749}&	 \MIIa{0.307}\\
&\MIIb{PEm}    	 & \MIIb{0.979}&	 \MIIb{0.463}\\
&\MIIa{5\_Foot}	 & \MIIa{-1.035}&	 \MIIa{0.000}\\
&\MIIb{PEc\#1} 	 & \MIIb{1.152}&	 \MIIb{0.228}\\
&\MIIa{PGm}    	 & \MIIa{2.188}&	 \MIIa{0.627}\\
&\MIIb{PECg}   	 & \MIIb{1.267}&	 \MIIb{0.399}\\
\hlineB{2}\hlineB{2}\multirow{1}{*}{OL    }
&\MIIa{V6A}    	 & \MIIa{0.231}&	 \MIIa{0.375}\\
\hlineB{2}\multirow{3}{*}{cing  }
&\MIIb{24d}    	 & \MIIb{-0.114}&	 \MIIb{0.290}\\
&\MIIa{23c}    	 & \MIIa{2.188}&	 \MIIa{0.563}\\
&\MIIb{TSA}    	 & \MIIb{-0.057}&	 \MIIb{0.355}\\
\hlineB{2}\multirow{1}{*}{Insula}
&\MIIa{Ri\#1}  	 & \MIIa{-0.632}&	 \MIIa{0.494}\\
\hlineB{2}\multirow{15}{*}{Thal  }
&\MIIb{Pcn}    	 & \MIIb{0.289}&	 \MIIb{0.650}\\
&\MIIa{CM\#2}  	 & \MIIa{0.231}&	 \MIIa{0.226}\\
&\MIIb{Csl}    	 & \MIIb{-0.287}&	 \MIIb{0.640}\\
&\MIIa{Ret}    	 & \MIIa{-0.287}&	 \MIIa{0.338}\\
&\MIIb{Pul.o}  	 & \MIIb{0.576}&	 \MIIb{0.291}\\
&\MIIa{X}      	 & \MIIa{-0.287}&	 \MIIa{0.427}\\
&\MIIb{VPS}    	 & \MIIb{-0.862}&	 \MIIb{0.000}\\
&\MIIa{VPM}    	 & \MIIa{-0.460}&	 \MIIa{0.278}\\
&\MIIb{VPLo}   	 & \MIIb{-0.517}&	 \MIIb{0.000}\\
&\MIIa{VPLc}   	 & \MIIa{-0.057}&	 \MIIa{0.194}\\
&\MIIb{VLm}    	 & \MIIb{-0.575}&	 \MIIb{0.180}\\
&\MIIa{VLps}   	 & \MIIa{-0.575}&	 \MIIa{0.320}\\
&\MIIb{VLo}    	 & \MIIb{0.058}&	 \MIIb{0.308}\\
&\MIIa{VLc}    	 & \MIIa{-0.114}&	 \MIIa{0.198}\\
&\MIIb{VApc}   	 & \MIIb{-0.460}&	 \MIIb{0.375}\\
\hlineB{2}\hlineB{2}\multirow{1}{*}{BG    }
&\MIIa{Pu\_c}	 & \MIIa{-0.460}&	 \MIIa{0.375}\\
\hlineB{2}\hlineB{2}\hlineB{2}\multicolumn{4}{V{2}cV{2}}{\MIIIa{Module \#3}}\\
\hlineB{2}
\Hd{Lobe/nuclei}&	 \Hd{region}&	 \Hd{zscore}&	 \Hd{pcoeff}\\
\hlineB{2}
\hlineB{2}\hlineB{2}\multirow{17}{*}{TL    }
&\MIIIa{TFM}    	& \MIIIa{-0.309}&	 \MIIIa{0.000}\\
&\MIIIb{TFL}    	& \MIIIb{0.190}&	 \MIIIb{0.254}\\
&\MIIIa{35}     	& \MIIIa{2.044}&	 \MIIIa{0.469}\\
&\MIIIb{36c}    	& \MIIIb{1.117}&	 \MIIIb{0.111}\\
&\MIIIa{36r}    	& \MIIIa{2.329}&	 \MIIIa{0.177}\\
&\MIIIb{36p}    	& \MIIIb{-0.452}&	 \MIIIb{0.000}\\
&\MIIIa{EI}     	& \MIIIa{0.547}&	 \MIIIa{0.453}\\
&\MIIIb{ER\#1}   	& \MIIIb{0.048}&	 \MIIIb{0.499}\\
&\MIIIa{28m}    	& \MIIIa{-0.594}&	 \MIIIa{0.180}\\
&\MIIIb{ECL}    	& \MIIIb{-0.309}&	 \MIIIb{0.408}\\
&\MIIIa{EC\#2}   	& \MIIIa{0.689}&	 \MIIIa{0.487}\\
&\MIIIb{Pros.}  	& \MIIIb{-0.166}&	 \MIIIb{0.430}\\
&\MIIIa{PaS}    	& \MIIIa{-0.737}&	 \MIIIa{0.406}\\
&\MIIIb{TH}     	& \MIIIb{3.327}&	 \MIIIb{0.666}\\
&\MIIIa{PrS}		& \MIIIa{-0.452}&	 \MIIIa{0.514}\\
&\MIIIb{CA1}		& \MIIIb{0.333}&	 \MIIIb{0.159}\\
&\MIIIa{DG}     	& \MIIIa{-0.808}&	 \MIIIa{0.245}\\
\hlineB{2}\hlineB{2}\multirow{1}{*}{cing  }
&\MIIIb{29d}    	& \MIIIb{-0.951}&	 \MIIIb{0.560}\\
\hlineB{2}\hlineB{2}\hlineB{2}\hlineB{2}
\multirow{4}{*}{BG}
&\MIIIa{Bla}    	& \MIIIa{0.261}&	 \MIIIa{0.442}\\
&\MIIIb{Abpc}   	& \MIIIb{0.832}&	 \MIIIb{0.320}\\
&\MIIIa{Bi}     	& \MIIIa{0.903}&	 \MIIIa{0.412}\\
&\MIIIb{ABd}    	& \MIIIb{0.261}&	 \MIIIb{0.000}\\
&\MIIIa{Bvl}    	& \MIIIa{-0.024}&	 \MIIIa{0.000}\\
&\MIIIb{ABv}    	& \MIIIb{0.261}&	 \MIIIb{0.000}\\
&\MIIIa{MB}     	& \MIIIa{-0.024}&	 \MIIIa{0.290}\\
&\MIIIb{ABvm}   	& \MIIIb{0.547}&	 \MIIIb{0.204}\\
&\MIIIa{ABmg}   	& \MIIIa{0.832}&	 \MIIIa{0.420}\\
&\MIIIb{A}      	& \MIIIb{-0.095}&	 \MIIIb{0.111}\\
&\MIIIa{I\#2}    	& \MIIIa{-0.879}&	 \MIIIa{0.278}\\
&\MIIIb{ME\#1}   	& \MIIIb{-0.166}&	 \MIIIb{0.430}\\
&\MIIIa{CE\#1}   	& \MIIIa{-0.095}&	 \MIIIa{0.360}\\
&\MIIIb{AHA}    	& \MIIIb{-0.594}&	 \MIIIb{0.180}\\
&\MIIIa{PAC2}   	& \MIIIa{-0.238}&	 \MIIIa{0.000}\\
&\MIIIb{COp}    	& \MIIIb{-0.808}&	 \MIIIb{0.000}\\
&\MIIIa{NLOT}   	& \MIIIa{-0.808}&	 \MIIIa{0.000}\\
&\MIIIb{COa}    	& \MIIIb{-0.523}&	 \MIIIb{0.397}\\
&\MIIIa{Ldi}    	& \MIIIa{1.117}&	 \MIIIa{0.216}\\
&\MIIIb{Ld\#2}   	& \MIIIb{-0.166}&	 \MIIIb{0.219}\\
&\MIIIa{Lv}     	& \MIIIa{0.974}&	 \MIIIa{0.229}\\
&\MIIIb{Lvl}    	& \MIIIb{0.618}&	 \MIIIb{0.137}\\
\hlineB{2}\hlineB{2}\hlineB{2}\multicolumn{4}{V{2}cV{2}}{\MIVa{Module \#4}}\\
\hlineB{2}
\Hd{Lobe/nuclei}&	 \Hd{region}&	 \Hd{zscore}&	 \Hd{pcoeff}\\
\hlineB{2}
\multirow{5}{*}{FL    }
&\MIVa{M9}    	 & \MIVa{0.227}&	 \MIVa{0.677}\\
&\MIVb{D9}     	 & \MIVb{-0.257}&	 \MIVb{0.602}\\
&\MIVa{46v}    	 & \MIVa{2.437}&	 \MIVa{0.741}\\
&\MIVb{46d}    	 & \MIVb{0.848}&	 \MIVb{0.477}\\
&\MIVa{8B}     	 & \MIVa{2.299}&	 \MIVa{0.629}\\
\hlineB{2}\hlineB{2}\multirow{14}{*}{TL    }
&\MIVb{A1}     	 & \MIVb{0.641}&	 \MIVb{0.340}\\
&\MIVa{STPg}   	 & \MIVa{-0.326}&	 \MIVa{0.418}\\
&\MIVb{ProK}   	 & \MIVb{-0.326}&	 \MIVb{0.231}\\
&\MIVa{paAc}   	 & \MIVa{0.089}&	 \MIVa{0.381}\\
&\MIVb{L\#1}   	 & \MIVb{0.019}&	 \MIVb{0.320}\\
&\MIVa{CL\#4}  	 & \MIVa{-0.671}&	 \MIVa{0.480}\\
&\MIVb{AL\#4}  	 & \MIVb{-0.464}&	 \MIVb{0.426}\\
&\MIVa{ST3}    	 & \MIVa{1.194}&	 \MIVa{0.370}\\
&\MIVb{ST2}    	 & \MIVb{0.641}&	 \MIVb{0.445}\\
&\MIVa{ST1}    	 & \MIVa{0.227}&	 \MIVa{0.469}\\
&\MIVb{Tpt}    	 & \MIVb{0.848}&	 \MIVb{0.461}\\
&\MIVa{TPOc}   	 & \MIVa{0.710}&	 \MIVa{0.571}\\
&\MIVb{TPOr}   	 & \MIVb{-0.257}&	 \MIVb{0.492}\\
&\MIVa{TAa}    	 & \MIVa{0.434}&	 \MIVa{0.441}\\
\hlineB{2}\hlineB{2}\hlineB{2}\hlineB{2}\multirow{3}{*}{Thal  }
&\MIVb{MG}     	 & \MIVb{-0.533}&	 \MIVb{0.000}\\
&\MIVa{SG}     	 & \MIVa{-0.188}&	 \MIVa{0.519}\\
&\MIVb{Li}     	 & \MIVb{0.019}&	 \MIVb{0.615}\\
\hlineB{2}\hlineB{2}\hlineB{2}\hlineB{2}\hlineB{2}\multicolumn{4}{V{2}cV{2}}{\MVa{Module \#5}}\\
\hlineB{2}
\Hd{Lobe/nuclei}&	 \Hd{region}&	 \Hd{zscore}&	 \Hd{pcoeff}\\
\hlineB{2}
\multirow{2}{*}{FL    }
&\MVa{45A}    	 & \MVa{0.385}&		\MVa{0.562}\\
&\MVb{8Ac}    	 & \MVb{-1.003}&	\MVb{0.000}\\
\hlineB{2}\multirow{4}{*}{PL    }
&\MVa{LIPe}   	 & \MVa{-0.268}&	\MVa{0.660}\\
&\MVb{LIPi}   	 & \MVb{-0.023}&	\MVb{0.568}\\
&\MVa{VIP}    	 & \MVa{0.793}&		\MVa{0.507}\\
&\MVb{PIP\#1}  	 & \MVb{-0.268}&	\MVb{0.000}\\
\hlineB{2}\multirow{9}{*}{TL    }
&\MVa{CITv}   	 & \MVa{0.466}&		\MVa{0.560}\\
&\MVb{TEm}    	 & \MVb{-0.350}&	\MVb{0.691}\\
&\MVa{PITd}  	 & \MVa{-0.023}&	\MVa{0.142}\\
&\MVb{PITv}   	 & \MVb{0.058}&		\MVb{0.500}\\
&\MVa{IPa}    	 & \MVa{0.385}&		\MVa{0.710}\\
&\MVb{MT}     	 & \MVb{2.997}&		\MVb{0.314}\\
&\MVa{FST}    	 & \MVa{1.446}&		\MVa{0.454}\\
&\MVb{MSTp}   	 & \MVb{0.140}&		\MVb{0.231}\\
&\MVa{MSTd}   	 & \MVa{1.446}&		\MVa{0.497}\\
\hlineB{2}\multirow{12}{*}{OL    }
&\MVb{V3A}    	 & \MVb{0.793}&		\MVb{0.159}\\
&\MVa{V3v}    	 & \MVa{0.711}&		\MVa{0.000}\\
&\MVb{V4t}    	 & \MVb{0.140}&		\MVb{0.338}\\
&\MVa{DLr}    	 & \MVa{-0.921}&	\MVa{0.000}\\
&\MVb{DLc}    	 & \MVb{-0.921}&	\MVb{0.000}\\
&\MVa{V4v}    	 & \MVa{-0.758}&	\MVa{0.000}\\
&\MVb{VPP}    	 & \MVb{-0.921}&	\MVb{0.000}\\
&\MVa{V6}     	 & \MVa{1.283}&		\MVa{0.447}\\
&\MVb{DP}     	 & \MVb{0.303}&		\MVb{0.443}\\
&\MVa{VOT}    	 & \MVa{-0.595}&	\MVa{0.000}\\
&\MVb{V1}     	 & \MVb{1.283}&		\MVb{0.250}\\
&\MVa{V2}     	 & \MVa{3.160}&		\MVa{0.387}\\
\hlineB{2}\hlineB{2}\hlineB{2}\multirow{9}{*}{Thal  }
&\MVb{LGN}    	 & \MVb{-0.595}&	\MVb{0.278}\\
&\MVa{PIl-s}  	 & \MVa{-0.921}&	\MVa{0.000}\\
&\MVb{PIp}    	 & \MVb{-0.840}&	\MVb{0.000}\\
&\MVa{PIm}    	 & \MVa{-0.513}&	\MVa{0.245}\\
&\MVb{PIl}    	 & \MVb{-0.431}&	\MVb{0.000}\\
&\MVa{PIc}    	 & \MVa{-0.431}&	\MVa{0.219}\\
&\MVb{PLa\#1}  	 & \MVb{-0.921}&	\MVb{0.000}\\
&\MVa{PLvl}   	 & \MVa{-0.758}&	\MVa{0.000}\\
&\MVb{PLvm}   	 & \MVb{-0.758}&	\MVb{0.000}\\
\hlineB{2}\hlineB{2}\multirow{1}{*}{BG    }
&\MVa{Cd\_g}   	 & \MVa{-0.105}&	\MVa{0.653}\\
\hlineB{2}\multirow{1}{*}{MB    }
&\MVb{MB\#2}   	 & \MVb{-0.187}&	\MVb{0.298}\\
\hlineB{2}\hlineB{2}
\end{longtable}


\clearpage
\FloatBarrier

\subsection{Modular decomposition of the cortical and sub-cortical subdivisions of the Macaque brain}
\begin{figure}[tbp]
\begin{tabular}{cc}
{\bf Frontal Lobe} & {\bf Parietal Lobe}\\
\includegraphics[width=0.47\linewidth]{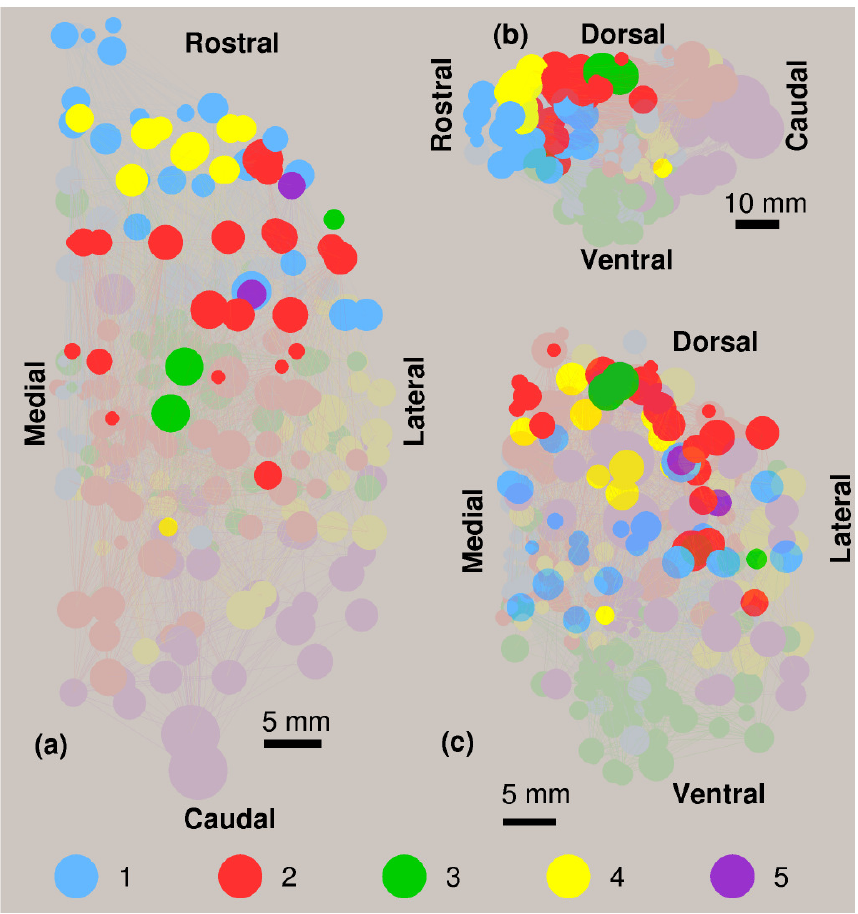}&
\includegraphics[width=0.47\linewidth]{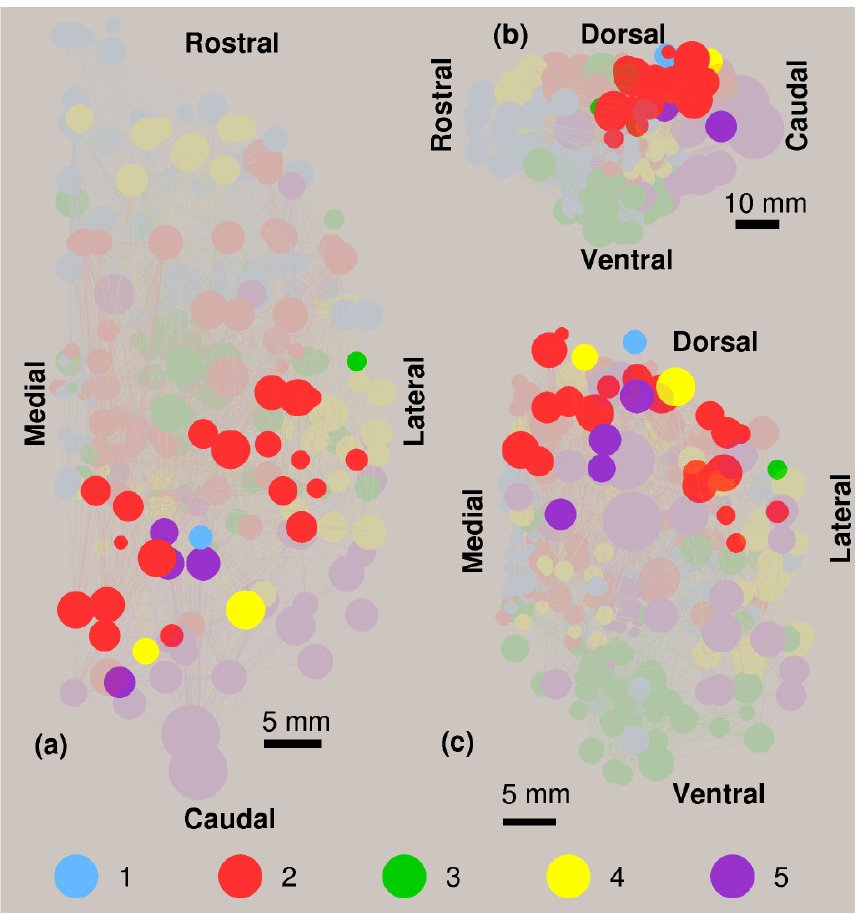}\\
 & \\
{\bf Temporal Lobe} & {\bf Occipital Lobe}\\
\includegraphics[width=0.47\linewidth]{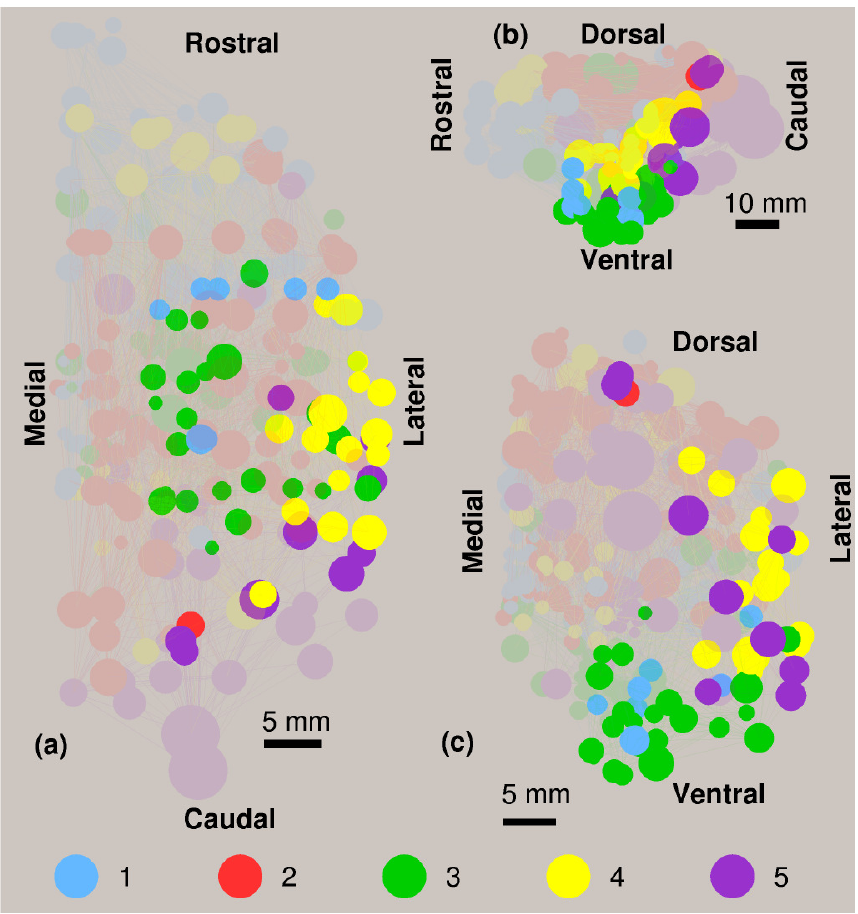}&
\includegraphics[width=0.47\linewidth]{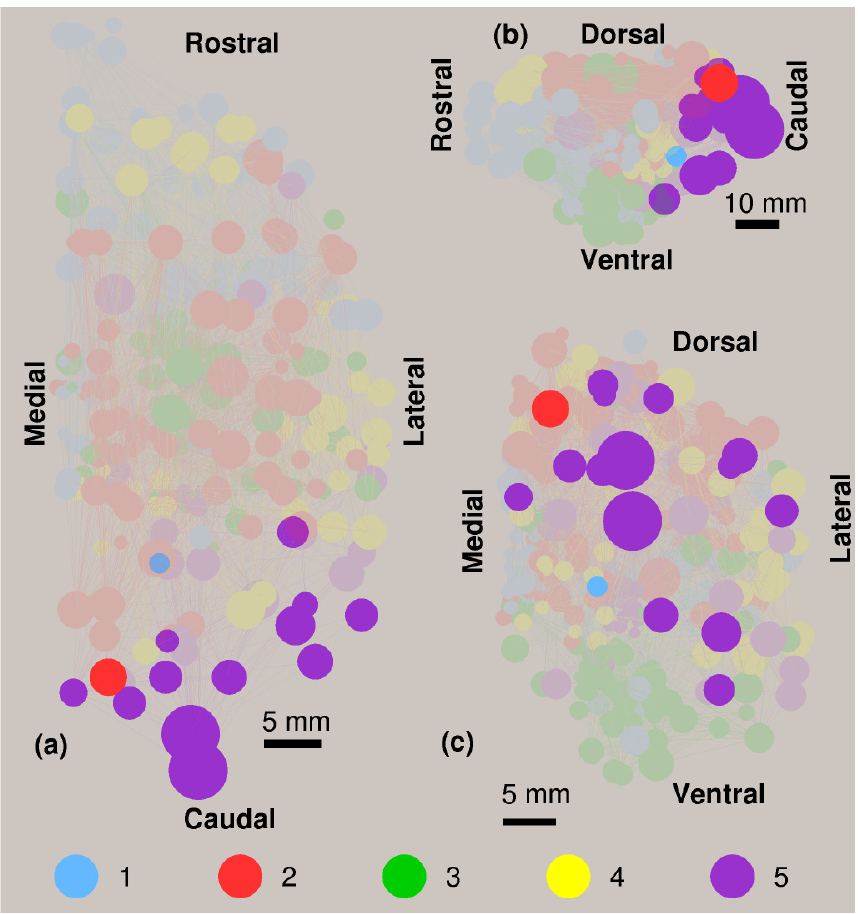}\\
\end{tabular}
\caption{The network of brain regions shown in (a) horizontal, (b) sagittal and (c) coronal projections, 
indicating the modular memberships of the regions (highlighted) that belong to the frontal (top left), parietal (top right), temporal (bottom left) and 
occipital (bottom right) lobes. As in Fig.~1~(a-c) in the main text, the modular membership of each node is represented by its color (see color key at the bottom of
each panel), the spatial positions of the nodes are specified by the three-dimensional stereotaxic coordinates of the corresponding regions,
and node sizes provide a representation of the relative volumes of the corresponding brain regions (the spatial scale
being indicated by the horizontal bar shown next to each projection).}
\label{fig:FigS5}
\end{figure}

\begin{figure}[tbp]
\begin{tabular}{cc}
{\bf Thalamus} & {\bf Basal Ganglia}\\
\includegraphics[width=0.47\linewidth]{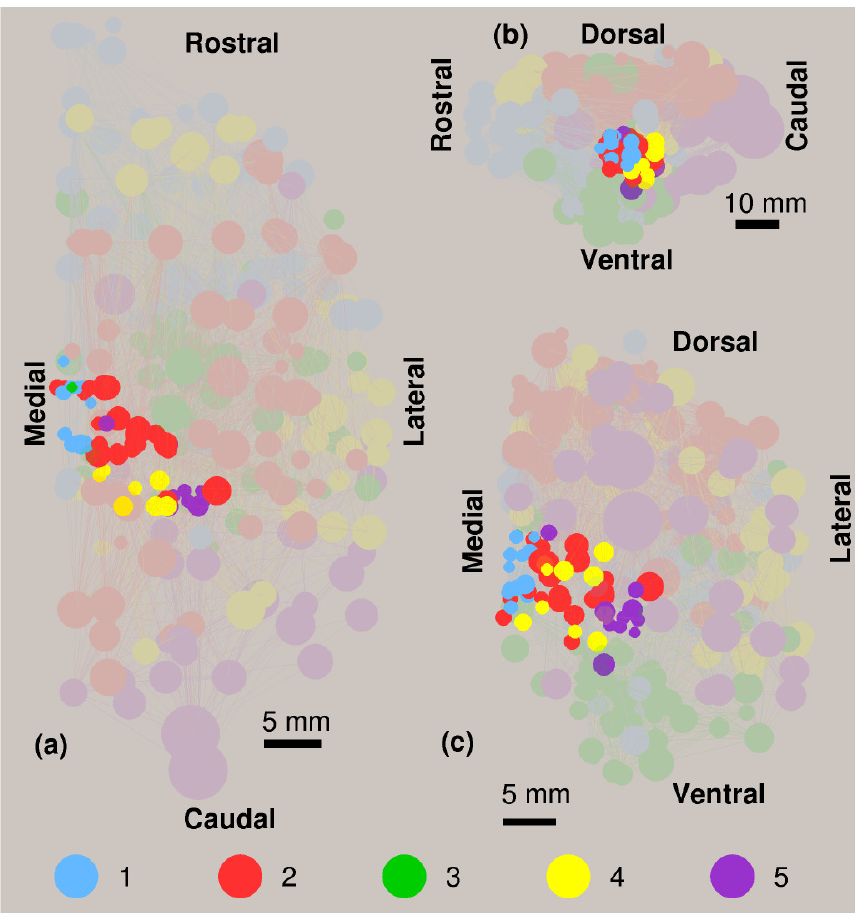}&
\includegraphics[width=0.47\linewidth]{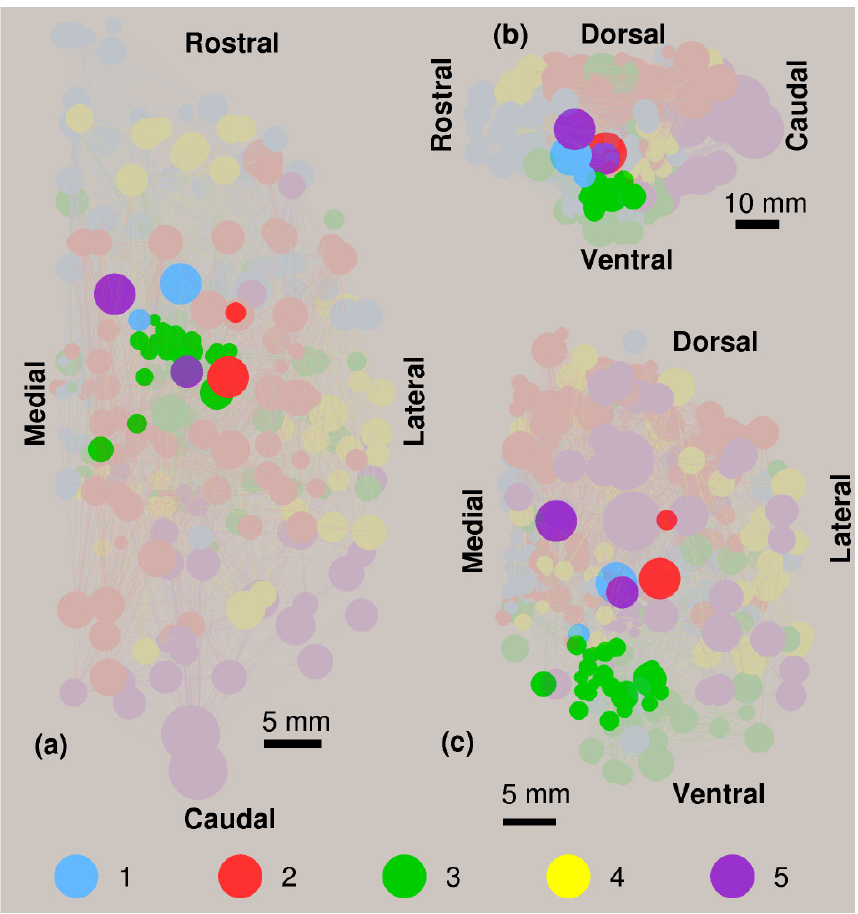}\\
 & \\
{\bf Cingulate} & {\bf Insula}\\
\includegraphics[width=0.47\linewidth]{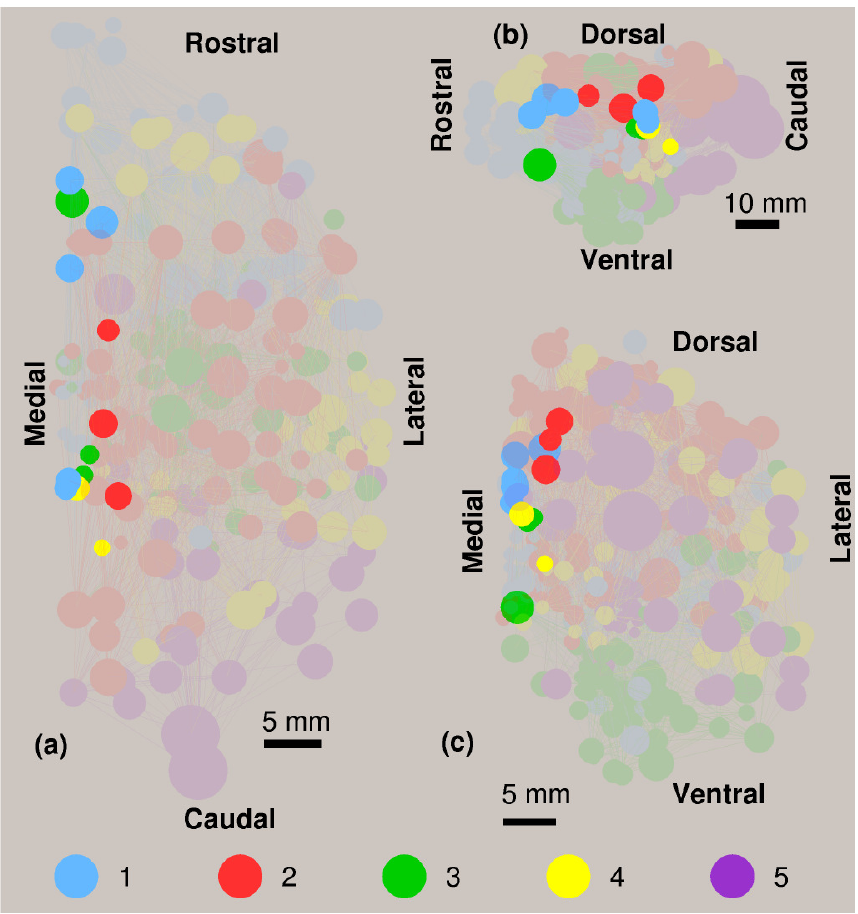}&
\includegraphics[width=0.47\linewidth]{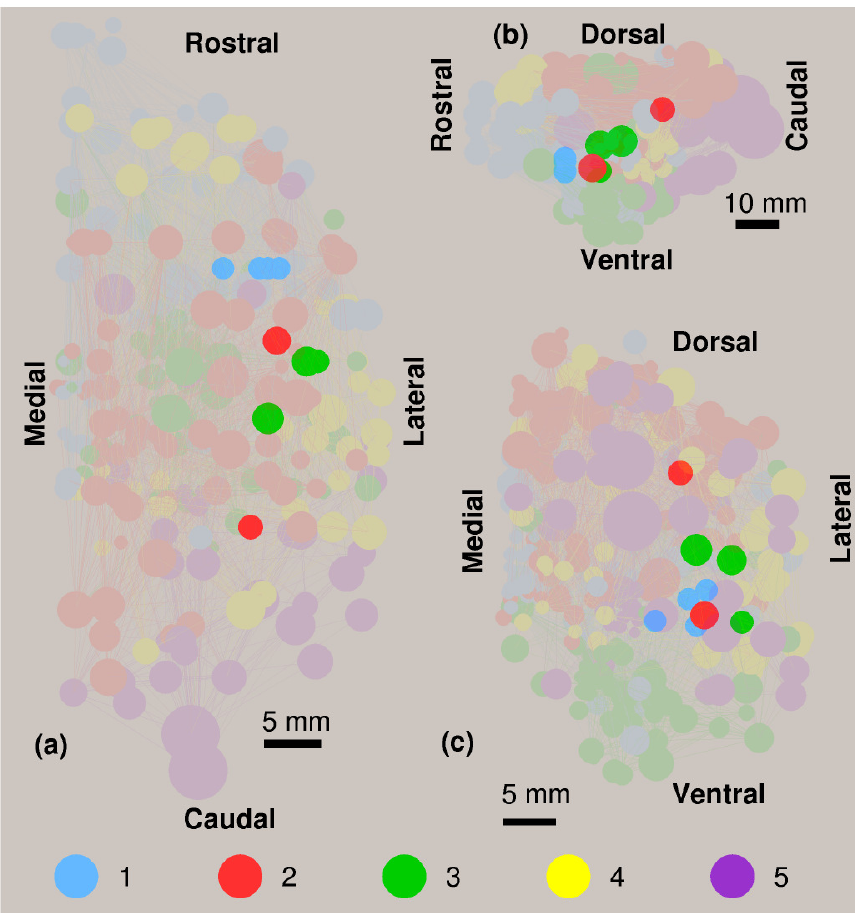}\\
\end{tabular}
\caption{The network of brain regions shown in (a) horizontal, (b) sagittal and (c) coronal projections, 
indicating the modular memberships of the regions (highlighted) that belong to the thalamus (top left), basal ganglia (top right), cingulate (bottom left) and 
insula (bottom right). As in Fig.~1~(a-c) in the main text, the modular membership of each node is represented by its color (see color key at the bottom of
each panel), the spatial positions of the nodes are specified by the three-dimensional stereotaxic coordinates of the corresponding regions,
and node sizes provide a representation of the relative volumes of the corresponding brain regions (the spatial scale
being indicated by the horizontal bar shown next to each projection).}
\label{fig:FigS6}
\end{figure}

As mentioned in the main text, there is no simple correspondence between the modules and the anatomical subdivisions of the brain.
The nodes of the connectome we have investigated are brain regions that belong to larger subdivisions, such as the prefrontal cortex, which in turn are
part of broader anatomical categories such as the frontal lobe. The association between the network modules and the largest subdivisions
have been shown in Fig.~1(d) in the main text. 
A more detailed representation of this relation is given in terms of the modular spectra of the anatomical subdivisions in Table~\ref{tab:tab2} which
indicates how the regions belonging to each subdivision are distributed among
the five modules. 
We note that some of the subdivisions constitute a single brain region in the connectome we consider (e.g., Visual area V1 in the Occipital
lobe), so that they belong exclusively to one of the modules. Larger subdivisions that comprise multiple regions, on the other hand, can
have their constituent regions distributed non-uniformly among several modules. In such cases, we highlight the dominant module(s) of the
subdivision, i.e., those amongst the five modules having the largest number of brain regions, in the table. The spatial layout of the 
brain regions belonging to these larger subdivisions, colored according to the modules to which they belong, are also shown in
Figs.~\ref{fig:FigS5} and \ref{fig:FigS6}. Note that, the regions belonging to the parietal lobe occur predominantly in module \#2, while those
in the occipital lobe occur predominantly in module \#5 (see Fig.~\ref{fig:FigS5}). Fig.~\ref{fig:FigS6} suggests that the regions belonging to
the basal ganglia mostly occur in module \#3.
 
\FloatBarrier

\begin{longtable}{V{2}C{3.4cm}V{2}>{\columncolor{LightPink2}}p{5.9cm}V{2}>{\columncolor{MistyRose1}}p{0.18cm}p{0.18cm}>{\columncolor{MistyRose1}}p{0.18cm}p{0.18cm}>{\columncolor{MistyRose1}}p{0.18cm}V{2}}
\caption{Modular decomposition of the brain regions in different anatomical subdivisions of the Macaque brain.}
\label{tab:tab2}
\endfirsthead
\endhead
\hlineB{2}
Lobe/Nuclei  & \centering{Subdivision [abbreviation]} & \multicolumn{5}{cV{2}}{\cellcolor{MistyRose1}modular distrib.}\\
(no. of brain regions)   &                                         & \textbf{1} & \textbf{2} & \textbf{3} & \textbf{4} & \textbf{5}\\
\hlineB{2}
\multirow{4}{*}{Frontal lobe (58)}
&{beltline of sensorymotor syst. [belt\_sm]} &   0    & \R{1}  &  0  &  0  &  0  \\
&\C{Prefrontal cortex [PFC]}                 & \R{18} &   5    &  0  & 11  &  2  \\
&{Supplementary motor area [Area 6]}         &   3    & \R{11} &  1  &  0  &  0  \\
&\C{Primary motor area [MI]}                 &   0    & \R{4}  &  2  &  0  &  0  \\
\hlineB{2}
\multirow{5}{*}{Temporal lobe (56)}
&{Ventral temporal cortex [TCV]}             &   4   &  0  & \R{6}  &   0    &   0   \\ 
&\C{Parahippocampal cortex [PHC]}            &   3   &  0  & \R{11} &   0    &   0   \\
&{Hippocampus [Hip]}                         &   0   &  0  & \R{3}  &   0    &   0   \\
&\C{Superior temporal gyrus [STG]}           &   0   &  0  &   0    & \R{11} &   0   \\
&{Inferotemporal area [TE]}                  &   0   &  0  &   3    &   0    & \R{4} \\
&\C{Superior temporal sulcus [STS]}          &   0   &  1  &   0    & \R{5}  & \R{5} \\
\hlineB{2}
\multirow{7}{*}{Parietal lobe (27)}
&{Primary somatosensory cortex [S1]}         &   0   & \R{4} &   0   &  0  &   0   \\
&\C{Secondary somatosensory cortex [S2]}     &   0   & \R{1} & \R{1} &  0  &   0   \\
&{beltline of sensory syst. [belt\_s]}       & \R{1} &   0   &   0   &  0  &   0   \\
&\C{Rostral parietal area [PR\#4]}           &   0   & \R{1} &   0   &  0  &   0   \\
&{Somatosensory association area [7\#1]}     &   0   & \R{5} &   0   &  2  &   0   \\
&\C{Cortex of intraparietal sulcus [PCip]}   &   0   &   2   &   0   &  0  & \R{4} \\
&{Dorsal parietal cortex [PCd\#2]}           &   0   & \R{6} &   0   &  0  &   0   \\
\hlineB{2}
\multirow{4}{*}{Occipital lobe (16)}
&\C{Visual anterior cortex [VAC]}            &   0   &  1  &  0  &  0  & \R{12} \\
&{Visual area V1 [V1]}                       &   0   &  0  &  0  &  0  & \R{1}  \\
&\C{Prostriate cortex [ProST]}               & \R{1} &  0  &  0  &  0  &   0    \\
&{Visual area V2 [V2]}                       &   0   &  0  &  0  &  0  & \R{1}  \\
\hlineB{2}                                 
\multirow{10}{*}{Thalamus (53)}
&\C{Anterior nuclei [AN]}                    & \R{2}  &   1    &  0  &   0   &   1   \\
&{Midline nuclei [ML]}                       & \R{5}  &   1    &  1  &   0   &   0   \\
&\C{Geniculate nucleus [GN]}                 &   0    &   0    &  0  & \R{1} & \R{1} \\
&{Intralaminar nuclei [IL2]}                 &   1    & \R{3}  &  0  &   1   &   0   \\
&\C{Massa intermedia [MI1]}                  &   0    & \R{1}  &  0  &   0   &   0   \\
&{Posterior nuclei [PN]}                     &   0    &   0    &  0  & \R{2} &   0   \\
&\C{Reticularis thalami [Ret]}               &   0    & \R{1}  &  0  &   0   &   0   \\
&{Pulvinaris thalami [Pul\#1]}               &   0    &   1    &  0  &   3   & \R{8} \\
&\C{Medial dorsal nucleus [MD]}              & \R{3}  &   2    &  0  &   1   &   0   \\
&{Ventrolateral nuclei [VN]}                 &   1    & \R{12} &  0  &   0   &   0   \\
\hlineB{2}
\multirow{7}{*}{Basal Ganglia (31)}
&\C{Amygdala [Amyg]}                         &   0   &   0   & \R{22} & 0  &   0   \\
&{Substantia nigra [SN]}                     &   0   &   0   & \R{1}  & 0  &   0   \\
&\C{Substantia innominata [SI\#2]}           & \R{1} &   0   &   0    & 0  &   0   \\
&{Nucleus subthalamus [Sub.Th]}              &   0   &   0   & \R{1}  & 0  &   0   \\
&\C{Globus pallidus [GPe]}                   &   0   &   0   &   0    & 0  & \R{1} \\
&{Striatum [STR]}                            & \R{1} & \R{1} & \R{1}  & 0  & \R{1} \\
&\C{Claustrum [Clau]}                        &   0   & \R{1} &   0    & 0  &   0   \\
\hlineB{2}
\multirow{4}{*}{Cingulate Gyrus (13)}
&{Area 24 [24]}                              & \R{3} &   1   &   0   &   0   & 0  \\
&\C{Area 23 [23]}                            & \R{2} & \R{2} &   0   &   0   & 0  \\
&{Area 26 [26]}                              &   0   &   0   & \R{2} & \R{2} & 0  \\
&\C{Area 25 [25]}                            &   0   &   0   & \R{1} &   0   & 0  \\
\hlineB{2}
\multirow{5}{*}{Insula (9)}
&{Granular insular cortex [Ig\#1]}           &   0   &   0   & \R{1} &  0  &  0  \\
&\C{Retroinsular cortex [Ri\#1]}             &   0   & \R{1} &   0   &  0  &  0  \\
&{Insular proisocortex [IPro]}               &   0   & \R{1} &   0   &  0  &  0  \\
&\C{Parainsular field [Pi\#1]}               &   0   &   0   & \R{1} &  0  &  0  \\
&{Anterior insula [IA]}                      & \R{4} &   0   &   1   &  0  &  0  \\
\hlineB{2}
Hypothalamus (1)
&\C{Hypothalamus [Hyp]}                      &   0   &   0   & \R{1} &  0  &   0  \\
\hlineB{2}
Midbrain (1)
&{Midbrain [MB]}                             &   0   &   0   &   0   &  0  & \R{1} \\
\hlineB{2}
Olfactory complex (1)
&\C{Olfactory complex [OFC]}                 & \R{1} &   0   &   0   &  0  &   0   \\
\hlineB{2}
\end{longtable}

\FloatBarrier

\FloatBarrier


\FloatBarrier
\subsection{Functional characterization of modules}
As mentioned in the main text, we have investigated a possible structure-function correlation in the mesoscopic organization of the connectome, 
which would be reflected in the modules being predominantly associated with certain functionalities. In Table~\ref{tab:tab3} we list the known functions (obtained from the literature) either of the brain regions belonging to each of the modules, or of the broader subdivisions to which such regions belong
(the first column indicating the lobe or nuclei, and the second specifying the areas comprising it). As the different regions belonging to a subdivision
may lie in distinct modules, the fraction of all the regions in a subdivision that are in a specific module are indicated in the third column.
The role that these regions play in terms intra- and inter-modular communication can be inferred from the average values (computed over all the 
regions in the subdivision that are in the same module) of the participation coefficient, $\langle p \rangle$, and the within-module 
degree $z$-score, $\langle z \rangle$, which are shown in the fourth and fifth columns, respectively.
Some of the brain regions in a subdivision that have been investigated relatively more extensively are mentioned in the sixth column, while
the seventh column provides a non-exhaustive list of the functions that are known to be associated with these regions and/or the subdivision to which
they belong (along with references to the relevant literature). As alluded to in the main text, regions belonging to the same module have certain functions that appear to complement each
other in carrying out a specific cognitive task, e.g., high-level multimodal sensory integration and decision-making (module $\#1$),
motor control and somato-sensory processing (module $\#2$), memory and emotion (module $\#3$), auditory processing (module $\#4$)
and visual processing (module $\#5$).

\FloatBarrier

\textcolor{white}{}
\vspace{-0.5cm}

\begin{longtable}{V{2}C{2.25cm}V{2}C{1.2cm}V{2}C{0.9cm}V{2}C{0.65cm}V{2}C{0.75cm}V{2}C{2.85cm}V{2}C{3.20cm}V{2}}
\caption{Functional characterization of modules}
\label{tab:tab3}
\endfirsthead
\endhead
\hlineB{2}
\multicolumn{7}{V{2}cV{2}}{\MIa{Module \#1}}\\
\hlineB{2}
\Hd{Lobe/Nuclei}
&\Hd{Subdivision}			&\Hd{Frac.}	&\Hd{$\langle p\rangle$} &\Hd{$\langle z\rangle$} &\Hd{Notable regions}	&\Hd{Known functions}\\
\hlineB{2}
\multirow{4}{*}{Frontal lobe}
&\MIa{{PFC}}		&\MIa{18/36}&\MIa{0.24}		&\MIa{0.63}		&\MIa{10, 11, 13, 14}	&\MIa{sensory integration, decision making \cite{Walton2010, Fellows2007, Walton2004, Izquierdo2004, Rolls2000}}\\
&\MIb{Area 6}		&\MIb{3/15}	&\MIb{0.48}		&\MIb{-0.18}	&\MIb{6Va, 6Vb, PrCo}	&\MIb{complex locomotion (e.g. climbing) \cite{Graziano2007, Gentilucci1988}}\\
\hlineB{2}
\multirow{2}{*}{Temporal lobe}
&\MIa{{TCv}}		&\MIa{4/10}	&\MIa{0.00}		&\MIa{-0.51}	&\MIa{}	&\MIa{}\\
&\MIb{{PHC}}		&\MIb{3/14}	&\MIb{0.17}		&\MIb{-0.68}	&\MIb{EO} & \MIb{olfaction~\cite{Carmichael1994}}\\
\hlineB{2}
\multirow{1}{*}{Parietal lobe}
&\MIa{belt\_s}		&\MIa{1/1}	&\MIa{0.64}		&\MIa{-0.70}	&\MIa{}					&\MIa{}\\
\hlineB{2}
\multirow{1}{*}{Occipital lobe}
&\MIb{ProSt}		&\MIb{1/1}	&\MIb{0.50}		&\MIb{-0.86}	&\MIb{}					&\MIb{}\\
\hlineB{2}
\multirow{5}{*}{Thalamus}
&\MIa{AN}			&\MIa{2/4}	&\MIa{0.52}		&\MIa{-0.62}	&\MIa{}					&\MIa{}\\
&\MIb{{ML}}		&\MIb{5/7}	&\MIb{0.39}		&\MIb{-0.78}	&\MIb{}					&\MIb{}\\
&\MIa{{IL\#2}}	&\MIa{1/5}	&\MIa{0.00}		&\MIa{-0.90}	&\MIa{}					&\MIa{}\\
&\MIb{{MD}}		&\MIb{3/6}	&\MIb{0.00}		&\MIb{-0.75}	&\MIb{}					&\MIb{}\\
&\MIa{VN}			&\MIa{1/12}	&\MIa{0.57}		&\MIa{-0.15}	&\MIa{}					&\MIa{}\\
\hlineB{2}
\multirow{4}{*}{Basal Ganglia}
&\MIb{{SI\#2}}	&\MIb{1/1}	&\MIb{0.41}		&\MIb{0.22}		&\MIb{}					&\MIb{}\\
&\MIa{{STR}}		&\MIa{1/4}	&\MIa{0.43}		&\MIa{0.22}		&\MIa{Pu\_r} & \MIa{motor skills, reinforcement learning~\cite{Apicella1991}}\\
\hlineB{2}
\multirow{4}{*}{\makecell{Cingulate\\ gyrus}}
&\MIb{Area 24}		&\MIb{3/4}	&\MIb{0.57}		&\MIb{0.60}		&\MIb{24a, 24b, 24c}	&\MIb{emotional behavioural control~\cite{Baleydier1980}}\\
&\MIa{Area 23}		&\MIa{2/4}	&\MIa{0.61}		&\MIa{-0.55}	&\MIa{23a,23b}			&\MIa{multi-sensory integration~\cite{Baleydier1980}}\\
\hlineB{2}
\multirow{1}{*}{Insula}
&\MIb{{IA}}		&\MIb{4/5}	&\MIb{0.33}		&\MIb{0.70}		&\MIb{Iam, Iai}	&\MIb{social cognition~\cite{Evrard2012}}\\
\hlineB{2}
\multirow{1}{*}{\makecell{olfactory\\ complex}}&\multirow{1}{*}{\MIa{{OFC}}}		&\multirow{1}{*}{\MIa{1/1}}	&\multirow{1}{*}{\MIa{0.40}}		&\multirow{1}{*}{\MIa{-0.68}}	&\multirow{1}{*}{\MIa{}}					&\multirow{1}{*}{\MIa{olfaction~\cite{Carmichael1994}}}\\
&\MIa{}&\MIa{}&\MIa{}&\MIa{}&\MIa{}&\MIa{}\\
\hlineB{2}
\multicolumn{7}{V{2}cV{2}}{\MIIa{Module \#2}}\\
\hlineB{2}
\Hd{Lobe/nuclei}
&\Hd{Subdivision}		&\Hd{Frac.}	&\Hd{$\langle p\rangle$} &\Hd{$\langle z\rangle$}	&\Hd{Notable regions}	&\Hd{Known function}\\
\hlineB{2}
\multirow{9}{*}{Frontal lobe}							
&\MIIa{belt\_sm}	&\MIIa{1/1}		&\MIIa{0.54}	&\MIIa{-0.52}			&\MIIa{}			&\MIIa{} \\
&\MIIb{{PFC}}	    &\MIIb{5/36}	&\MIIb{0.48}	&\MIIb{-0.46}			&\MIIb{45B, 8Ad}	&\MIIb{saccadic guidance (frontal eye field) \cite{Schall2004}}\\
&\MIIa{{Area 6}}	&\MIIa{11/15}	&\MIIa{0.36}	&\MIIa{0.66}			&\MIIa{}							&\MIIa{complex locomotion (climbing etc.)~\cite{Graziano2007}}\\
&\MIIb{{MI}}		&\MIIb{4/6}		&\MIIb{0.27}	&\MIIb{0.20}			&\MIIb{M1-FL, M1-HL}	&\MIIb{voluntary movement (primary motor area)~\cite{Kandel2000}}\\
\hlineB{2}
\multirow{1}{*}{Temporal lobe}							
&\MIIa{STS}			&\MIIa{1/11}	&\MIIa{0.44}	&\MIIa{-0.98}			&\MIIa{}		&\MIIa{}\\
\hlineB{2}
\multirow{12}{*}{Parietal lobe}							
&\MIIb{{S1}}		&\MIIb{4/4}		&\MIIb{0.21}	&\MIIb{0.46}			&\MIIb{}							&\MIIb{primary somatosensory cortex~\cite{Kandel2000}}\\
&\MIIa{S2}			&\MIIa{1/2}		&\MIIa{0.00}	&\MIIa{-0.92}			&\MIIa{SII-f}  &\MIIa{secondary somatosensory area (face representation)~\cite{Manzoni1986}} \\
&\MIIb{PR\#4}		&\MIIb{1/1}		&\MIIb{0.00}	&\MIIb{-0.69}			&\MIIb{}							&\MIIb{} \\
&\MIIa{{Area 7}}	&\MIIa{5/7}		&\MIIa{0.10}	&\MIIa{-0.28}			&\MIIa{PF\#1, PFG\#1}							&\MIIa{visual-motor coordination~\cite{Andersen1990}}\\
&\MIIb{{PCip}}	&\MIIb{2/6}		&\MIIb{0.24}	&\MIIb{0.12}			&\MIIb{AIP, MIP}					&\MIIb{visual control of reaching \& pointing~\cite{Sakata1995, Eskandar1999}}\\
&\MIIa{{PCd\#2}}	&\MIIa{6/6}		&\MIIa{0.39}	&\MIIa{0.91}			&\MIIa{PEm, PEc\#1}		&\MIIa{somesthesia \& motor control~\cite{Caminiti1996}} \\
\hlineB{2}
\multirow{1}{*}{Occipital lobe}							
&\MIIb{{VAC}}		&\MIIb{1/13}	&\MIIb{0.38}	&\MIIb{0.23}			&\MIIb{}							&\MIIb{}\\
\hlineB{2}
\multirow{10}{*}{Thalamus}							
&\MIIa{AN}			&\MIIa{1/4}		&\MIIa{0.72}	&\MIIa{-0.86}			&\MIIa{}							&\MIIa{}\\
&\MIIb{ML}			&\MIIb{1/7}		&\MIIb{0.59}	&\MIIb{-0.63}			&\MIIb{}							&\MIIb{}\\
&\MIIa{IL\#2}		&\MIIa{3/5}		&\MIIa{0.50}	&\MIIa{0.07}			&\MIIa{}							&\MIIa{}\\
&\MIIb{MI\#1}		&\MIIb{1/1}		&\MIIb{0.00}	&\MIIb{-1.09}			&\MIIb{}			&\MIIb{}\\
&\MIIa{{Ret}}		&\MIIa{1/1}		&\MIIa{0.34}	&\MIIa{-0.29}			&\MIIa{}			&\MIIa{}\\
&\MIIb{{Pul\#1}}	&\MIIb{1/12}	&\MIIb{0.29}	&\MIIb{0.58}			&\MIIb{}						&\MIIb{}\\
&\MIIa{MD}			&\MIIa{2/6}		&\MIIa{0.60}	&\MIIa{-0.50}			&\MIIa{}					&\MIIa{}\\
&\MIIb{{VN}}		&\MIIb{12/13}	&\MIIb{0.19}	&\MIIb{-0.45}			&\MIIb{}							&\MIIb{somatosensory information relay~\cite{Kandel2000}}\\
\hlineB{2}
\multirow{1}{*}{Basal Ganglia}
&\MIIa{{STR}}		&\MIIa{1/4}		&\MIIa{0.37}	&\MIIa{-0.50}			&\MIIa{Pu\_c}	&\MIIa{motor skills, reinforcement learning~\cite{Apicella1991}}\\
\hlineB{2}
\multirow{2}{*}{\makecell{Cingulate\\ gyrus}}							
&\MIIb{{Area 24}}	&\MIIb{1/4}		&\MIIb{0.29}	&\MIIb{-0.11}			&\MIIb{}							&\MIIb{}\\
&\MIIa{{Area 23}}	&\MIIa{2/4}		&\MIIa{0.46}	&\MIIa{1.06}			&\MIIa{}						&\MIIa{}\\
\hlineB{2}
\multirow{2}{*}{Insula}
&\MIIb{{Ri\#1}}	&\MIIb{1/1}		&\MIIb{0.49}	&\MIIb{-0.63}			&\MIIb{}							&\MIIb{}\\
&\MIIa{Ipro}		&\MIIa{1/1}		&\MIIa{0.00}	&\MIIa{-1.03}			&\MIIa{}							&\MIIa{}\\
\hlineB{2}
\multicolumn{7}{V{2}cV{2}}{\MIIIa{Module \#3}}\\
\hlineB{2}
\Hd{Lobe/nuclei}
&\Hd{Subdivision}			&\Hd{Frac.}		&\Hd{$\langle p\rangle$} &\Hd{$\langle z\rangle$}	&\Hd{Notable regions}		&\Hd{Known function}\\
\hlineB{2}
\multirow{2}{*}{Temporal lobe}							
&\MIIIa{{TCv}}	&\MIIIa{6/10}	&\MIIIa{0.16}	&\MIIIa{0.82}	&\MIIIa{35, 36c, 36r}	&\MIIIa{visual perception \& memory of objects~\cite{Murray2007}}\\
&\MIIIb{{PHC}}	&\MIIIb{11/14}	&\MIIIb{0.46}	&\MIIIb{0.09}	&\MIIIb{TH}							&\MIIIb{spatial memory~\cite{Malkova2003}}\\
&\MIIIa{{Hip}}	&\MIIIa{3/3}	&\MIIIa{0.28}	&\MIIIa{-0.38}	&\MIIIa{}							&\MIIIa{spatial cognition~\cite{Matsumura1999, Courellis2019} and recognition memory~\cite{Jutras2010}}\\
&\MIIIb{TE}			&\MIIIb{3/8}	&\MIIIb{0.65}	&\MIIIb{0.17}	&\MIIIb{}			&\MIIIb{}\\
\hlineB{2}
\multirow{1}{*}{Parietal lobe}							
&\MIIIa{S2}			&\MIIIa{1/2}	&\MIIIa{0.50}	&\MIIIa{-1.16}	&\MIIIa{}  &\MIIIa{}\\
\hlineB{2}
\multirow{4}{*}{Basal Ganglia}							
&\MIIIb{{Amyg}}	&\MIIIb{22/22}	&\MIIIb{0.20}	&\MIIIb{0.1}	&\MIIIb{}					&\MIIIb{emotional response~\cite{Weiskrantz1956, Davis1992}}\\
&\MIIIa{SN}			&\MIIIa{1/1}	&\MIIIa{0.61}	&\MIIIa{-0.88}	&\MIIIa{}					&\MIIIa{}\\
&\MIIIb{STR}		&\MIIIb{1/4}	&\MIIIb{0.57}	&\MIIIb{-0.80}	&\MIIIb{Cd\_t}	&\MIIIb{reinforcement learning~\cite{Apicella1991}}\\
\hlineB{2}
\multirow{2}{*}{\makecell{Cingulate\\ gyrus}}							
&\MIIIa{Area 26}	&\MIIIa{2/2}	&\MIIIa{0.57}	&\MIIIa{-0.88}	&\MIIIa{}				&\MIIIa{}\\
&\MIIIb{Area 25}	&\MIIIb{1/1}	&\MIIIb{0.65}	&\MIIIb{2.48}	&\MIIIb{}				&\MIIIb{}\\
\hlineB{2}
\multirow{3}{*}{Insula}							
&\MIIIa{Ig\#1}		&\MIIIa{1/1}	&\MIIIa{0.60}	&\MIIIa{0.83}	&\MIIIa{}				&\MIIIa{}\\
&\MIIIb{Pi\#1}		&\MIIIb{1/1}	&\MIIIb{0.64}	&\MIIIb{0.19}	&\MIIIb{}				&\MIIIb{}\\
&\MIIIa{IA}			&\MIIIa{1/5}	&\MIIIa{0.70}	&\MIIIa{0.90}	&\MIIIa{}			&\MIIIa{}\\
\hlineB{2}
\multirow{1}{*}{Hypothalamus}							
&\MIIIb{Hyp}		&\MIIIb{1/1}	&\MIIIb{0.69}	&\MIIIb{-0.80}	&\MIIIb{}				&\MIIIb{}\\
\hlineB{2}
\multicolumn{7}{V{2}cV{2}}{\MIVa{Module \#4}}\\
\hlineB{2}
\Hd{Lobe/nuclei}
&\Hd{Subdivision}		&\Hd{Frac.}	&\Hd{$\langle p\rangle$} &\Hd{$\langle z\rangle$}	&\Hd{Notable region}		&\Hd{Known function}\\
\hlineB{2}
\multirow{1}{*}{Frontal lobe}							
&\MIVa{{PFC}}	&\MIVa{11/36}	&\MIVa{0.55}	&\MIVa{0.25}	&\MIVa{46d, 46v}	&\MIVa{working memory~\cite{Petrides1991, Petrides1995}}\\
\hlineB{2}
\multirow{2}{*}{Temporal lobe}							
&\MIVb{{STG}}	&\MIVb{11/11}	&\MIVb{0.39}	&\MIVb{0.17}	&\MIVb{A1}						&\MIVb{auditory cortex~\cite{Morel1993, Romanski2009}}\\
&\MIVa{STS}		&\MIVa{5/11}	&\MIVa{0.52}	&\MIVa{-0.04}	&\MIVa{TPOc, TAa, Pga}	&\MIVa{complex sound processing~\cite{Barraclough2005, Romanski2009}} \\
\hlineB{2}
\multirow{1}{*}{Parietal lobe}							
&\MIVb{Area 7}	&\MIVb{2/7}		&\MIVb{0.69}	&\MIVb{1.40}	&\MIVb{PG\#1}	&\MIVb{somato-motor coordination~\cite{Andersen1990}} \\
\hlineB{2}
\multirow{5}{*}{Thalamus}							
&\MIVa{{GN}}	&\MIVa{1/2}		&\MIVa{0.00}	&\MIVa{-0.53}	&\MIVa{MG}		&\MIVa{auditory information relay~\cite{Kandel2000}} \\
&\MIVb{IL\#2}	&\MIVb{1/5}		&\MIVb{0.66}	&\MIVb{-0.67}	&\MIVb{}			&\MIVb{} \\
&\MIVa{PN}		&\MIVa{2/2}		&\MIVa{0.56}	&\MIVa{-0.08}	&\MIVa{}				&\MIVa{} \\
&\MIVb{Pul\#1}	&\MIVb{3/12}	&\MIVb{0.00}	&\MIVb{-1.22}	&\MIVb{}			&\MIVb{} \\
&\MIVa{MD}		&\MIVa{1/6}		&\MIVa{0.66}	&\MIVa{-0.60}	&\MIVa{}			&\MIVa{} \\
\hlineB{2}
\multirow{2}{*}{\makecell{Cingulate\\ gyrus}}							
&\multirow{1}{*}{\MIVb{Area 26}}	&\multirow{2}{*}{\MIVb{2/4}}		&\multirow{2}{*}{\MIVb{0.30}}	&\multirow{2}{*}{\MIVb{-0.81}}	&\multirow{2}{*}{\MIVb{}}			&\multirow{2}{*}{\MIVb{}} \\
& \MIVb{}& \MIVb{}& \MIVb{}& \MIVb{}& \MIVb{}&\MIVb{} \\
\hlineB{2}
\multicolumn{7}{V{2}cV{2}}{\MVa{Module \#5}}\\
\hlineB{2}
\Hd{Lobe/nuclei}
&\Hd{Subdivision}		&\Hd{Frac.}		&\Hd{$\langle p\rangle$} &\Hd{$\langle z\rangle$}	&\Hd{Notable region}	&\Hd{Known function}\\
\hlineB{2}
\multirow{1}{*}{Frontal lobe}							
&\MVa{{PFC}}			&\MVa{2/36}	&\MVa{0.28}	&\MVa{-0.30}	&\MVa{45A, 8Ac}			& \MVa{saccadic guidance (frontal eye field)~\cite{Schall2004}}\\
\hlineB{2}
\multirow{3}{*}{Temporal lobe}
&\MVb{{TE}}			&\MVb{4/7}	&\MVb{0.47}	&\MVb{0.03}	&\MVb{CITv, PITd, PITv,TEm} & \MVb{ventral visual pathway~\cite{Goodale1992, Goodale2018}}\\
&\MVa{{STS}}			&\MVa{5/11}	&\MVa{0.44}	&\MVa{1.28}	&\MVa{MT, MST, FST}		& \MVa{dorsal visual pathway~\cite{Goodale1992, Goodale2018}}\\
\hlineB{2}
\multirow{1}{*}{Parietal lobe}							
&\MVb{{PCip}}		&\MVb{4/6}		&\MVb{0.43}	&\MVb{0.06}	&\MVb{LIP, VIP, PIP}		&\MVb{visual attention~\cite{Duhamel1998, Pesaran2002}}\\
\hlineB{2}
\multirow{1}{*}{Occipital lobe}							
&\MVa{{VAC}}		&\MVa{12/13}	&\MVa{0.11}	&\MVa{-0.23}	&\MVa{V3A, V3B, V6}		&\MVa{visual cortex~\cite{Kandel2000}}\\
&\MVb{{V1}}		&\MVb{1/1}		&\MVb{0.25}	&\MVb{1.28}	&\MVb{}						&\MVb{primary visual cortex~\cite{Kandel2000}}\\
&\MVa{{V2}}		&\MVa{1/1}		&\MVa{0.39}	&\MVa{3.15}	&\MVa{}						&\MVa{secondary visual cortex~\cite{Kandel2000}}\\
\hlineB{2}
\multirow{1}{*}{Thalamus}							
&\MVb{AN}			&\MVb{1/4}		&\MVb{0.69}	&\MVb{-0.51}	&\MVb{}					&\MVb{} \\
&\MVa{{GN}}		&\MVa{1/2}		&\MVa{0.28}	&\MVa{-0.59}	&\MVa{LGN}					&\MVa{visual information relay~\cite{Kandel2000}}\\
&\MVb{Pul\#1}		&\MVb{8/12}		&\MVb{0.06}	&\MVb{-0.70}	&\MVb{}	&\MVb{visual processing~\cite{Smelser2001}} \\
\hlineB{2}
\multirow{3}{*}{Basal Ganglia}
&\MVa{Gpe}			&\MVa{1/1}		&\MVa{0.67}	&\MVa{-0.84}	&\MVa{}				&\MVa{} \\
&\MVb{STR}			&\MVb{1/4}		&\MVb{0.65}	&\MVb{-0.10}	&\MVb{Cd\_g}		&\MVb{reinforcement learning~\cite{Apicella1991}} \\
\hlineB{2}	
\multirow{1}{*}{Mid brain}							
&\MVa{{MB}}		&\MVa{1/1}		&\MVa{0.30}	&\MVa{-0.19}	&\MVa{} 				&\MVa{} \\
\hlineB{2}
\end{longtable}

\begin{figure}[tbp]
\centering
    \includegraphics[scale=1.1]{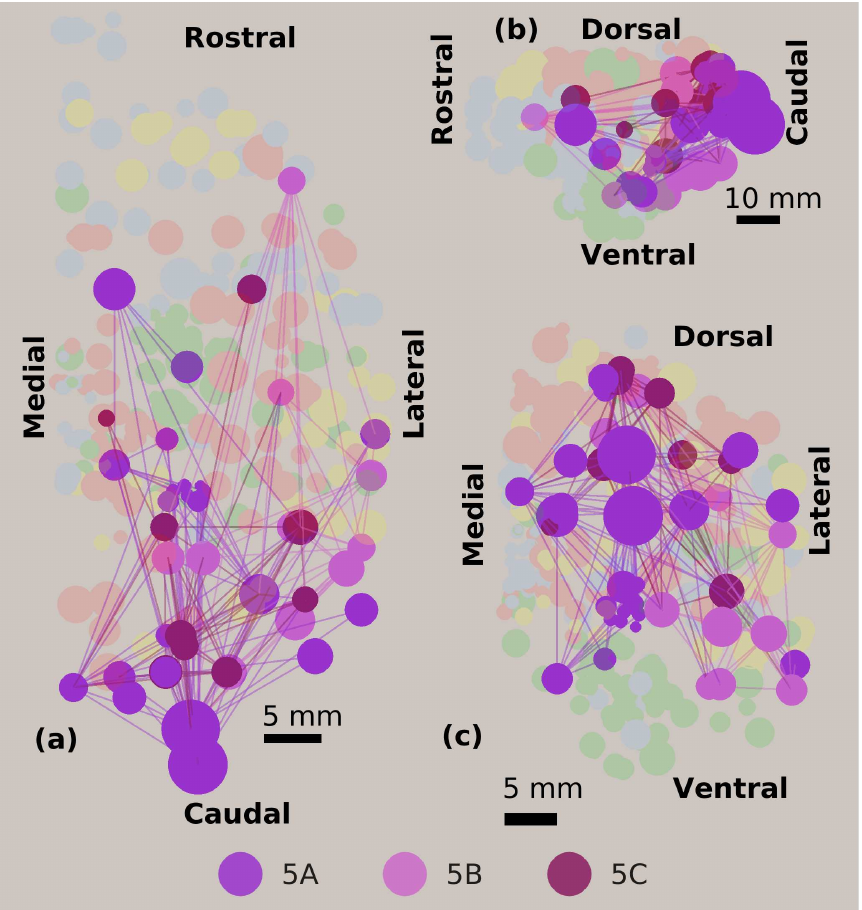}
    \qquad
    \begin{tabular}[b]{|c|c|c|} \hline
 \Hd{5A} & \Hd{5B} & \Hd{5C} \\ \hline
  \MVa{LGN} & \MVa{8Ac} & \MVa{45A} \\
  \MVb{PIl-s} & \MVb{VIP} & \MVb{LIPe} \\
  \MVa{PIp} & \MVa{FST} & \MVa{LIPi} \\
  \MVb{PIm} & \MVb{MSTp} & \MVb{CITv} \\
  \MVa{PIl} & \MVa{MSTd} & \MVa{PITd} \\
  \MVb{PIc} & \MVb{V3A} & \MVb{PITv} \\
  \MVa{PLa\#1} & \MVa{V4t} & \MVa{IPa} \\
  \MVb{PLvl} & \MVb{DP} & \MVb{V3v} \\
  \MVa{PLvm} & \MVa{LD\#1} & \MVa{V4v} \\
  \MVb{PIP\#1} & \MVb{} & \MVb{VOT} \\
  \MVa{TEm} & \MVa{} & \MVa{} \\
  \MVb{MT} & \MVb{} & \MVb{} \\
  \MVa{V3d} & \MVa{} & \MVa{} \\
  \MVb{DLr} & \MVb{} & \MVb{} \\
  \MVa{DLc} & \MVa{} & \MVa{} \\
  \MVb{VPP} & \MVb{} & \MVb{} \\
  \MVa{DI\#1} & \MVa{} & \MVa{} \\
  \MVb{V6} & \MVb{} & \MVb{} \\
  \MVa{V1} & \MVa{} & \MVa{} \\
  \MVb{V2} & \MVb{} & \MVb{} \\
  \MVa{MB\#2} & \MVa{} & \MVa{} \\
  \MVb{Cd\_g} & \MVb{} & \MVb{} \\
  \MVa{GPe} & \MVa{} & \MVa{} \\
   \hline
    \end{tabular}
\caption{The network of brain regions shown in (a) horizontal, (b) sagittal and (c) coronal projections, indicating that the nodes in module \#5 (highlighted)
can be further grouped into three sub-modules. 
The sub-modular membership of each node of module \#5 is represented by its color (see
color key at the bottom) with the list of brain regions belonging to each of the three sub-modules shown in the table in the right. 
Sub-module \#5A is seen to comprise primary visual regions and subcoritcal regions, 
while sub-modules \#5B and \#5C contain regions that belong to the ventral and dorsal visual pathways, respectively.
The node sizes provide a representation of the relative volumes of the corresponding brain regions
(the spatial scale being indicated by the horizontal bar in each panel). The spatial positions of the nodes are specified by the three-dimensional stereotaxic coordinates of the
corresponding regions. Links indicate the directed nerve tracts connecting pairs of brain regions, and are colored in accordance with their source nodes.}
\label{fig:FigS7}
\end{figure}

\FloatBarrier
\clearpage
The structure-function correlation associated with the mesoscopic organization, can be seen not only at the level of modules (as indicated by the Table~ref{tab:tab3} above)
but can be extended even further. As mentioned in the main text, we have sbjected module $\#5$ to further partitioning which yields three
sub-modules. Fig.~\ref{fig:FigS7} shows the nodes in module $\#5$ that belong to these sub-modules. We find that they ae associated with
distinct functionalities, with 5A containing the visual
cortex and almost all the sub-cortical components,
while the regions
identified with different visual processing pathways, viz.,
the ventral and dorsal streams belong to 5B and 5C, respectively.

\clearpage
\FloatBarrier

\subsection{Categorization of nodes in terms of inter- intra-modular connectivity}
\begin{figure}[hbp]
\centering
\includegraphics[width=0.99\linewidth]{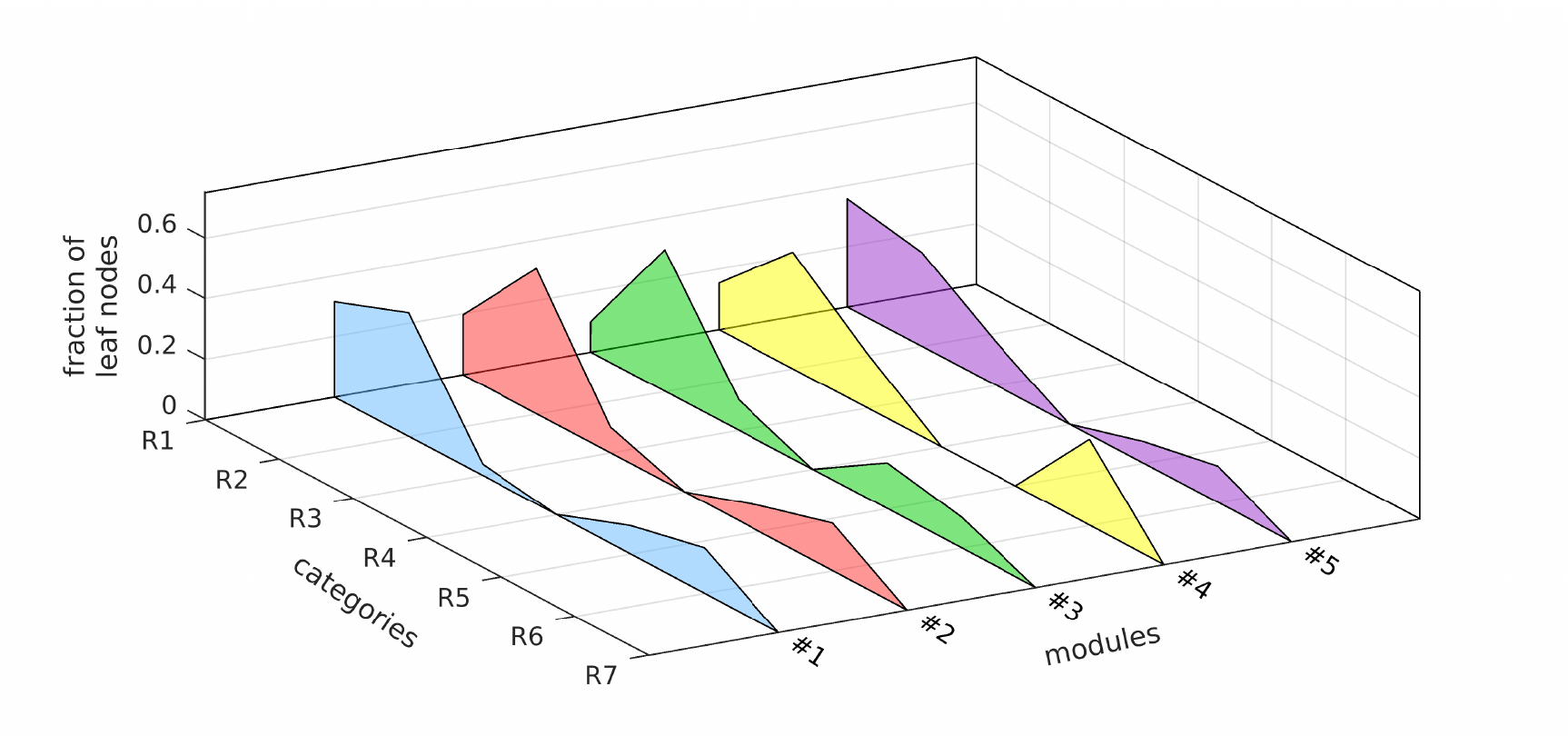}
\includegraphics[width=0.99\linewidth]{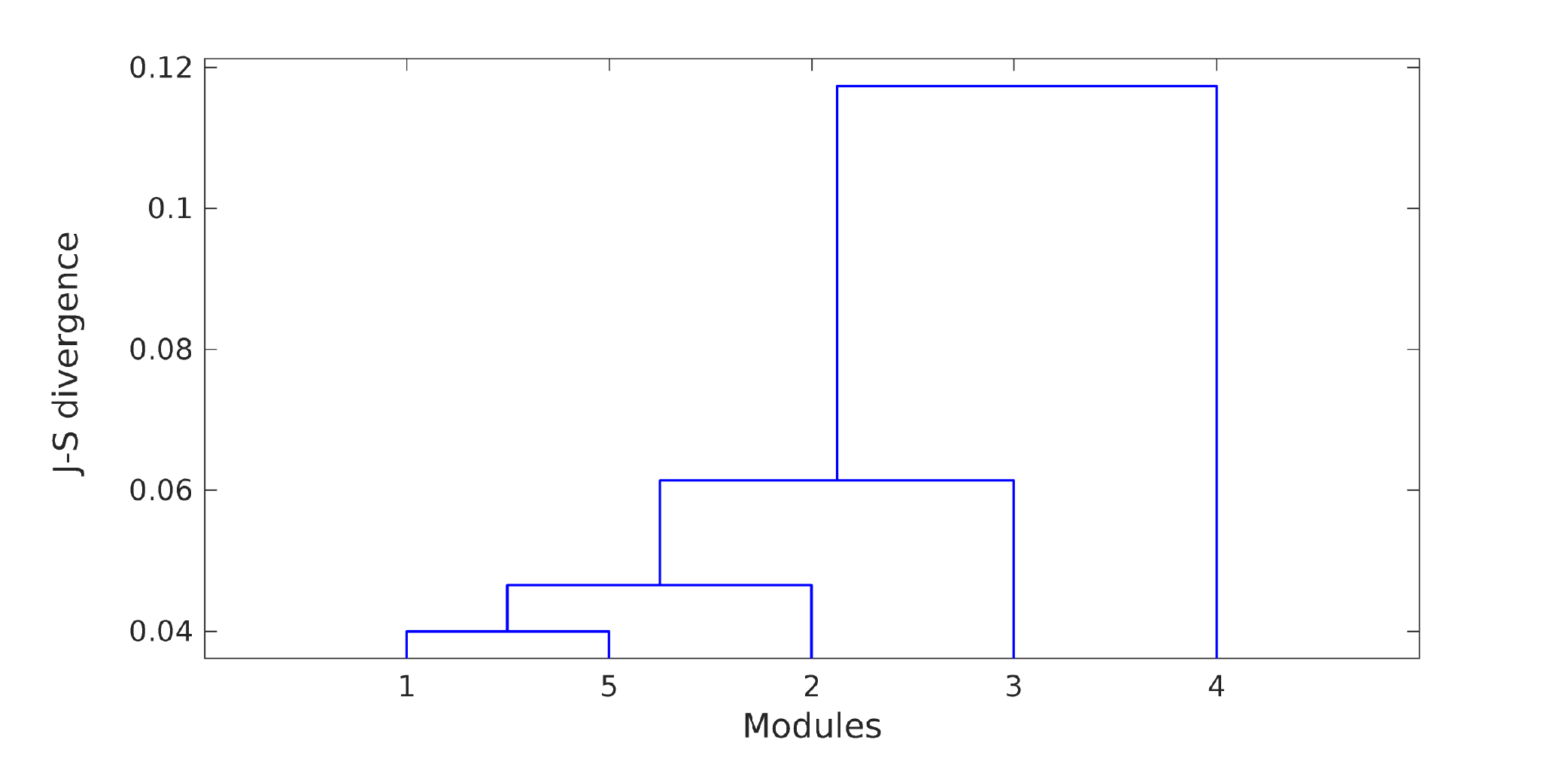}
\caption{The distribution of the regions across the different categories R1-R7 (see Fig. 2 in main text) is similar for different modules (top),
with the sole exception of module $\#4$ which does not possess any provincial hub (R5) nodes. This is illustrated in the dendrogram (bottom)
that represents the extent of similarity between these distributions, quantified by the Jensen-Shannon divergence, for the different modules.
}
\label{fig:FigS8}
\end{figure}

As described in the main text, the role played by each of the brain regions in the mesoscopic organization of the connectome can be
classified into seven categories according to their intra- and inter-modular connectivity,  viz., R1: ultra-peripheral, R2: peripheral, R3:
satellite connector, R4: kinless, R5: provincial hub, R6: connector hub, and R7: global hub (note that 
there are no regions in the Macaque brain belonging to the categories R4 and R7). With the exception of module
$\#4$ which has no region playing the role of a provincial hub, each module has a qualitatively similar distribution of its regions across
these categories (Fig.~\ref{fig:FigS8},~top).

\begin{figure}[hbp]
\includegraphics[width=0.99\linewidth]{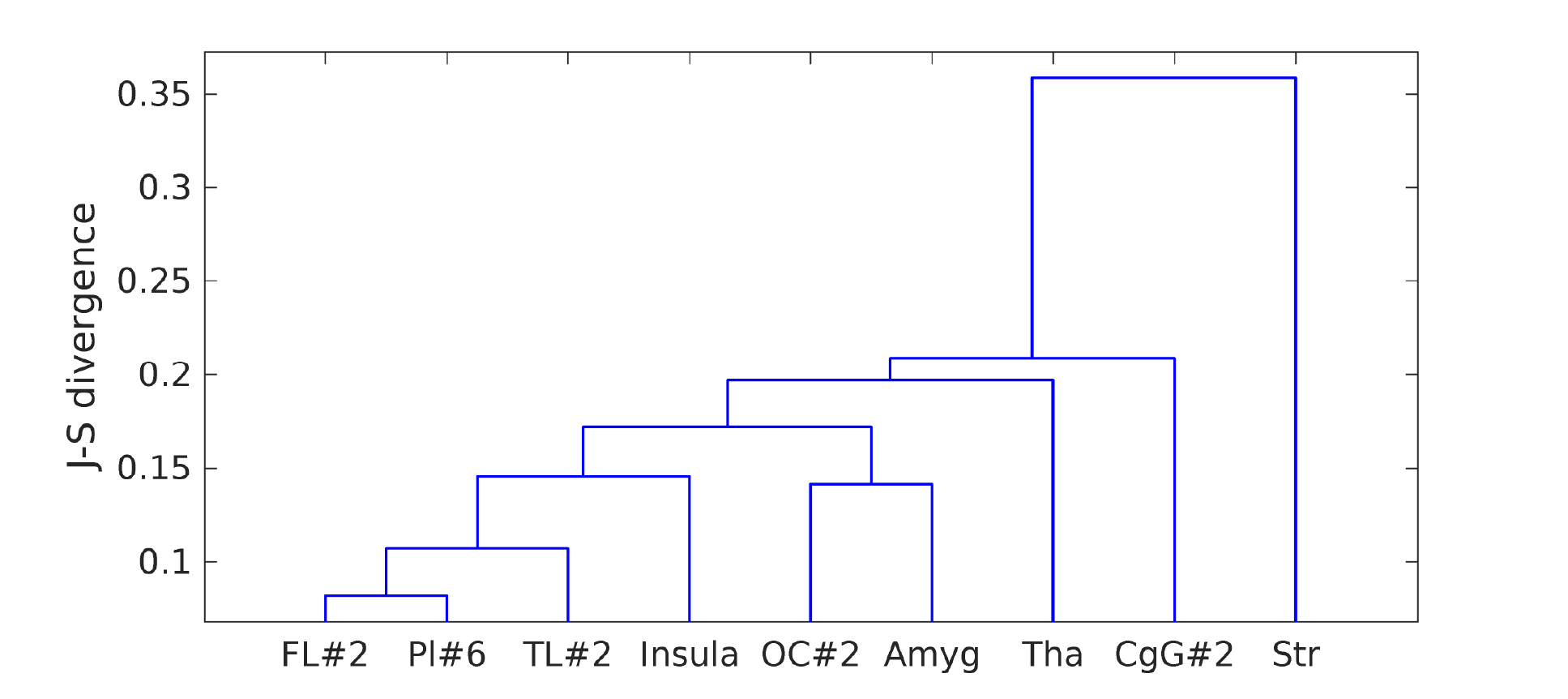}
\caption{Dendrogram illustrating the extent of similarity between several anatomical subdivisions of the brain,
viz., {\em Tha}: Thalamus, {\em FL\#2}: Frontal Lobe, {\em P1\#6}: Parietal Lobe, {\em CgG\#2}: Cingulate Gyrus, Insula, {\em TL\#2}: 
Temporal Lobe, {\em OC\#2}: Occipital Lobe, {\em Amyg}:
Amygdala and {\em STR}: Striatum, in terms of the distribution across the categories R1-R7 of their constituent regions (see
Fig.~2~(b) in main text). As in Fig.~\ref{fig:FigS8}~(bottom), the difference between the distributions corresponding to two subdivisions
is measured using the Jensen-Shannon divergence.
}
\label{fig:FigS9}
\end{figure}

\FloatBarrier
The extent of similarity between the modules is represented by the dendrogram shown in
Fig.~\ref{fig:FigS8}~(bottom) in which the
distance between the distributions across $x\in \{{\rm R1, \ldots, R7}\}$ for two modules $A$ and $B$ , viz., $P_A (x)$ and 
$P_B (x)$, is measured in terms 
of the Jensen-Shannon divergence: 
$$JSD (P_A, P_B)  = \frac{1}{2} \sum_x \left[ P_A(x) \ln P_A(x) + P_B(x) \ln P_B(x)
- \left\{P_A(x)+P_B(x)\right\} \ln \left(\frac{P_A(x)+P_B(x)}{2} \right) \right].$$
This quantification of the difference between a pair of distributions is also employed in Fig.~\ref{fig:FigS9} to indicate the extent of similarity between
the various anatomical subdivisions of the brain, the corresponding distributions of whose regions across the categories R1-R7 is
shown in Fig.~2~(b) in the main text. 

\begin{figure}[tbp]
\begin{center}
\includegraphics[width=0.5\linewidth]{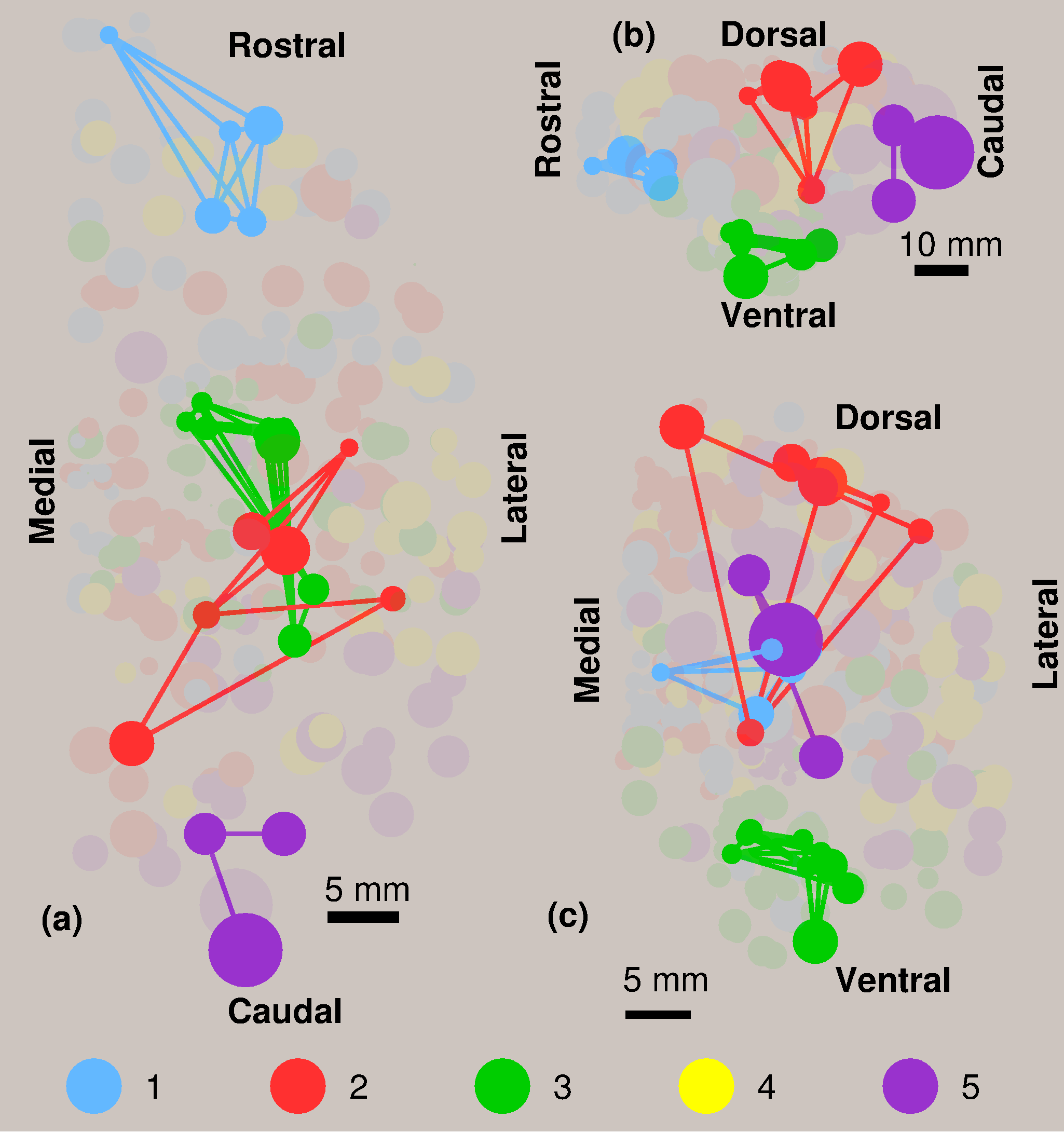}
\end{center}
\caption{The network of brain regions in (a) horizontal, (b) sagittal and (c) coronal projections, showing that connections between 
provincial hubs (highlighted nodes) are
localized within each module [Figure 2~(d) in the main text is identical to panel (b) above].
}
\label{fig:FigS10}
\end{figure}

\FloatBarrier
Next, we focus on how regions belonging to specific categories connect to each other. In the main text, 
we mention that within each module, the provincial  hubs (R5) connect with each other significantly more often
than expected by chance. This intra-modular connectivity between the R5 nodes can be clearly seen from Fig.~\ref{fig:FigS10}, where these
nodes are highlighted and their colors indicate the modules to which they belong (see color key at the bottom). 
Note that one of the projections shown here is identical to Fig.~2~(d).

\FloatBarrier

\begin{figure}[tbp]
\includegraphics[width=1\linewidth]{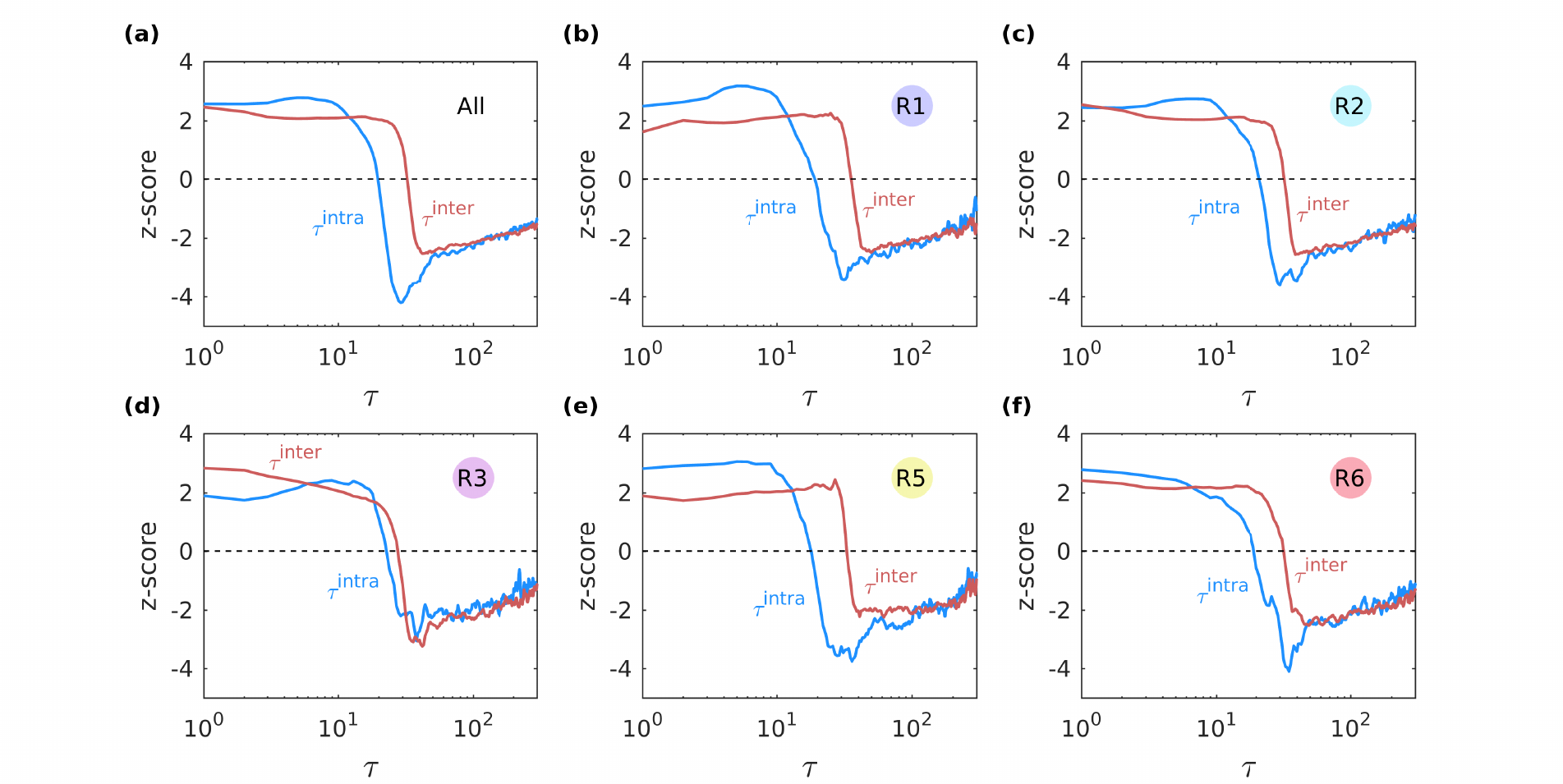}
\caption{To see how the different categories R1-R7 of brain regions allow spreading to occur faster in the empirical brain network than 
in equivalent randomized networks, we compare the case where the source node can belong to any category (a) with those where the 
source is either ultra-peripheral R1 (b), peripheral R2 (c), satellite connector R3 (d), provincial hub R5 (e), or global hub R6 (f). 
The $z$-score indicates that there is a statistically significant shift in the empirical distribution towards lower values of $\tau$ in
all cases. However, while for R3 the increase in the rate of spreading is similar irrespective of whether the target is in the same module or 
in a different one, we observe that there is a relatively larger shift at lower values for 
$\tau^{intra}$ as compared to $\tau^{inter}$ for most of the other categories (in particular, R1 and R5 ).
Indeed, the latter behavior dominates when we consider sources across all categories [see panel (a)].  
Note that panels (d) and (e) are identical to see Fig.~2~(g-h) in the main text. 
}
\label{fig:FigS11}
\end{figure}

\FloatBarrier
In the main text we have described our investigation of the role played by regions belonging to different
categories R1-R7 in facilitating information transmission. For this we simulate diffusive propagation within
and between modules and obtain the distribution of first passage times for random walks
between a source node and a target node. Fig.~\ref{fig:FigS11}~(a) shows that the rate of diffusion
in the connectome is enhanced both within a module and between modules (as indicated by the
statistically significant shift - measured in terms of $z$-score - in the empirical distributions for both $\tau^{intra}$ and $\tau^{inter}$ 
towards  lower values) as compared to that seen in equivalent randomized networks.

Fig.~\ref{fig:FigS11}~(b-f) show how nodes belonging to categories R1, R2, R3, R5 and R6 (respectively), which have 
distinct intra- and
inter-modular connectivity roles, contribute to enhancing communication
in the connectome. In each case the source node belongs to the respective category and
we quantify the difference in the distributions of both $\tau^{intra}$ and $\tau^{inter}$ from that obtained
from randomized surrogates. We observe that for source nodes of most categories, with the exception of satellite connectors R3, 
the  increase in the rate of diffusion within a module, compared to that in the
surrogate networks, is even higher than the increase in the
rate of diffusion across modules.

\section{Spatial dependence of connectivity and modular organization}
In the main text, we have stated that modular organization of the connectome is not primarily driven by constraints imposed by
the physical distance between the brain regions. This is established by using three classes of surrogate 
random network ensembles to investigate how spatial embedding affects the modular decomposition of a
network, with all the regions occupying the same positions in physical space as in the Macaque connectome. 
The three ensembles we have chosen for our investigation are specified by the dependence of the connection
probability $P$ between regions on the physical distance $d$ between them, viz., (i) $P \sim d^0$, i.e., independent of 
the distance, (ii) $P \sim 1/d$, i.e., power-law dependence as in the empirical network, and (iii) $P \sim \exp(-d)$, i.e., exponential
dependence, for which the constraint of distance
most strongly affects the probability of connection. For each category, we have generated $100$ different networks that
have identical numbers of nodes and links as the empirical connectome.  Subsequently, we subject these networks to community
detection techniques using information about the connection topology alone, as well as space-independent modular
decomposition which explicitly accounts for the dependence of $P$ on $d$ (see main text for details). 

\begin{figure}[tbp]
\includegraphics[width=1\linewidth]{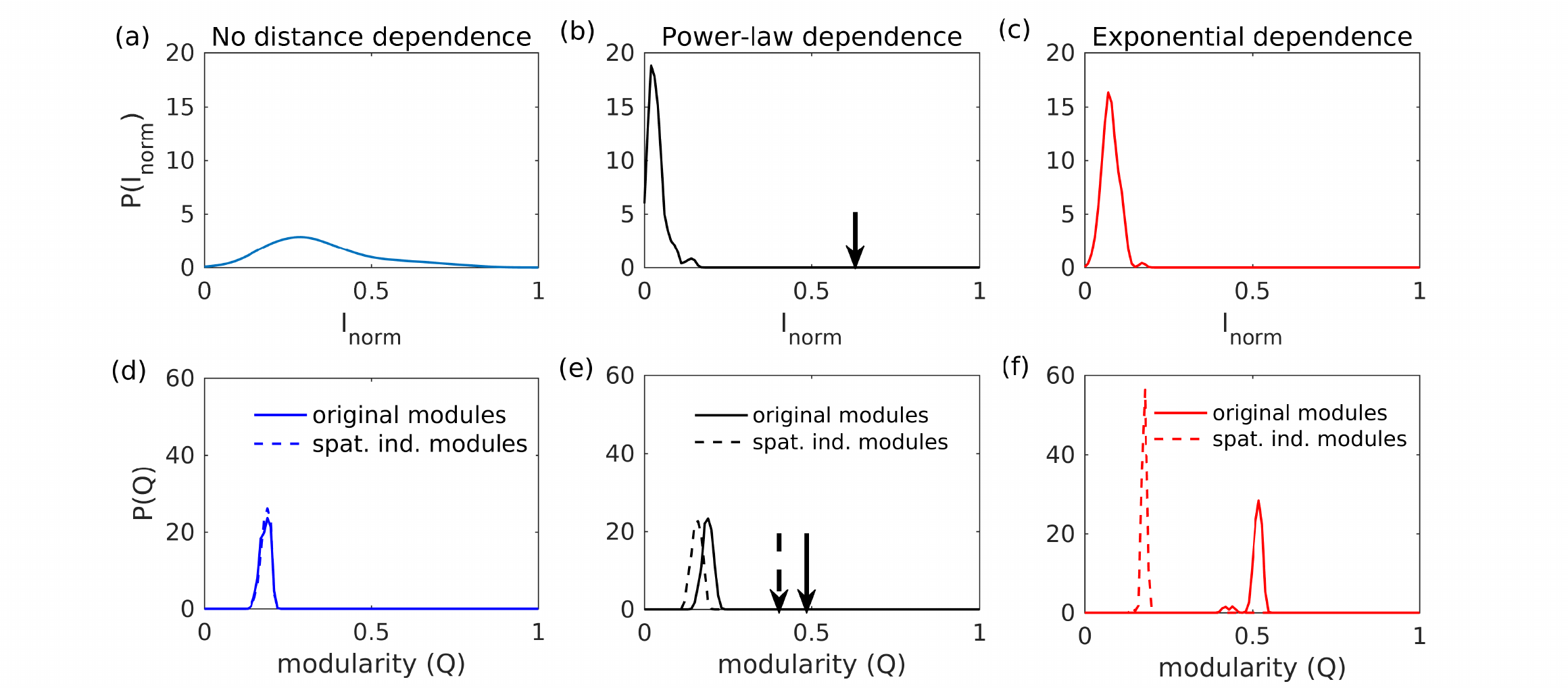}
\caption{The distributions of (top row) the degree of similarity between the topological and space-independent modular partitionings
of a network as measured by normalized mutual information $I_{norm}$ between them, and (bottom row) the corresponding
values for the modularity $Q$ obtained using the two methods, for 
three types of random surrogate network ensembles. These are distinguished by the dependence of connection probability $P$
between a pair of brain regions on the physical distance $d$ between them, viz.,  $P \sim d^0$ [(a) and (d)],  $P \sim 1/d$
[(b) and (e)] and $P \sim \exp(-d)$ [(c) and (f)].
As the Macaque connectome we have investigated also exhibits a power-law dependence, viz., $P \sim 1/d$, similar to that 
examined in (b) and (e), we have indicated in those panels the corresponding values for the empirical network (arrows).
} 
\label{fig:FigS12}
\end{figure}

Fig.~\ref{fig:FigS12} shows how the modular nature of the networks belonging to each of the three ensembles described above
vary upon two  approaches for identifying the modules, viz., (i) using the topological information about the connections alone, and (ii) 
employing a space-independent partitioning
that takes into account the dependence of the probability of connections between regions on the physical distance between them.
The similarity between the modules obtained using these two methods is measured using normalized mutual information $I_{norm}$
(see Methods in the main text). Note that, if the two types of partitionings yield identical modules 
then $I_{norm}=1$, while $I_{norm}=0$ implies maximal dissimilarity. 
Without any spatial dependence, the identified modules arise through fluctuations alone, and hence the similarity between the partitions
obtained by the two methods will be entirely stochastic in nature, resulting in the broad distribution for
$I_{norm}$ seen in panel (a).
In contrast, the ensemble underlying the distribution shown in panel (b) has an inverse relation between connection
probability and physical distance, as in the empirical network. The value of $I_{norm}$ obtained for 
the empirical network (indicated by the arrow) is seen to be significantly larger than those for the random ensemble.
This suggests that had the modules arisen exclusively from a distance-dependent constraint on connections, the topological and
space-independent approaches would have yielded highly dissimilar partitionings. 
Qualitatively similar results are obtained when the dependence of connection probability on physical distance is even stronger, 
viz., $P$ decaying exponentially with $d$ as in the case of the ensemble whose $I_{norm}$ distribution is shown in panel (c).  
The fact that partitioning the empirical network using either the topological or the space-independent approach results 
in relatively similar modular decompositions suggest that constraints other than those
related to physical distance plays a significant role in shaping the mesoscopic organization of the Macaque connectome.  
The results described above are supported by the corresponding distributions of the modularity $Q$ measured for the 
different partitionings obtained using each of the two approaches
(broken and solid curves in panels d-f). Thus, in the absence of any spatial dependence, the distributions of $Q$ obtained using the topological and the
space-independent approaches completely overlap [as seen in (d)]. When $P \sim d^{-1}$, the relatively weak spatial dependence
gives rise to marginally lower values of $Q$ for the partitionings obtained using the space-independent method, as
compared to those obtained using the topological information alone. This is seen to be true for both the
empirical network (broken and solid arrows) and the random ensembles [panel (e)].
With the stronger spatial dependence inherent in an exponentially decaying functional relation, we expect to see much larger
differences in the $Q$ values for the two types of partitionings, and this is indeed observed in the distributions shown in panel (f). 
Therefore, the more dominant the role of the constraint on physical distance in determining the connections, the more dissimilar 
the partitionings obtained by the two methods
and the larger the difference in the corresponding $Q$ values.

\begin{sidewaysfigure}[tbp]

\includegraphics[width=0.33\textwidth]{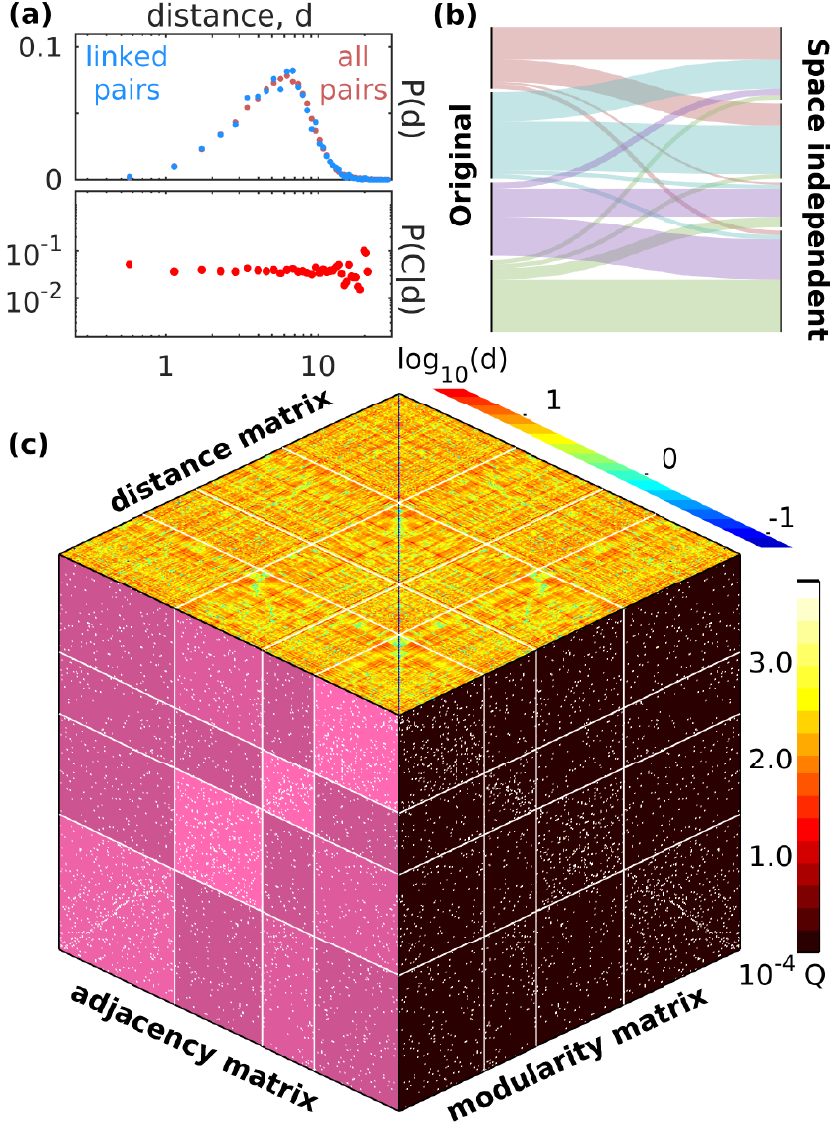}
\includegraphics[width=0.31\textwidth]{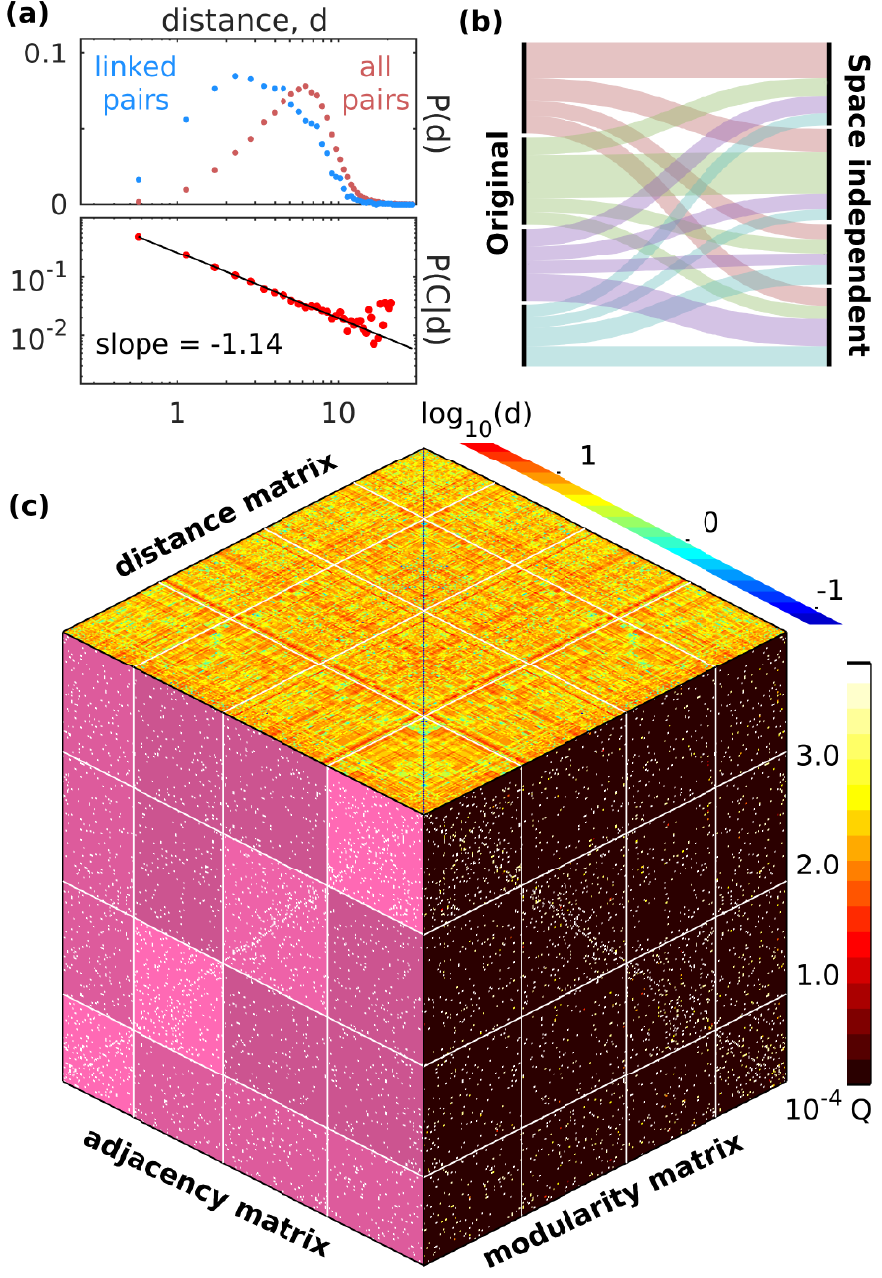}
\includegraphics[width=0.32\textwidth]{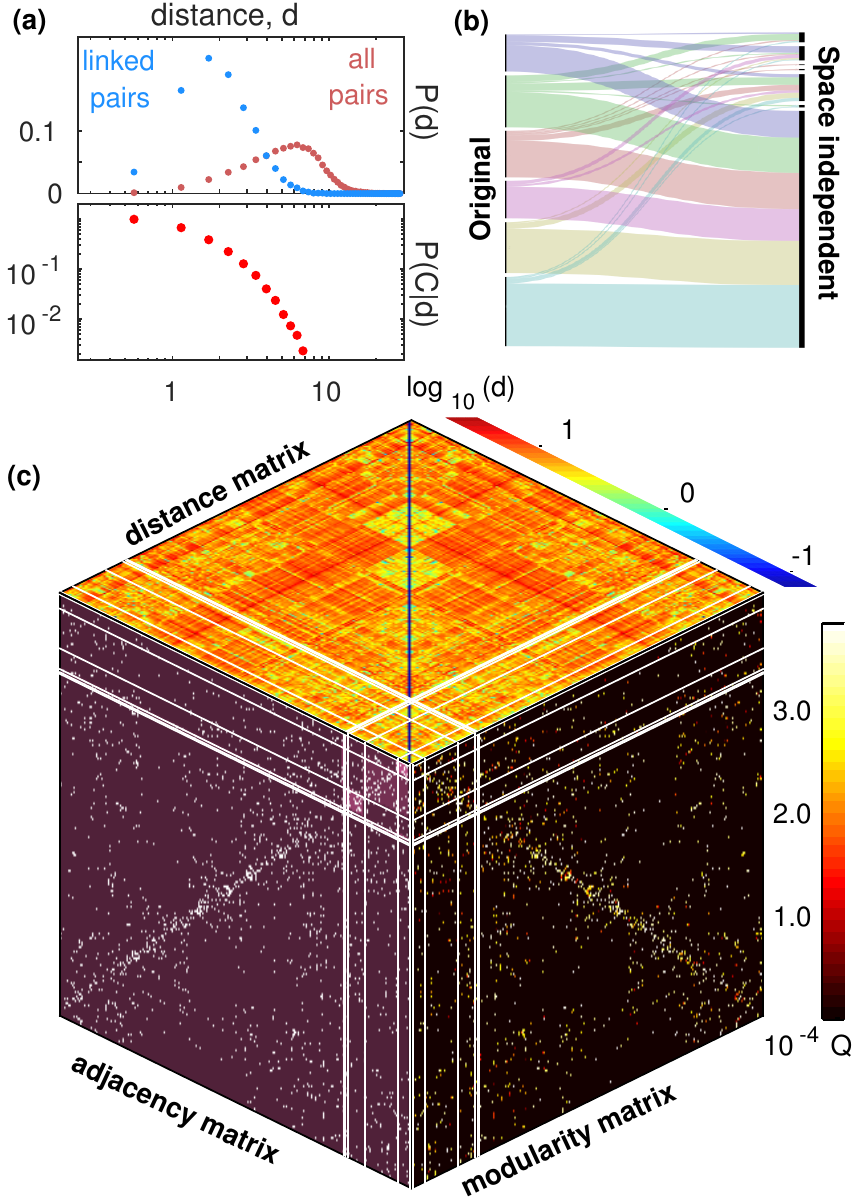}
\caption{The dependence of the modular organization of a network on the three-dimensional spatial arrangement
of the nodes shown for three different functional relations between the physical distance $d$ between a pair of nodes
and the connection probability $P$ between them, viz., (left) $P \sim d^0$, (center) $P \sim 1/d$, and (right) $P \sim \exp(-d)$.
For each functional relation, we show (a, top) the probability distribution of the physical distances $d$ between all pairs 
of nodes (red) contrasted with that of the connected pairs (blue), as well as (a, bottom) the variation of the connection probability
$P(C|d)$ with physical distance $d$ which characterizes the functional relation (the best fit power-law is indicated with a line, along with
its slope, in the central panel). The panels also contain
(b) visual
representations of the correspondence between the network modules determined
using exclusively information about the connection topology (``Original") and those
obtained from space-independent partitioning of the network into communities. on
the right). This alluvial diagram has been created using the online visualization tool
RAW~\cite{Mauri2017}.
Joint representation of the space-independent modular organization of each class of network
is shown in (c) using three matrices indicating adjacency $\{A_{ij}\}$ (left surface),
modularity $\{B_{ij}\}$ (normalized by total number of links $L$, right surface) and physical
distance $\{d_{ij}\}$ (top surface). As in Fig.~3~(b) of the main text,  the nodes are grouped into partitions corresponding to the
space-independent modules of the network with the boundaries indicated by solid
lines.
}
\label{fig:FigS13}
\end{sidewaysfigure}

Fig.~\ref{fig:FigS13} illustrates the space independent modular organization of random networks with the three different types of spatial constraints as described above, employing the representation used in Fig.~3. 
The distributions of the physical distances $d$ between the nodes and the nature of variation of the connection probability
between nodes with $d$ are shown in the panels (a) for networks the role of spatial constraint on connectivity is (left) absent, viz., $P \sim d^0$, 
(center) weak, viz., $P \sim 1/d$, and (right) strong, viz., $P \sim \exp(-d)$.
Comparison of the modules obtained using the information about the connection topology alone and those determined using the 
space-independent method (shown in the panels (b) for each of the networks) indicate that in the absence of any dependence of $P$ on $d$
(left) the partitions overlap to a large extent. Introducing a role for the spatial constraint in determining the connections result in the two types
of partitionings differing substantially. This is seen for the power-law dependence of $P$ on $d$ (center), but most prominently when
$P$ decays exponentially with $d$ (right). For the latter case, a single large module is seen to encompass the bulk of the network.
As in this case the topological modules arise primarily from the spatial constraint on connections between nodes, on taking this
dependence on $d$ into account in the space-independent method the mesoscopic structure becomes relatively homogeneous.  
The panels (c) show the joint representation of the adjacency, modularity and physical distance matrices for each of the the three networks
[as per the convention used in Fig.~3~(b) of the main text]. The partitions obtained by using the space-independent method are indicated 
by bounding lines in each matrix. Note that when we take into account the constraint that physical distance places on connectivity between regions,
the partitioning results in modules that exhibit only a marginally higher density of connections within them (compared to the overall connection
density). This is expected as the modules observed in the topological arrangement of connections in these random networks arise exclusively from the spatial constraint, and therefore the space-independent
method should render the networks relatively homogeneous. Thus, the observation
of non-trivial modules in the empirical network upon partitioning it with the space-independent method suggests that the observed mesoscopic
organization of the Macaque connectome cannot be explained exclusively by the spatial layout of the regions. 

\FloatBarrier
\bibliography{references}
\bibliographystyle{pnas-new}
